%% file: GKPchannel_final.tex
\documentclass[aps
,pra
,twocolumn
,letterpaper
,10pt
,tightenlines
,superscriptaddress
,amsmath
,amssymb
,amsfonts
,eqsecnum
,nofootinbib
,tightenlines
,longbibliography
,nobalancelastpage
,notitlepage]{revtex4-2}

\usepackage{nag}
\usepackage{graphicx}
\usepackage{amsthm}
\usepackage{tabulary}
\usepackage[colorlinks=true,
citecolor=blue,
    linkcolor=blue,
    filecolor=magenta,      
    urlcolor=cyan]{hyperref}
\usepackage{qcircuit}
\usepackage{mathtools}
\usepackage{bm}
\usepackage{xcolor}
\usepackage{braket}
\usepackage{datetime}
\usepackage{stackrel}
\usepackage{stmaryrd}
\usepackage{dsfont}
\usepackage{qcircuit}
\usepackage[english]{babel}
\usepackage[normalem]{ulem}
\usepackage{xargs}
\usepackage{cleveref}
\usepackage[linesnumbered,ruled,vlined]{algorithm2e}
\usepackage{tikz}
\usepackage{soul}

\let\originaladdtocontents\addtocontents
\newcommand{\notoc}{\renewcommand{\addtocontents}[2]{}}
\newcommand{\toc}{\let\addtocontents\originaladdtocontents}

\newcommand{\damppar}[0]{\beta}
\newcommand{\corrvec}[0]{\vec{c}(\mu)} %
\newcommand{\corrvect}[1]{\vec{c}(\mu_{#1})} %
\newcommand{\bitsvec}[0]{\vec{b}}

\newcommand{\Nop}[0]{\op{N}}

\newcommand{\EPR}{\text{EPR}}
\newcommand{\projGKP}{\bar{\Pi}}

\newcommand{\tel}{\text{tel}}

\newcommand{\custommeter}[2][]{ %
    \raisebox{-0.9cm}{ %
        \begin{tikzpicture}
            \draw[] (0,0) 
                -- (0.4,0) 
                arc[start angle=270, end angle=360, radius=0.2cm] 
                -- (0.6,0.23)
                arc[start angle=0, end angle=90, radius=0.2cm] 
                -- (0,0.43) -- cycle;
            \node at (0.3,0.2) {#2};
              \node[anchor=north west] at (0.48, 0.22) {\scriptsize #1};
        \end{tikzpicture}
    }
}

\newcommand{\custommeterwide}[2][]{ %
    \raisebox{-0.9cm}{ %
        \begin{tikzpicture}[xscale=1.1,yscale=1.2]
            \draw[] (0,0) 
                -- (0.80,0) 
                arc[start angle=270, end angle=360, radius=0.2cm] 
                -- (1.00,0.23)
                arc[start angle=0, end angle=90, radius=0.2cm] 
                -- (0,0.43) -- cycle;
            \node at (0.45,0.2) {#2};
              \node[anchor=north west] at (0.88, 0.22) {\scriptsize #1};
        \end{tikzpicture}
    }
}

\newcommand{\PT}{\text{PTd}}

\newcommand{\oldQ}{\bar{Q}}
\newcommand{\newQ}{\bar{Q}^{\PT}}

\newcommand{\GKPPauli}{\bar{\sigma}}
\newcommand{\GKPPaulicorr}[1]{\bar{\sigma}_{\vec{a}(\mu){#1}}}
\newcommand{\twirlnorm}{\mathfrak{n}_\damppar}

\newcommand{\te}{\text{equiv}}

\DeclareMathOperator*{\sumint}{%
\mathchoice%
{\ooalign{$\displaystyle\sum$\cr\hidewidth$\displaystyle\int$\hidewidth\cr}}
  {\ooalign{\raisebox{.14\height}{\scalebox{.7}{$\textstyle\sum$}}\cr\hidewidth$\textstyle\int$\hidewidth\cr}}
  {\ooalign{\raisebox{.2\height}{\scalebox{.6}{$\scriptstyle\sum$}}\cr$\scriptstyle\int$\cr}}
  {\ooalign{\raisebox{.2\height}{\scalebox{.6}{$\scriptstyle\sum$}}\cr$\scriptstyle\int$\cr}}
}

\usepackage{soul}

\date{\today}

\input{preamble}

\newcommand{\GKP}{\text{GKP}}

\begin{document}

\title{Logical channels in approximate Gottesman-Kitaev-Preskill error correction}

\author{Mahnaz Jafarzadeh}
\affiliation{Xanadu Quantum Technologies, Toronto, ON M5G 2C8, Canada}
\author{Jonathan Conrad}
\affiliation{Institute of Computer and Communication Sciences, EPFL, Lausanne CH-1015, Switzerland}
\author{Rafael N. Alexander}
\affiliation{Xanadu Quantum Technologies, Toronto, ON M5G 2C8, Canada}
\author{Ben Q. Baragiola}
\affiliation{Xanadu Quantum Technologies, Toronto, ON M5G 2C8, Canada}
\affiliation{Centre for Quantum Computation and Communication Technology, School of Science, RMIT University, Melbourne, VIC 3000, Australia}
\affiliation{Yukawa Institute for Theoretical Physics, Kyoto University, Kitashirakawa Oiwakecho, Sakyo-ku, Kyoto 606-8502, Japan}

\begin{abstract}
    The Gottesman-Kitaev-Preskill (GKP) encoding is a top contender among bosonic codes for fault-tolerant quantum computation. However, analysis of the code is complicated by the fact that finite-energy code states leak out of the ideal GKP code space and are not orthogonal. 
    We analyze a variant of the GKP stabilizer measurement circuit that virtually projects onto the ideal GKP code space between rounds of error correction even when damped, finite-energy GKP states are used. This allows us to identify logical maps between projectors;
    however, due to finite-energy effects, these maps fail to resolve completely positive, trace-preserving (CPTP) channels on the logical GKP code space. 
    We present two solutions to this problem based on channel twirling the damping operator. 
    The first uses symmetries of standard binning (SB) decoding to \emph{passively} twirl over the full stabilizer group. Doing so converts small amounts of damping into stochastic Gaussian random noise (GRN). 
    The second uses \emph{active} twirling over a minimal set of representative Pauli shifts that keeps the energy in the code finite and allows for arbitrary decoding.
    This approach is not limited to small damping 
    and allows the study of decoding more general than SB, which can be optimized to the noise in the circuit.
    Focusing on damping, we compare decoding strategies tailored to different levels of effective squeezing. While our results indicate that SB decoding is suboptimal for finite-energy GKP states, we observe
    that the advantage of optimized decoding shrinks as the energy in the code increases. Moreover, the performance of both decoding strategies converges to that of the stabilizer-twirled GRN logical channel. 
    These studies provide stronger arguments for two commonplace procedures in the analysis of GKP error correction in the fault-tolerance regime: $(i)$ using stochastically shifted GKP states in place of coherently damped ones, and $(ii)$ the use of SB decoding. 
    
\end{abstract}

\maketitle
\tableofcontents

\section{Introduction}
Multiple pathways lead towards fault-tolerant quantum computing. One promising approach involves encoding logical quantum information into bosonic modes~\cite{chuang1997bosoniccode,cochrane1999cat,albert2018performance,grimsmo2020rotation,eczoo_oscillators,cai2021bosoniccodes}, which are ubiquitous in nature. Among the many ways to do so, the Gottesman-Kitaev-Preskill (GKP) code stands out for its all-Gaussian gate set, Pauli measurements via homodyne detection~\cite{gottesman2001encoding}, tolerance to loss~\cite{albert2018performance,noh2019capacity,Zheng2024pureloss}, and seamless integration into optical cluster states~\cite{menicucci2014fault,tzitrin2021fault,pantaleoni2021hidden,walshe2024totl,ostergaard2025octorail}. Although the GKP code was developed almost 20 years before technology was prepared for it, recent advances in quantum control across platforms has ushered an era of experimental bosonic-code research. GKP states are routinely prepared in trapped ions~\cite{fluhmann2019encoding,matsos2024bosonic}, superconducting cavities~\cite{campagne2020quantum,Sivak2023GKPbreakeven,lachance2024GKPautonomous}, and optics~\cite{asavanant2024opticalGKP}, and various implementations have been proposed for other physical systems~\cite{baranes2023freeelectron,kendell2024GKPcoldatoms}. 

Working with GKP codes presents the challenge that ideal code words contain infinite energy, and finite-energy approximations come with a penalty: physical code words are not orthogonal. 
Fortunately, this does not appear to be a hindrance to fault-tolerant operation~\cite{menicucci2014fault,Matsuura24} when GKP codes are used in conjunction with other error correcting codes in fixed-qubit~\cite{vuillot2019quantum, noh2019fault} or measurement-based~\cite{bourassa2020blueprint, larsen2020architecture, fukui2017analog, fukui2019high} settings. 
However, typical analyses of fault tolerance assume that approximate GKP states can be treated as ideal states with additive Gaussian noise, which we call Gaussian random noise (GRN) GKP states. This assumption, while not physical, is commonly argued~\cite{menicucci2014fault,noh2019fault} as safe when dealing with high-quality GKP states, since it does not significantly alter the measurement statistics
obtained for translation-invariant observables in short prepare-and-measure circuits. The real dynamics of CV circuits however is more complicated, and the treatment of approximate GKP states under the GRN approximation requires more scrutiny.

This is a reflection of a key issue that arises when assuming physicality of a QEC protocol: the GKP stabilizers cannot be measured exactly, such that finite-energy GKP error correction fails to ``digitize'' CV noise into elementary displacements. This issue is observed in implementations of stabilizer measurements via phase estimation, where a strongly coupled qubit in superconducting or ion-based systems is used to measure modular position and momentum one bit at a time~\cite{Terhal_2016,valahu2024sensing}, and a continuous syndrome cannot be extracted in finite time and space. 
A more comprehensive technique involves preparation of auxiliary modes in GKP states and using these to read out the full modular syndromes in one fell swoop. Here, a consequence of finite energy is that the coherent noise from the auxiliary GKP states pollutes subsequent measurement outcomes, and syndrome extraction weakly measures logical information. Together, these induce potentially non-Pauli and non-unitary logical effects on the encoded information. 
Moreover, this noise drives the state out of the GKP code space, which is an obstacle to describing dynamics on the logical qubit alone.

Ultimately, the logically encoded information in the GKP code is our primary concern; the fact that finite-energy GKP codes reside in the larger CV Hilbert space is a nuisance we yearn to avoid.
In this work, we present a procedure that sheds the CV description and describes the logical effects of performing GKP error correction with finite-energy GKP states.
Using teleportation-based GKP error correction supplemented with state preparation and readout circuits, the procedure yields completely-positive, trace-preserving (CPTP) quantum channels on the encoded qubit information for all values of the damping parameter, $\damppar$, that quantifies the quality of the auxiliary states. 
In the limit $\damppar \rightarrow 0$, the logical channel arising from high-quality GKP auxiliary states approximates the identity channel one expects from ideal GKP states.
The beauty of using logical channels is that the infinite dimensionality of the CV space can be ignored --- all the dynamics are described in the greatly reduced Hilbert space of the qubit(s)~\cite{Doherty2002logicalchannel, Beale2021logicalchannel}. 
A faithful GKP-logical description of a CV circuit simplifies its analysis and 
can be used for various tasks including optimized decoding for various CV noises, logical randomized benchmarking~\cite{combes2017LRB}, and large-scale simulations of GKP-based quantum computers~\cite{Xanadu2025scaling}.

The key to embedding logical channels is first identifying exact GKP projectors lurking within repeated finite-energy error correction, see Sec.~\ref{sec:GKP-logicalmaps}. 
However, direct projection into the ideal GKP code space fails to produce a CPTP channel due both to the trace-decreasing nature of the approximation map (\emph{e.g.} the non-unitary damping operator)~\cite{conrad2019masters,harris2024} and to correlations between adjacent rounds of correction. 
We provide two distinct resolutions to these issues by channel twirling the damping operator in two different ways, one passive and one active.

The first method, presented in Sec.~\ref{sec:GRNGKPchannel}, uses the fact that standard binning (SB) decoding is invariant under stabilizer shifts to channel twirl the damping operator over the full stabilizer group. 
Replacing the damping operator by its channel-twirled version is possible due to the symmetry of the measurements and decoder and does not require modifying the circuit; in this sense it is a \emph{passive} twirl. A result is that the binned homodyne outcomes are identical in either case, which is the primary concern for decoded GKP measurements.
When the damping is small, this twirl converts damped GKP states in the teleportation circuit into GRN GKP states~\cite{noh2019fault}, from which we derive a CPTP logical Pauli channel.

The second method, presented in Sec.~\ref{sec:logical_channel}, is based on a finite-energy channel twirl using a representative set of four shifts that covers the Pauli group. Using a recovery that is aware of the twirling, we show that this minimal set suffices to promote the logical maps to genuine Kraus operators for a CPTP logical channel without relying on stabilizer symmetries (inherent to the SB decoder), which allows freedom to choose a logical decoder.
This procedure requires modifying the circuit by shifting the location of the damping operators in the GKP Bell pairs; in this sense, it is an \emph{active} twirl.

Equipped with these tools, we compare the average gate fidelity for three logical channels arising from GKP Bell pairs composed of: GRN GKP states with SB decoding (arising from the stabilizer twirl), finite-energy twirled GKP states with SB decoding, and finite-energy twirled GKP states with optimal lookup decoding. 
When the damping is low, we find that all three perform nearly identically, with the GRN model slightly outperforming the others. This conclusion is not unexpected, but our analysis places the use of GRN GKP states for fault-tolerance analyses on firmer footing. That is, while previous studies used stabilizer twirling phenomenologically to convert damped GKP states to GRN GKP states~\cite{noh2019fault}, we show that it is in fact a passive byproduct when SB decoding is employed.

Finally, we provide a host of useful calculations in the various Appendices. Notably, the physical construction in the main text uses teleportation-based error correction with beam splitters serving as the entangling gate. In Appendix~\ref{Sec:otherflavors}, we provide the recipe to adapt this construction to other methods of GKP error correction that use controlled gates instead. We show a direct equivalence between Knill and Steane error correction with controlled gates and identify the noisy GKP projectors within.

\section{The GKP code} \label{sec:GKPcode}

A single bosonic mode is a quantum system with an infinite dimensional Hilbert space. Quantum states can be described over continuous or discrete bases. Here we will make use of the continuum of (improper) eigenstates of position $\op{q}$ and momentum $\op{p}$ operators, satisfying $[\op{q}, \op{p}] = i$ (with $\hbar =1$), and the related creation/annihilation operators $\op{a} = (\op{q} + i \op{p})/\sqrt{2}$ and $\op{a}^\dagger = (\op{q} - i \op{p})/\sqrt{2}$ satisfying $[\op{a}, \op{a}^{\dagger}]=1$. Position eigenstates are denoted $\ket{s}_{q}$ and satisfy $\op{q}\ket{s}_{q} = s\ket{s}_{q}$, $\prescript{}{q}{\braket{t\vert s}}_{q} = \delta(s-t)$ with $s, t \in \mathbb{R}$. Momentum eigenstates are related to these by the Fourier transform operation $\op F \coloneqq e^{i\pi (\op{q}^{2} + \op{p}^{2})/4}$, via $\ket{s}_{p} \coloneqq \op F \ket{s}_{q}$, where $s\in\mathbb{R}$. 

Ideal, computational-basis, square-lattice GKP code states $\ket{\bar{\psi}}$ can be expressed as infinite superpositions of position or momentum eigenstates~\cite{gottesman2001encoding} 
\begin{align} \label{eq:defwords}
    \ket{\overline{j}} &\coloneqq \sum_{n \in \mathbb{Z}}\ket{(2n+j)\sqrt{\pi}}_q 
     =  \sum_{n \in \mathbb{Z}} (-1)^{nj} \ket{n\sqrt{\pi}}_p \blk
\end{align}
up to arbitrary normalization factors.
GKP states are stabilized by integer powers of the operators
\begin{subequations} \label{GKPstabs}
\begin{align}
    \op S_{X} &= \op D (\sqrt{2\pi}) = e^{-i 2\sqrt{\pi} \op p}\, , \\
    \op S_{Z} &= \op D (i\sqrt{2\pi}) = e^{i 2\sqrt{\pi} \op q}
    \, ,
\end{align}
\end{subequations}
where $\op D(\alpha)$ with $\alpha = \alpha_R + i \alpha_I \in \mathbb{C}$ is a standard Glauber displacement operator,
\begin{align}
    \op D(\alpha) &\coloneqq e^{\alpha \op{a}^{\dagger} - \alpha^{*} \op a} 
    = e^{i \sqrt{2} (-\alpha_R \op{p} + \alpha_I \op{q})} 
    \, ,
\end{align}
satisfying the composition and commutation rule
\begin{align} \label{eq:displacementbraiding}
    \op D(\alpha) \op D(\beta) &= e^{ \frac{1}{2} \omega(\alpha, \beta) } \op D(\alpha + \beta)
    =e^{\omega(\alpha, \beta) } \op D( \beta) \op D(\alpha),
\end{align}
where $\omega(\alpha, \beta) \coloneqq \alpha \beta^{*} - \alpha^{*} \beta = 2 i \Im (\alpha\beta^*)$. 

The projector onto the two-dimensional GKP codespace, $\projGKP$, can be equivalently expressed in several useful ways: in terms of ideal basis states, the stabilizers in Eq.~\eqref{GKPstabs}, or Dirac combs in the quadrature operators~\cite{walshe2020continuousvariable},
\begin{align}
    \projGKP &= \ket{\overline{0}}\!\!\bra{\overline{0}}+\ket{\overline{1}}\!\!\bra{\overline{1}} 
    \propto 
    \sum_{m, n \in \mathbb{Z} } (\op S_{X})^{m} (\op S_{Z})^{n} 
    \\
    & \propto \Sh_{\sqrt{\pi}}(\op{q}) \Sh_{\sqrt{\pi}}(\op{p})
    = \Sh_{\sqrt{\pi}}(\op{p}) \Sh_{\sqrt{\pi}}(\op{q}), \label{eq:Shacommutation}
\end{align}
such that $\projGKP^2 = \projGKP$ fixes the code space.\footnote{The non-normalizability of ideal GKP states implies $\op \projGKP^2 = c \projGKP$, where $c = \braket{\overline{0}|\overline{0}}$ is an infinite constant. In physical settings, we generally ignore global constants. One can be more methodical by defining parametrized, damped projectors $\tilde{\Pi}^{\damppar} \coloneqq \Nop \projGKP \Nop$, such that $\| \tilde{\Pi}^2 - \tilde{\Pi} \|\leq f(\damppar) $ for small $\damppar$ and some function $f$.} 
The expressions on the final line follow from the stabilizer representation using the Fourier series representation of a Dirac comb, 
    \begin{align} \label{eq:DiracComb}
        \Sh_{T}(x) \coloneqq \sqrt{T} \sum_{k=-\infty}^\infty \delta(x-kT) = \frac{1}{ \sqrt{T} }\sum_{k=-\infty}^\infty e^{i \frac{2\pi}{T} k x } .
    \end{align}
Note that in this work, we use an overbar for states, operators, and maps that act only in the GKP subspace, \emph{e.g.} $\projGKP$.

\subsection{Finite-energy GKP states} \label{sec:EncodingIntro}

Producing exact ideal GKP states is physically impossible---the code words in Eq.~(\ref{eq:defwords}) have infinite energy and cannot be normalized. Any physical realization of the GKP code use a set of states that approximates the ideal ones. 
While ideal GKP states are all alike; every type of approximate GKP state is approximate in its own way.\footnote{``\emph{Happy families are all alike; every unhappy family is unhappy in its own way.}'' --Lev Tolstoy.} 
Various methods for generating different types of approximate GKP states have been proposed, and some have been implemented in physical systems~\cite{fluhmann2019encoding, campagne2020quantum, vasconcelos2010all, su2019conversion, tzitrin2020progress, weigand2018generating, eaton2019non, motes2017encoding, pirandola2006generating}.  
In this work, we focus on a convenient theoretical model that applies a non-unitary damping operator
    \begin{align} \label{eq:Nop}
        \Nop \coloneqq e^{-\damppar \hat{n}}
    \end{align}
to an ideal GKP state and then normalizes it.\footnote{We focus on finite energy GKP states with this simple, non-unitary damping operator largely for purposes of clarity and notational convenience. The results described here apply more generally to the case where $\hat{N}$ is a single-mode Gaussian CP map that preserves the parity operator (equivalently, the origin of phase space), and our results can be straightforwardly adapted to include all Gaussian CP maps. This includes single-mode Gaussian unitaries that change the square-GKP lattice to any other single-mode lattice.} This procedure regularizes the states in two intertwined ways: it damps support on large number states and smears out each delta function in the wave function (and in the Wigner function) into a narrow Gaussian. A more subtle effect of the envelope is a shifting of the peak locations towards the origin, such that peaks are no longer separated by multiples of $\sqrt{\pi}$.

Consider an ideal GKP qubit state $  \ket{\psi_\text{qubit}} = c_0 \ket{\bar 0} + c_1 \ket{\bar 1}$ with $|c_0|^2 + |c_1|^2 = 1$.
A standard damping-operator encoding, $\ket{\psi_\text{qubit}} \rightarrow e^{-\damppar \op{n}} \big( c_0 \ket{\bar 0} +  c_1 \ket{\bar 1} \big)$, produces a state that is not normalized, because $e^{-\damppar \op n}$ is trace decreasing.\footnote{One way to generate this map is to mix the state with an auxiliary vacuum mode on a beamsplitter. Detecting vacuum in the auxiliary mode applies the energy damping operator to the state~\cite{harris2024}} 
The state is normalized by its trace, $\mathfrak{n}_\psi = \sum_{j,k} c_j^* c_k \bra{ \bar j}  \op{N}^2 \ket{\bar k}$, giving
    \begin{align} \label{eq:finitestates}
        \ket{\bar{\psi}_\damppar} 
        \coloneqq 
        \tfrac{1}{\sqrt{\mathfrak{n}_\psi}} \Nop \ket{\bar{\psi}}
        \, .
    \end{align}
The normalization depends on the encoded qubit state itself, as is apparent in the appearance of the coefficients $c_j$ in $\mathfrak{n}_\psi$. This means that the encoding (from states to states) is not a linear map, and orthogonality is not preserved: $\inprod{\bar{\psi}_\damppar}{\bar{\phi}_\damppar} \neq 0$ even when the unencoded states are orthogonal $\inprod{\psi}{\phi} = 0$.

\subsubsection{Relating damping to random displacement noise}

The translational symmetry of GKP states provides a natural robustness to small displacement errors.  More concretely, a generic single-mode bosonic noise channel $\mathcal{E}$ can be expanded in the displacement basis as \cite{gottesman2001encoding, conrad2020twirling}
\begin{align}
    \mathcal{E}  = \int d^{2}\alpha d^{2} \alpha' \, 
 c(\alpha, \alpha') \op D(\alpha) \cdot \op D^{\dagger}(\alpha') \label{eq:dispE}
 \, ,
 \end{align}
 with characteristic function $c(\alpha, \alpha')$. 
Consider an ideal GKP state subject to a channel where $c(\alpha, \alpha')$ is non-zero only when $\sqrt{2}\max\{\abs{\alpha_R}, \abs{\alpha_I}\}\leq \frac{\sqrt{\pi}}{2}$ and $\sqrt{2}\max\{\abs{\alpha'_R}, \abs{\alpha'_I}\}\leq \frac{\sqrt{\pi}}{2}$; 
that is, $\mathcal{E}$ only results in small displacements relative to the lattice spacing. 
When $c(\alpha,\alpha')$ is only approximately zero outside the domain $\sqrt{2}[-\frac{\sqrt{\pi}}{2},\frac{\sqrt{\pi}}{2}]^{\times 4}$, error correction succeeds in recovering the state with high probability and leaves only a slim chance of introducing a logical error~\cite{Matsuura24}.  This is the case for weak Gaussian noise of the form $c(\alpha, \alpha') \propto \exp(- \tfrac{1}{2} \vec{z}^{\text{T}} \mat \Sigma^{-1} \vec{z} )$,
where $\vec{z}^{\text{T}} \coloneqq (\alpha_R, \alpha'_R, \alpha_I, \alpha'_I)$, when the operator norm of the covariance matrix is small, \emph{e.g.} when $\norm{\boldsymbol\Sigma} < \frac{\pi}{8}$. 

Using the characteristic function for the damping operator, Eq.~\eqref{eq:charfun_damping}, the damping map $\mathcal{E}_\text{damp} \coloneqq \Nop \cdot \Nop$, is
    \begin{align}
    \mathcal{E}_\text{damp}
    &=
     \frac{1}{\pi^2(1-e^{-\damppar})^2}
     \int d^{2} \alpha d^{2} \alpha' \, e^{-\frac{\abs{\alpha}^{2} + \abs{\alpha'}^{2}}{2 \tanh \frac{\damppar}{2}} } \op{D}(\alpha) \cdot \hat{D}^{\dagger}(\alpha')
    \, , 
    \end{align}
where $\tanh \frac{\damppar}{2}= \frac{\damppar}{2} + O(\damppar^3)$
characterizes the variances in the characteristic function.
In phase space, this map transforms a Wigner function
by applying a Gaussian convolution in the position and momentum variables and a joint Gaussian envelope with zero mean. 
For an ideal GKP state $\bar{\rho}$ with Wigner function $W_{\bar{\rho}}$, this yields (see Appendix~\ref{appendix:dampingoperatorphasespace})
\begin{align}
    & W_{\mathcal{E}_\text{damp}( \bar{\rho} )} (q, p) \nonumber \\
    & \propto \label{eq:wigenvelope}
     G_{\frac{1}{2} t_{\damppar}^{-1}}(q,\, p) \int d^{2} \boldsymbol{\tau} \, G_{\frac{1}{2} t_{\damppar}} (\tau_1, \tau_2) W_{\overline{\rho}} (s_{\damppar}q-\tau_1, s_{\damppar}p-\tau_2) 
    \, ,
\end{align}
 where the variances are governed by $t_{\damppar} \coloneqq \tanh \damppar$ and its inverse, and we define a two-dimensional Gaussian function
    \begin{equation} 
    G_{\sigma^2} (x ,y) \coloneqq \frac{1}{2 \pi \sigma^2} e^{-\frac{x^{2} + y^{2}}{2\sigma^2} }.
    \label{eq:GaussianFunc}
    \end{equation}
Additionally, the factor $s_{\damppar} \coloneqq \sech\damppar=1+O(\damppar^2)$ indicates a slight shift of the GKP spike centers before the blurring and envelopes occur~\cite{matsuura2020equivalence}. 
The translational symmetry of the ideal GKP state is broken by the Gaussian envelopes, only being recovered in the limit of $\damppar\rightarrow 0$.

For many applications, such as the analysis of fault-tolerant quantum computing architectures~\cite{menicucci2014fault, bourassa2020blueprint, larsen2020architecture, fukui2017analog, fukui2019high}, 
it has been convenient to ignore this broken symmetry by modelling imperfect GKP states using a Gaussian random noise (GRN) channel, given by Eq.~\eqref{eq:dispE} with a diagonal, Gaussian kernel,
    \begin{align} \label{eq:GRN}
    \mathcal{E}_\text{GRN} 
    &= \frac{1}{2\pi \sigma^2} \int ds dt \, G_{\sigma^2}(s,t) e^{i t \op{q}} e^{-i s\op{p}} \cdot e^{i s \op{p}} e^{-i t \op{q}}
    \\
     &= \frac{1}{\pi \sigma^2} \int d^{2} \alpha \, G_{\sigma^2/2}(\alpha_R, \alpha_I) \op D(\alpha) \cdot  \op D^{\dagger}(\alpha) \label{eq:GRN_2}
    \, ,
    \end{align}
parameterized such that $\sigma^2$ characterizes the variance added to position and to momentum.
The Wigner function for $\bar{\rho}$ under the GRN channel,
\begin{align}
    W_{\mathcal{E}_\text{GRN}(\bar{\rho})} (q, p) 
    \propto
    \int d^{2} \boldsymbol{\tau} \, G_{\sigma^2} (\tau_1, \tau_2) W_{\overline{\rho}} (q-\tau_1, p-\tau_2) , 
     \label{eq:wignoenv}
\end{align}
has its translational symmetry preserved, and thus it retains infinite energy and non-normalizability. Setting the variance in the GRN channel to be 
    \begin{equation} \label{eq:GRNtoenvelope_comparison}
        \sigma^2 = \frac{1}{2} \tanh \damppar ,
    \end{equation} 
a comparison with Eq.~\eqref{eq:wigenvelope} reveals that the Wigner functions are identical up to the overall Gaussian envelope---which is indeed the broken translational symmetry under the damping channel. 
For high quality GKP states ($s_{\damppar} \approx 1$), one may approximate damped GKP states with GRN-blurred GKP states by twirling over the full set of GKP stabilizers~\cite{Noh2022lowoverhead, mensen2020phase, conrad2020twirling}.  
GRN-blurred GKP states are used in noise and threshold studies, because translational invariance simplifies the analysis. Since every spike in the Wigner function is an identical Gaussian, the noise properties of entire state can be extracted from the covariance matrix of a single spike~\cite{menicucci2014fault, walshe2021streamlined}. A full analysis using finite-energy, damped states, however, requires understanding the influence of the envelope, which is one of the main goals of this work.

\subsection{GKP error correction with finite-energy states}\label{sec:equivalence}

There are several ways of measuring the GKP stabilizers in Eqs.~(\ref{GKPstabs}). One method relies on coupling the continuous Hilbert space of the mode to the discrete Hilbert space of a qudit via, \emph{e.g.}, a conditional displacement~\cite{Terhal_2016, fluhmann2019encoding, campagne2020quantum, motes2017encoding}. Measurement of the qudit allows extraction of partial information about the syndrome via truncated quantum phase estimation and imprints digital information onto the oscillator. Another method is laid out in the original GKP paper~\cite{gottesman2001encoding}: entangle the mode with another continuous Hilbert space (an auxiliary mode), where either the interaction~\cite{terhal_review} or the auxiliary state~\cite{gottesman2001encoding} are structured such that the modular quadratures are measured by homodyne detection. 

Using auxiliary modes, two strategies are employed, Steane-type and Knill-type error correction (EC). In Steane EC, the data mode is sequentially coupled to two auxiliary states, which are then measured to extract the syndrome information. Steane EC requires non-logical corrections to restore the code space in addition to logical corrections as determined by a decoder. In Knill EC, the data mode is entangled with an encoded GKP Bell pair. 
A decoded logical Bell-measurement then teleports the noisy input state into the logical subspace on the output mode, which, up to a logical correction, acts as filter for correctable errors.

The logical CNOT gate for the GKP code requires active squeezing and is typically not native to CV platforms (such as optics).
However, it was discovered that the humble beam splitter, despite being a linear-optical element, performs the entangling operation between two modes needed to produce a GKP Bell pair~\cite{ walshe2020continuousvariable}
and more generally can be used to generate fault-tolerant cluster states for computation in combination with the GKP code%
~\cite{larsen2020architecture, tzitrin2021fault, walshe2023equivalent, walshe2024totl}.
We proceed here using beam splitter-based Knill EC, because it has the nice property that the damping operators decouple from the GKP projector~\cite{walshe2020continuousvariable}, discussed below. Steane and Knill EC based on CNOT gates are discussed in Appendix~\ref{Sec:otherflavors}.

The foundation of beam-splitter based Knill EC is a teleportation circuit acting on a (potentially noisy) input state $\op{\rho}_\text{in}$: 
\begin{equation} \label{Knillcircuitimage}
    \begin{split}
    \includegraphics[scale=0.3]{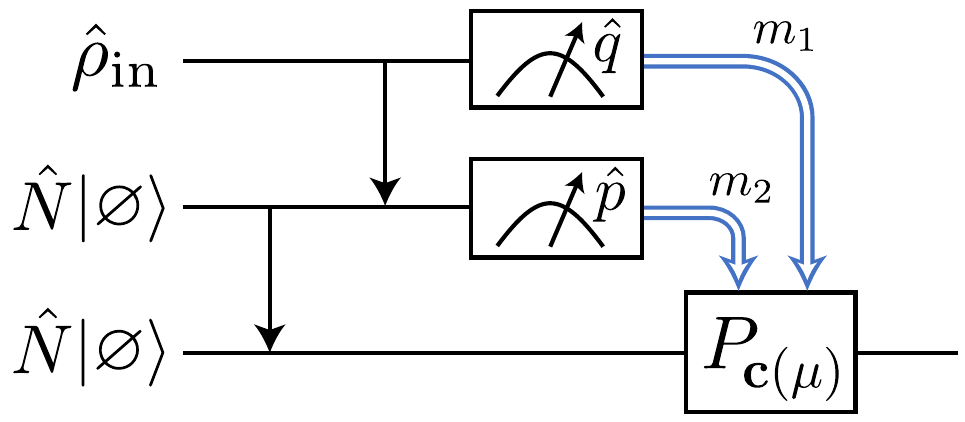}
    \end{split}
\end{equation}
where a balanced (50:50) beam splitter between mode $j$ and mode $k$ is defined as
    \begin{align}
        \begin{split}
        \raisebox{-1.3em}{$\op{B}_{jk} \coloneqq e^{-i \frac{\pi}{4} (\op{q}_j \op{p}_k - \op{p}_j \op{q}_k)}$ = \quad }
          \Qcircuit @C=1em @R=2.3em @! 
         {
         	& \bsbal{1} & \qw \\
         	& \qw       & \qw
  		  } 
        \end{split} \, .
    \end{align}
The outcomes of the position and momentum homodyne measurements, $m_1$ and $m_2$, extract the stabilizer syndromes.  We group them into a single complex number
    \begin{equation} \label{eq:mu}
        \mu \coloneqq m_1 + i m_2
        \,
    \end{equation}
to be fed into an unspecified decoder that determines the correction, discussed below. 

Due to the inherent squeezing in the beamsplitter, two grid states with periodicity $\sqrt{2\pi}$ in both position and momentum, often called \emph{qunaught} or \emph{sensor} states,
    \begin{align}
        \ket{\varnothing} \coloneqq \sum_{n \in \mathbb{Z}} \qket{\sqrt{2\pi}n} 
    \end{align}
are employed to make an ideal GKP Bell pair.
By applying the damping operator, Eq.~\eqref{eq:finitestates}, directly to these states before they are entangled, we create a noisy, finite-energy GKP Bell pair $\ket{\bar{\Phi}^+} \coloneqq \frac{1}{\sqrt{2}}(\ket{\bar 0} \otimes \ket{\bar 0} + \ket{\bar 1} \otimes \ket{\bar 1})$~\cite{walshe2020continuousvariable},
    \begin{equation} 
        \begin{split}
\label{noisybeamsplitterBellpair}
    \raisebox{-1em}{ $\ket{\bar{\Phi}_\damppar^+} \coloneqq$  }
    \qquad \qquad
    \Qcircuit @C=1em @R=2em 
    {
 &\lstick{\frac{1}{\sqrt{\mathfrak{n}_\varnothing}}\Nop \ket{\varnothing} } &\bsbal{1} &\qw  \\
 &\lstick{\frac{1}{\sqrt{\mathfrak{n}_\varnothing}}\Nop \ket{\varnothing}} &\qw       &\qw 
		} \, 
  \raisebox{-1em}{ $\displaystyle = \frac{1}{\mathfrak{n}_\varnothing}
 \op{N} \otimes \Nop \ket{\bar{\Phi}^+}$ }
         \end{split}
    \end{equation}
with normalization $\mathfrak{n}_\varnothing = \bra{\varnothing} \Nop^2 \ket{\varnothing}$ for each input state. The above relation follows from a feature of beam splitters that is not true of general CV gates: identical zero-mean Gaussian operations\footnote{Zero-mean refers to the fact that the operation preserves the phase space origin. Displacements can be included in $\hat{N}$ to extend Eq.~\eqref{eq:beamsplitter_gaussianopcommutation} to more general Gaussian CP maps, though the displacements before and after the beamsplitter will differ.} (such as $\op{N} = e^{-\damppar \op{n}}$) commute with a beamsplitter~\cite{fukui2021alloptical},
    \begin{equation} 
    \begin{split}
\label{eq:beamsplitter_gaussianopcommutation}
       \quad \quad 
       \Qcircuit @C=1em @R=0.8em 
        {
        &\gate{N} &\bsbal{1} &\qw  \\
        &\gate{N} &\qw       &\qw 
	} \, 
        &\raisebox{-1em}{\quad = \quad}
        \Qcircuit @C=1em @R=0.8em 
        {
        &\bsbal{1} &\gate{N} &\qw \\
        &\qw       &\gate{N} &\qw 
	} 
    \end{split} .
    \end{equation}

The final step in Circuit \eqref{Knillcircuitimage} is an outcome-dependent logical Pauli correction. To realize this correction, the decoder chooses a  shift
\begin{align} \label{eq:paulishifts}
    \hat{P}_{\vec \ell}
    &\coloneqq
    e^{
    i \sqrt{\pi} (-  \ell_1 \hat{p}
    +  \ell_2 \hat{q})
    } 
   = \op{D}(\vec{\iota}^\tp \vec{\ell}) 
\end{align}
where the vector $\vec{\ell} \in \mathbb{Z}^2$ is integer and $\vec{\iota}^\tp = \sqrt{\pi/2} \; (1\, i)$.
The shifts satisfy $\op{P}^\dagger_{\vec{\ell}} = \op{P}_{-\vec{\ell}}$ and $\op{P}_{\vec{\ell}} \op{P}_{\vec{m}} = i^{\Im (\vec{\ell}^\tp \vec{m}^*)}\op{P}_{\vec{\ell} + \vec{m}}$, with the resulting phase being precisely that for multiplying Pauli operators. 
The vector $\vec{\ell}$ (redundantly) labels the logical action in the GKP subspace via
    \begin{align} \label{eq:Pauliconnections}
        \GKPPauli_{\vec a}
     & \coloneqq \frac{1}{\sqrt{\pi}} \hat{P}_{\vec \ell}\sum_{ \vec{n} \in \mathbb{Z}^2} (\op{S}_X)^{n_1}(\op{S}_Z)^{n_2} 
     =\op{P}_{\vec \ell}\projGKP \\
     & \nonumber \quad \quad  \forall \,  \vec{\ell} \text{ mod 2} = \vec{a},
    \end{align}
where $\GKPPauli_{\vec a}$ are the \emph{GKP subspace Paulis},
\begin{subequations} \label{eq:ideal-paulis}
\begin{align}
     \GKPPauli_{0,0} & \coloneqq \bar{I} = \projGKP = \ket{\bar{0}}\bra{\bar{0}}+\ket{\bar{1}}\Bra{\bar{1}},\\
     \GKPPauli_{1,0} & \coloneqq \bar{X}  =   \ket{\bar{0}}\Bra{\bar{1}}+\ket{\bar{1}}\bra{\bar{0}},\\
     \GKPPauli_{1,1} & \coloneqq \bar{Y}  = -i\ket{\bar{0}}\Bra{\bar{1}}+i\ket{\bar{1}}\bra{\bar{0}},\\
     \GKPPauli_{0,1} & \coloneqq \bar{Z}  =   \ket{\bar{0}}\bra{\bar{0}}-\ket{\bar{1}}\Bra{\bar{1}}.
\end{align}
\end{subequations}
The difference between these operators is that $\GKPPauli_{\vec a}$ act only on the GKP codespace, while the shifts $\hat{P}_{\vec \ell}$ have non-trivial action on the whole CV Hilbert space. We emphasize that $\hat{P}_{\vec \ell}$ and $\hat{P}_{\vec \ell'}$ are distinct operators, but they realize the same logical action when $\vec \ell \text{ mod }2 = \vec \ell' \text{ mod }2$.
In this work, we employ two notations for the GKP subspace Paulis, $\{ \GKPPauli_{\vec{a}} \}$ and $\{\bar{I}$, $\bar{X}$, $\bar{Y}$, $\bar{Z}\}$, depending on the setting.

In the course of round of QEC, the decoder's job is to take the syndrome outcomes $\mu$, choose a GKP Pauli $\GKPPauli_{\vec{a}(\mu)}$, and then decide on a $\corrvec \in \mathbb{Z}$ to implement the corrective shift $\op{P}_{\corrvec}$ corresponding to $\GKPPauli_{\vec{a}(\mu)}$
\begin{align}
    \mu \longrightarrow \boxed{\text{decoder}}\longrightarrow \corrvec  .
\end{align}

The Kraus operator for the EC circuit in Eq.~\eqref{Knillcircuitimage} is 
\begin{align} 
    \op{K}(\mu) 
    & = \frac{1}{ \mathfrak{n} \sqrt{\pi} } \op{P}_{\corrvec} \Nop \projGKP \Nop \op{D}^\dagger(\mu) \label{eq:telKrausOp_OG}
    \\ 
    & = \frac{1}{ \mathfrak{n} \sqrt{\pi} }  \Nop_{\corrvec} 
    \GKPPaulicorr{}
    \projGKP \Nop \op{D}^\dagger(\mu). \label{eq:telKrausOp_shifted}
\end{align}
with $\mathfrak{n} = \sqrt{\Tr[\Nop^2 \projGKP \Nop^2 \projGKP]} = \Tr[\Nop \projGKP \Nop]$ such that $\int d^2 \mu \, \op{K}^\dagger(\mu) \op{K}(\mu) = \op{I}_\text{CV}$.
In the second line, we moved the corrective Pauli shift through the damping operator, which resulted in a shifted damping operator
    \begin{equation} \label{eq:shifteddamping}
        \op{N}_{\vec \ell} \coloneqq \op{P}_{\vec \ell} \op{N} \op{P}^\dagger_{\vec \ell},
    \end{equation}
following the corrective shift. Equations~\eqref{eq:Pauliconnections} then let us express the shift as a subspace GKP Pauli operator, $\op{P}_{\corrvec} \projGKP = \GKPPaulicorr{} \projGKP$.
The two expressions describe the fact that performing a correction \emph{after} an unshifted damping operator is equivalent to performing the same correction \emph{before} a shifted damping operator. 
From this perspective, the Knill error-correction circuit applies a damping operator shifted by $\corrvec$ to an ideal, error-corrected logical state. 

The CV channel for one round of teleportation is given by averaging over the outcomes, $\mathcal{E} = \int d^2 \mu \, \op{K}(\mu) \cdot \op{K}^\dagger(\mu)$. Inserting the first form of the Kraus operators from above gives the Knill EC teleportation channel,
    \begin{align} \label{eq:CVchannel}
        \mathcal{E}_\tel 
        &= \frac{1}{\pi \mathfrak{n}^2  } \int d^2 \mu \, \op{P}_{\corrvec} \Nop \projGKP  \Nop  \op{D}^\dagger(\mu) \cdot  \op{D}(\mu)  \Nop   \projGKP \Nop \op{P}^\dagger_{\corrvec}.
    \end{align}

\subsection{Standard binning decoder} \label{Sec:standardbinningdecoder}

The standard binning (SB) decoder~\cite{gottesman2001encoding}, originally designed to correct small random displacements, relies on separating the real-valued syndrome outcomes into integer and fractional parts with respect to $\sqrt{\pi}$. In Steane-style error correction, a corrective shift by the fractional parts of the syndrome re-centers the state on the proper GKP grid and leaves behind a byproduct effective Pauli correction.
In Knill-style error correction, the grid is already properly centered after syndrome extraction, and SB corrections are purely logical operations determined by the \emph{integer} part of the scaled syndrome outcomes with respect to $\sqrt{\pi}$.

For the beam-splitter Knill circuit, syndrome extraction implicitly implements a displacement
    \begin{equation} \label{eq:knilldisplacement}
        \op{D}^\dagger(\mu) = e^{i m_q m_p} e^{i \sqrt{2} m_q \op{p}} e^{-i \sqrt{2} m_p \op{q}}
    \end{equation}
which describes a shift by $-\sqrt{2}m_p$ in momentum followed by a shift by $-\sqrt{2}m_q$ in position and an unimportant global phase. 
These shifts arise from randomness in the teleportation and can induce logical Pauli operations to the input state, which we aim to undo with a correction.
 Formally, a real number $x$ can be decomposed into a centered integer multiple of $\sqrt{\pi}$ and a fractional remainder, $x = \sqrt{\pi} \left\lfloor x \right \rfloor_{\sqrt{\pi}} + \{ x \}_{\sqrt{\pi}}$~\cite{PantaleoniSubsystem21, harris2024}, where
    \begin{align} \label{eq:SBdecoder}
        \left\lfloor x \right\rfloor_{\sqrt{\pi}} &\coloneqq  \left\lfloor\frac{x}{\sqrt{\pi}} + \frac{1}{2} \right\rfloor 
        , \\
        \{ x \}_{\sqrt{\pi}} &\coloneqq x - \sqrt{\pi} \left\lfloor x \right \rfloor_{\sqrt{\pi}},
    \end{align}
and $\left\lfloor \cdot \right\rfloor$ is the floor function. 
The SB decoder chooses a logical Pauli $\GKPPauli_{\vec{a}(\mu)}$ depending on whether the integer parts of the $\sqrt{2}$-scaled syndromes fall into even or odd bins of $\sqrt{\pi}$:  
    \begin{equation} \label{eq:SBdecoder_Pauli}
        \vec{a}(\mu) = 
        \Big( \lfloor \sqrt{2} m_q \rfloor \text{ mod 2},
         \lfloor \sqrt{2} m_p \rfloor_{\sqrt{\pi}}  \text{ mod 2} \Big).
    \end{equation} 
From Eq.~\eqref{eq:Pauliconnections}, any corrective shift with shift vector $\vec{c}(\mu)$ can be used to implement this Pauli if $c_1(\mu) \text{ mod } 2 = a_1(\mu)$ and $c_2(\mu) \text{ mod } 2 = a_2(\mu)$.

When working with finite-energy states, one can be conscientious by selecting a shift to minimize the energy of the state (for example, to bring the envelope as close as possible to the origin)~\cite{terhal_review, conrad2019masters}. 
Although this choice does not have an immediate logical effect, it can provide protection from future noise that has detrimental effects on higher-energy states, notably loss and dephasing. Further, it can reduce logical bias in the stabilizer measurements, see below. Since Knill syndrome extraction automatically applies the unshifted damping operator (\emph{i.e.} centers the envelope), one should choose the smallest corrective shift.
By restricting the above SB shift vector to only the minimal-energy representatives, we get: 
    \begin{equation} \label{eq:SBdecoder_minshift}
    \vec{c}(\mu) = 
    \Big( \pm   \lfloor \sqrt{2} m_q \rfloor_{\sqrt{\pi}} \text{ mod 2},
    \pm \lfloor \sqrt{2} m_p \rfloor_{\sqrt{\pi}} \text{ mod 2} \Big).
    \end{equation}
The independent $\pm$ factors arise because shifting in either direction is energetically equivalent applied to an unshifted damping operator. The above describes 7 possible shifts, one for logical identity and 2 for each non-trivial Pauli.

\subsection{Weak logical measurements in approximate error correction} \label{sec:weakmeasurements}

When applied to ideal GKP states, the previously defined Kraus operators smear each Wigner-function spike into a narrow Gaussian and introduce a broad envelope whose center is located at point $\corrvec$ in phase space. This is evident from yet a third description, $\op{K}(\mu) = \frac{1}{ \mathfrak{n}_\varnothing } \op{P}_{\corrvec} \op{D}^\dagger(\mu) \widetilde{\Pi}^{\damppar}(\mu)$, where $\widetilde{\Pi}^{\damppar}(\mu) \coloneqq \op{D}(\mu)\Nop \projGKP \Nop \op{D}^\dagger(\mu)$ is a damped codespace projector centered at phase-space location $\mu$. 
As the damping is reduced, $\op{K}(\mu)$ limits to displaced ideal projectors, $ \lim_{\damppar \to 0}\widetilde{\Pi}^{\damppar}(\mu) = \projGKP(\mu)$ that satisfy translational invariance $ \projGKP    (\mu+\vec{\iota}^\tp \vec{\ell}) = \projGKP(\mu)$.
In this limit, syndrome measurements are perfect and carry no information about the logical content of an input state.
An ideal projection can be interpreted as the measurement-backaction that projects an input state onto the $\mu$-displaced GKP code space labeled by position and momentum mod $\sqrt{\pi}$~\cite{ketterer2016quantum}.
Recovery then returns the state to the code space where modular position and momentum are both zero.

For finite energy ($\damppar >0$), the syndrome measurements are only approximate, and the measurement backaction is not a projection. 
Instead, states with consistent support are amplified by locally Gaussian factors.
On top of this, the globally decaying structure (the envelope) of $\widetilde{\Pi}^{\damppar}(\mu)$ biases the outcome to be \textit{globally} supported around the outcome $\mu$. This effect breaks the translational/logical invariance of $\bar{\Pi}(\mu)$, such that the use of approximate GKP states in the stabilizer measurement protocol also weakly measures \emph{logical} information in addition to the intended stabilizers, which drives logical decoherence when averaged over $\mu$.

\section{GKP-Logical maps for finite-energy states} \label{sec:GKP-logicalmaps}

Our ultimate goal is to identify local, composable logical CPTP channels for qubit information encoded in the GKP code beginning from logical state preparation/encoding and proceeding through a series of teleportation circuits until logical readout. 
This sections provides the first step towards this goal: we show that identifying logical-GKP maps is made possible by the appearance of a GKP codespace projector in the Kraus operator for the Knill EC teleportation circuit, Eq.~\eqref{eq:telKrausOp_shifted}. We supplement this with gadgets for logical state preparation and readout, both of which similarly contain implicit codespace projectors. Our construction is such that GKP projectors are distributed throughout the physical evolution in a manner that allows the interpretation that steps in between projectors are logical  maps. 
However, these logical maps are not \textit{local} in that they depend on multiple teleportation outcomes, and they fail to serve as Kraus operators for a CPTP channel even for the global qubit map from state preparation to readout. Later sections are dedicated to additional machinery, based on channel twirling the damping operator, that transforms these maps into genuine logical CPTP channels.

\subsubsection{State preparation}

As discussed in Sec.~\ref{sec:EncodingIntro}, encoding an ideal GKP state into damped codewords is not a linear map and does not preserve orthogonality. A physical approximation to an intended qubit state $\op \rho$ with Bloch four-vector $\vec{r} = (1, r_{{X}}, r_{{Y}}, r_{Z})$ is a damped GKP state,
\begin{equation}\label{eq:physical-state}
    \bar{\rho}_\damppar 
        = \frac{1}{\mathfrak{n}} \op{N} \bar \rho \op{N} =  \frac{1}{2 \mathfrak{n}} 
 \sum_{\vec{a} \in \mathbb{Z}_2^{\times 2}} r_{\vec{a}} \op{N} \GKPPauli_{\vec{a}} \op{N},
\end{equation}
with normalization
        $\mathfrak{n} %
        = \frac{1}{2} \sum_{\vec{a} \in \mathbb{Z}_2^{\times 2}} r_{\vec{a}} \Tr[ \op{N} \GKPPauli_{\vec{a}} \op{N} ].$
The damped GKP subspace Paulis, $\op{N} \GKPPauli_{\vec{a}} \op{N}$, are not orthogonal to each other due to the leakage of their support outside the GKP subspace. Thus, the intended Bloch vector $\vec{r}$ is not exactly recoverable from the physical state $ \bar{\rho}_\damppar$; \emph{i.e.} encoding into damped GKP states leads to a loss of (qubit) information. 

\begin{figure*}[t]
    \centering
    \includegraphics[scale=.275]{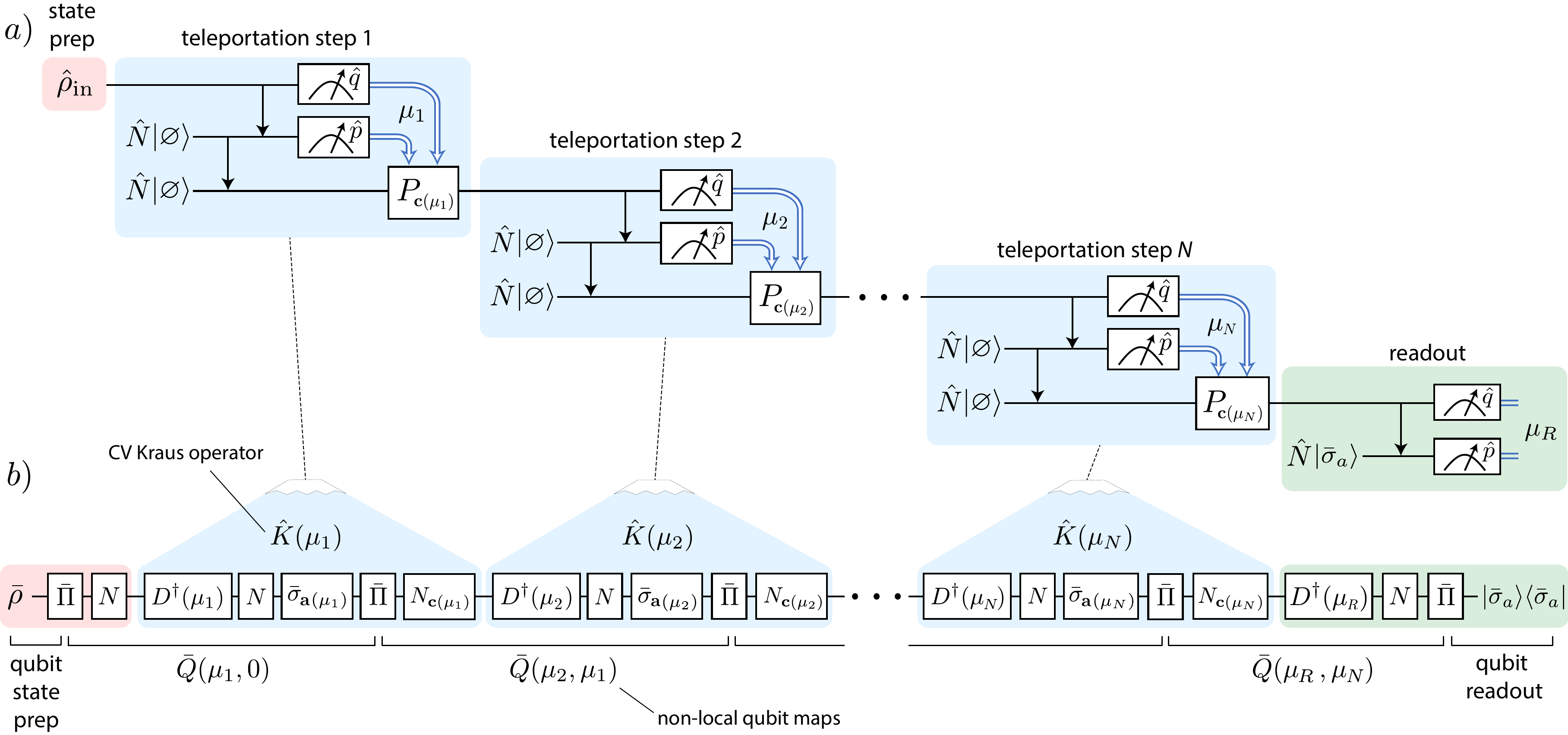}
    \caption{(a) Circuit showing the composition of state preparation, $N$ rounds of teleportation, and auxiliary state-assisted readout. (b) Associated global map given an input state of the form in Eq.~\eqref{eq:physical-state}. Above the global map, we label the CV Kraus operators, Eq.~\eqref{eq:telKrausOp_shifted}. Below the global map, we label the quasi-local qubit maps that appear between GKP projectors, Eq.~\eqref{eq:correlatedqubitop}. The corrections commute with GKP projectors, so we reorder them here to better indicate the qubit maps, and we use Eq.~\eqref{eq:Pauliconnections} to write the corrections as GKP subspace Paulis, noting that $\vec{a}(\mu_t) = \vec{c}(\mu_t) \text{ mod } 2$.
    Note that normalizations have been excluded throughout. }
    \label{fig:general-teleportation-scheme}
\end{figure*}

\subsubsection{Logical measurement gadget}

Destructive Pauli measurements for the square-lattice GKP code are often described by performing homodyne detection in one of the three bases corresponding to logical $X$, $Y$, and $Z$: $\op{p}$, $\tfrac{1}{\sqrt{2}}(\op{q} - \op{p})$, and $\op{q}$, respectively. The logical readout is determined by a decoder --- for example, the SB procedure rounds the homodyne outcome to the closest integer multiple of $\sqrt{\pi}$. 
However, when this naive strategy is used for logical-state reconstruction/tomography, the reconstructed operator is not guaranteed to be positive~\cite{shawSSD}.
We propose an alternative readout method by supplementing an auxiliary (noisy) GKP state,
        $\frac{1}{\sqrt{ \mathfrak{n}_{\vec{a}} }} 
        \op{N} \ket{\GKPPauli_{\vec{a}}} $,
where $\ket{\GKPPauli_{\vec{a}}}$ is a +1 eigenstate of $\GKPPauli_{\vec{a}}$, 
and $\mathfrak{n}_{\vec{a}} = \bra{\GKPPauli_{\vec{a}}} \Nop^2 \ket{\GKPPauli_{\vec{a}}}$. Readout is done by performing an EPR measurement between the data mode and the auxiliary mode. This can be realized by projecting onto an EPR state, $\ket{\EPR} := \frac{1}{\sqrt{2 \pi} } \int ds \, \qket{  s} \otimes \qket{ s}$, using a beam splitter followed by a position and a momentum measurement on the two modes~\cite{walshe2020continuousvariable},
    \begin{align}
        \qbra{m_1} \otimes \pbra{m_2} \op{B}_{12}  = \sqrt{2} \bra{\EPR} \op{D}^\dagger (\mu) 
        ,
    \end{align}  
where $\mu = m_1 + i m_2$. 
The circuit describing the readout gadget~\cite{shaw2024gates}
\begin{equation} \label{fig:povm}
    \includegraphics[scale=.35]{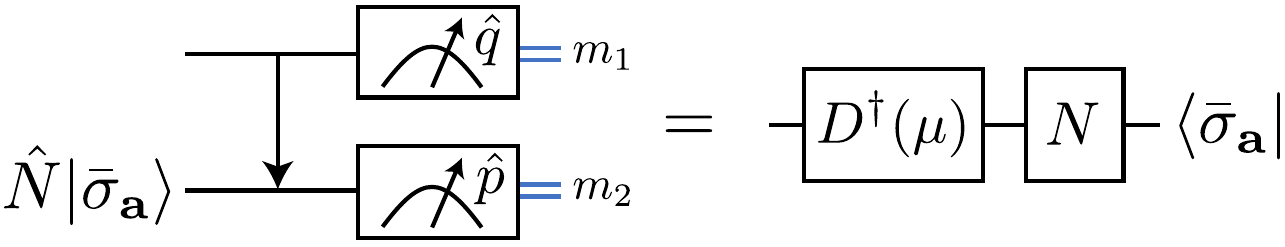}
\end{equation}
implements the operation
    \begin{align} \label{eq:logicalmeasurement}
        \bra{\EPR} \op{D}_1^\dagger(\mu) \otimes \op{N}_2 \ket{ \GKPPauli_{\vec{a}} }_2 
        =
        \bra{ \GKPPauli_{\vec{a}}} \projGKP \op{N} \op{D}^\dagger(\mu)
    \end{align}
up to normalization, with all of the objects on the right-hand side of the equation being on the data mode where the state to be measured lies. In the above, we bounced the noise operator from the auxiliary mode to the data mode~\cite{walshe2020continuousvariable} and used $\inprod{\EPR}{\GKPPauli_{\vec{a}}} = \bra{\GKPPauli_{\vec{a}}}$. Additionally, observe that we have extracted a GKP projector from the GKP state for later use, $\ket{\GKPPauli_{\vec{a}}} = \projGKP \ket{\GKPPauli_{\vec{a}}}$.
This circuit is preferred to homodyne detection of the data mode alone, because it measures both a  logical Pauli and a conjuguate stabilizer.

From the projection in Eq.~\eqref{eq:logicalmeasurement}, we define a logical, two-outcome POVM by dividing the space of outcomes, $\mu \in \mathbb{C}$, into two disjoint sets $\Omega_0$ and $\Omega_1$, determined by the decoder, each corresponding to a \emph{logical} outcome in the qubit basis given by $\ket{\GKPPauli_{\vec{a}}}$.
The POVM elements $\op{\Omega}_j$ for $j \in \{0,1 \}$  
\begin{equation} \label{eq:povm}
 \op \Omega_{j}
 \propto
 \frac{1}{2 \pi \mathfrak{n}_{\vec{a}}}
 \int_{\Omega_j}d^2 \mu \, \op D(\mu) \Nop  \projGKP
 \outprod{\GKPPauli_{\vec{a}}}{\GKPPauli_{\vec{a}}} 
 \projGKP \, \Nop \op D^{\dagger}(\mu)
\end{equation}
satisfy $\op{\Omega}_0 + \op{\Omega}_1 \propto \op I_\text{CV}$.
For example, given $\ket{\GKPPauli_{\vec{a}}} = \ket{\bar{0}}$, the SB decoder for logical $Z$ measurement defines $\Omega_0$ as all $m_1$ values in even bins of $\sqrt{\pi}$ and $\Omega_1$ contains $m_1$ values in odd bins of $\sqrt{\pi}$; $m_2$ is not needed. More sophisticated decoders may take into account finite-energy effects from the damping operator and other relevant noise processes. We leave $\Omega_0$ and $\Omega_1$ unspecified such that the decoder is left arbitrary.

\subsection{Logical GKP-qubit maps} \label{sec:logicalmaps}

We now combine state preparation, teleportation, and readout into a sequence of GKP-logical maps. 
By taking into account state preparation, we incorporate the first qubit map into the framework. Consider a qubit state $\ket{\psi_\text{qubit}}$ encoded into a damped GKP state, Eq.~\eqref{eq:physical-state} and then immediately measured in some GKP Pauli basis,
\begin{align} \label{eq:basicqubitmap}
     \bra{\GKPPauli_{\vec{a}}} \underbrace{ \projGKP \op{N} \op{D}^\dagger (\mu) \op{N} \projGKP}_{ \coloneqq \oldQ(\mu) } \ket{\bar{\psi}} 
\end{align}
with normalization $(\mathfrak{n}_{\vec{a}} \mathfrak{n}_\psi)^{-\frac{1}{2}}$,
and we used the fact that a logical state is implicitly accompanied by a code-space projector, $\ket{\bar{\psi}} = \projGKP \ket{\bar{\psi}}$.
We identify the operation between the two projectors as a state-dependent, trace-decreasing GKP qubit map $\oldQ(\mu)$, with the state dependence carried by the normalization factor.
A way to understand the state-dependence is that, for e.g. $\ket{\GKPPauli_{\vec{a}}}=\ket{\overline{0}}$, the distribution 
   $ P(\mu)  \propto |\bra{\GKPPauli_{\vec{a}}} \oldQ(\mu) \ket{\bar{\psi}}|^2 $
defined by this inner product 
is concentrated around the measurement values $\mu$ that create the largest overlap between the displaced approximate $\op{N}\ket{\GKPPauli_{\vec{a}}}$ state and the input state $\bar{\rho}$. The measurement is facilitated by the POVM elements in Eq.~\eqref{eq:povm}, which are simply the displaced approximate GKP $\ket{\GKPPauli_{\vec{a}}}$ states, such that Eq.~\eqref{eq:basicqubitmap} can be interpreted as fidelity between the input state and the POVM reference state $\ket{\GKPPauli_{\vec{a}}}$.

In a more general setting, we have $N$ rounds of teleportation,
 $   \op K(\mu_N)
    \dots
    \op K(\mu_2)
    \op K(\mu_1) $,
each defined by Eq.~\eqref{eq:telKrausOp_shifted}.
Inserting these between state preparation and readout allows us to express the quantum operation in terms of $N+1$ operators acting on the GKP subspace,
\begin{equation}\label{eq:concatenated-Ks}
    \bra{\GKPPauli_{\vec{a}}} 
    \oldQ(\mu_R, \mu_N)
    \oldQ(\mu_N, \mu_{N-1})
    \dots
    \oldQ(\mu_2, \mu_1)
    \oldQ(\mu_1,0)
     \ket{\bar{\psi}} ,
\end{equation}   
normalized by $(\mathfrak{n}_a \mathfrak{n}_\varnothing^{2N} \mathfrak{n}_\psi)^{-\frac{1}{2}}$. A graphical description is given in Fig.~\ref{fig:general-teleportation-scheme}. 
Here, we have defined \emph{quasi-local} qubit operators,
    \begin{align} \label{eq:correlatedqubitop}
        \oldQ(\mu_{t}, \mu_{t-1}) 
        & \coloneqq 
        \GKPPauli_{\vec{a}(\mu_t)} 
        \projGKP \Nop \op{D}^\dagger(\mu_t) \Nop_{\corrvect{t-1}} \projGKP 
    \end{align}
with $t \in \{1,2,\dots N, R\}$ and normalization ignored. The first map in Eq.~\eqref{eq:concatenated-Ks}, $\oldQ(\mu_{1},0)$, has no predecessor, and the last one, $\oldQ(\mu_{R},\mu_{N})$, has no correction, as it occurs just before readout. 
For every qubit operator 
in between, a shifted damping operator contains the information about the correction from the previous outcome $\mu_{t-1}$ through its shift vector $\corrvect{t-1}$.

We take a moment to discuss the properties of the associated qubit maps,
    \begin{align} \label{eq:dampedanc_maps}
        \mathcal{M}_{t,t-1} (\cdot)&\coloneqq \oldQ(\mu_t,\mu_{t-1}) \cdot \oldQ^\dagger (\mu_t,\mu_{t-1}) .
    \end{align}
It would be desirable if this map could be interpreted a physical (i.e. CPTP) qubit-level channel. This is indeed not the case as we discuss as follows.
Observe that each $\oldQ$, other than the first, depends on the measurement outcomes from one teleportation step, $\mu_t$, as well as the outcomes from the previous step, $\mu_{t-1}$, through the correction. The latter determines the location of the shifted envelope through $\Nop_{\corrvect{t-1}}$, see Eq.~\eqref{eq:telKrausOp_shifted}.
This dependence on multiple outcomes prevents the qubit maps from being specified independently---successive maps are correlated by outcome-dependent corrections, and are thus only quasi-local in time.
Further, 
when averaged over outcomes, the qubit maps are not proportional to the identity, indicating that they do not form a Kraus representation for a CPTP channel. This effect arises even when there are no corrections (syndrome extraction alone). For example, two rounds of adjacent syndrome extraction yield a qubit map $\oldQ^\text{syn}(\mu_{t}) =  \projGKP \Nop \op{D}^\dagger(\mu_t) \Nop \projGKP $ (the outcome $\mu_{t-1}$ participates in the prior map), where $\oldQ^\text{syn}(\mu_{t})^{\dagger}\oldQ^\text{syn}(\mu_{t})$ describes the POVM element that induces the probability to measure $\mu_t$. Averaged over outcomes, we obtain
    \begin{equation} \label{eq:baremap_outcomeavg}
    \int d^2 \mu_t \big[\oldQ^\text{syn}(\mu_{t}) \big]^\dagger \oldQ^\text{syn}(\mu_{t}) 
    \propto 
     \projGKP \op{N}^2  \projGKP.     
    \end{equation}
The RHS evaluates to the Knill-Laflamme conditions for QEC. As $\op{N}$, however, is not a correctable error, we recognize that $\oldQ^\text{syn}(\mu_{t})$ cannot be understood as a Kraus operator element for a logical CPTP channel.

In summary, the qubit maps exhibit the following issues, which need to be resolved in order to identify local CPTP GKP-qubit channels:
\begin{enumerate} \label{issues}
        \item \label{item1}
        Each qubit map depends on the outcomes from two successive time steps through the corrections, meaning that these maps are not local, nor are they composable.
        \item \label{item2}
        The qubit maps do not resolve the identity when averaged over outcomes, even when no corrections are applied.
\end{enumerate}
These issues appear to be in conflict with the fact that the CV Kraus operators for teleportation, $\op{K}(\mu_t)$, are complete and composable CPTP channels at the CV level.
By selectively shifting the window of time in order to identify qubit maps, we spoil locality, and by focusing on the qubit subspace, we spoil TP and introduce the bias described above. 
Nevertheless, we may define a global qubit map that extends from state preparation all the way to readout and depends on the full string of $N+1$ outcomes ($N$ from teleportation and one from readout),
    \begin{align} \label{eq:globalmap}
        \mathcal{M}_\text{global} &= \mathcal{M}_{R,t_N} \circ \dots \circ\mathcal{M}_{t_3,t_2} \circ \mathcal{M}_{t_2,t_1} \circ \mathcal{M}_{t_1,t_0}.
    \end{align}
Note that even the global qubit map is not TP owing to the fact that the damping in state preparation and readout are included as part of the qubit evolution. 

In the following sections, we show that two distinct types of channel twirling, one over the full stabilizer group and the other over a minimal set of Pauli representatives, can each resolve the above issues in order to define qubit CPTP channels. The latter has the beneficial property that it only makes use of finite-energy GKP states.

\section{Stabilizer twirling: logical channel for stochastically displaced GKP states} \label{sec:GRNGKPchannel}

In this section, we show that the symmetries of standard-binning (SB) decoding naturally provide a mechanism to passively channel twirl the damping operator over the GKP stabilizer group. When the damping parameter is small, the twirling damping operator becomes a GRN channel, Eq.~\eqref{eq:GRN}~\cite{noh2019fault}. 
Thus, we provide a physical justification for using GRN GKP states in place of coherently damped ones; a practice that is already routinely used in fault-tolerance studies with the GKP code.
With this in hand, we replace the damping operators in the GKP Bell pairs with GRN channels and construct a logical channel for repeated teleportation that is free from the issues with the maps in the previous section; \emph{i.e.} it is local, composable, and complete, and further, we show that it is a Pauli channel.

\subsection{Stabilizer twirling from standard binning decoding} 

In logical settings, we are interested in decoded measurements.
Since the SB decoder relies only on the parity of the integer parts of the two syndrome measurements to choose a corrective Pauli, Eq.~\eqref{eq:SBdecoder_Pauli}, we define a two-mode SB POVM $\{ \op{\Omega}_{jk} \}$, similar to Eq.~\eqref{eq:povm}, based on binned homodyne outcomes for stabilizer readout. It contains four elements,
        $\op \Omega_{jk} = \op \Omega^q_j \otimes \op \Omega^p_k$
where  $j,k \in \{0,1\}$ indicate even/odd bins of size $\sqrt{\pi}$ in position and momentum over each of the two modes, and
    \begin{align} \label{eq:standbinningPOVM}
        \op \Omega^{q/p}_j 
        &\coloneqq \sum_{n \in \mathbb{Z}} \int_{(2n+j)\sqrt{\pi} - \frac{\sqrt{\pi}}{2}}^{(2n+j)\sqrt{\pi} + \frac{\sqrt{\pi}}{2} }
        d m'_{q/p} \, \outprodsubsub{m'_{q/p}}{ m'_{q/p} }{q/p}{} \,
    \end{align}
and $m'_{q/p} \coloneqq \sqrt{2} m_{q/p}$.\footnote{The factor of $\sqrt{2}$ arises from squeezing in the beam splitter and is included because we prefer the bins to be size $\sqrt{\pi}$. This factor is not present in GKP EC using controlled gates.}
Each binned homodyne POVM is invariant under stabilizer shifts,
    \begin{align} \label{eq:POVMstabilizerinvariance}
        \big(\op{S}_Z\big)^{n_1} \big(\op{S}_X\big)^{n_2}  \op \Omega^{q/p}_{jk} \big(\op{S}^\dagger_X\big)^{n_2} \big(\op{S}^\dagger_Z\big)^{n_1} = \op \Omega_{jk}^{q/p}, 
    \end{align}
and thus so is $\op \Omega_{jk}$.
A consequence of this invariance is that, given arbitrary state $\op \rho$, the probability for decoded bin-parity outcome $j,k$ is
        $\text{Pr}_{j,k} = \Tr [\op \Omega_{jk} \op \rho] 
        \propto 
        \Tr [\op \Omega_{jk}\tilde{\rho}],$
where $\tilde{\rho} \coloneqq \sum_{\vec{n} \in \mathbb{Z}_2} 
\big(\op{S}^\dagger_Z\big)^{n_1} \big(\op{S}^\dagger_X\big)^{n_2} \op \rho \big(\op{S}_X\big)^{n_2} \big(\op{S}_Z\big)^{n_1}  $ is the stabilizer-twirled input state. 
This state has infinite energy, so probabilities are found by integrating the outcome space over a single unit cell
of size $\sqrt{\pi} \times \sqrt{\pi}$. 
More details about stabilizer-invariant observables and twirling can be found in Appendix~\ref{app:stabilizertwirling_observables}. Use of the twirled state in place of the original state can be used for observables that are invariant under stabilizer shifts just as the binned homodyne probabilities are. 

Using the above stabilizer invariance of
SB decoding allows us to perform a channel twirl of each damping operator $\op{N}$ by the full stabilizer group, see Appendix~\ref{app:stabilizertwirling}. That is, we are free to make the replacement
    \begin{equation} \label{eq:displacementchannelgeneric}
       \mathcal{E}_\text{damp}  = \Nop \cdot \Nop \longrightarrow \tilde{\mathcal{E}}_\text{damp}  \coloneqq \sum_{\vec{n} \in \mathbb{Z}^2} \Nop_{2\vec{n}} \cdot \Nop_{2\vec{n}} ,
    \end{equation}   
keeping in mind that expectation values are calculated over only a single unit cell (described above). Importantly, this replacement does not require an active change to the circuit; rather, the twirling is passive and stems directly from the symmetries in SB decoding. 
An analytic expression for $\tilde{\mathcal{E}}_\text{damp}$ is given in Eq.~\eqref{eq:twirleddampingmap} revealing that it is, in general, not diagonal in the displacement-operator basis and thus is not a GRN channel unless the damping parameter is small.\blk

Recognize that performing the channel twirling of $\Nop$ here is equivalent to \emph{state twirling} the GKP Bell pair in the teleportation circuit,\footnote{This circuit equivalence requires that $\Nop \otimes \Nop$ is applied directly to the entangled Bell pair or that it commutes with the beam splitter, which is the case for $\Nop = e^{- \damppar \op n}$.} 
    \begin{equation} 
    \begin{split}
\label{RDbeamsplitterBellpair}
        \Qcircuit @C=0.75em @R=0.7em 
    {
 &\lstick{ \ket{\varnothing}} &\bsbal{1} & \gate{ N } &\qw  \\
 &\lstick{ \ket{\varnothing}} &\qw       & \gate{ N} & \qw 
		} \, 
\quad \raisebox{-1em}{ $\longrightarrow$ }\quad \quad \,
     \Qcircuit @C=0.75em @R=0.6em 
    {
 &\lstick{\ket{\varnothing}} &\bsbal{1} &\gate{\tilde {\mathcal{E}}_\text{damp}} &\qw  \\
 &\lstick{\ket{\varnothing}} &\qw       &\gate{\tilde {\mathcal{E}}_\text{damp}} &\qw 
		} \, 
    \end{split}
    \end{equation}
As our derivation has closely followed Appendix A of Noh \emph{et al.}~\cite{noh2019fault}, this perspective reveals the connection between channel and state twirling of noisy GKP states.

\subsection{Logical channel in the GRN approximation}

In the limit of small enough damping, $\damppar \lesssim 0.2$, the twirled damping map gives an effective GRN channel, $\tilde{\mathcal{E}}_\text{damp} \approx \mathcal{E}_\text{GRN}$ with kernel $G_{\sigma^2/2} (\alpha_R, \alpha_I)$ and $\sigma^2 = \tanh \frac{\damppar}{2}$.
From Eq.~\eqref{RDbeamsplitterBellpair}, we may replace each damped GKP Bell pair by its GRN counterpart,
    \begin{equation}
    \begin{split}
\label{GRNbeamsplitterBellpair}
     \Qcircuit @C=0.75em @R=0.6em 
    {
 &\lstick{\ket{\varnothing}} &\bsbal{1} &\gate{\mathcal{E}_\text{GRN}} &\qw  \\
 &\lstick{\ket{\varnothing}} &\qw       &\gate{\mathcal{E}_\text{GRN}} &\qw 
		} \, 
  \raisebox{-1em}{ $=
\mathcal{E}_\text{GRN} \otimes \mathcal{E}_\text{GRN} \big( \outprod{\bar{\Phi}^+}{\bar{\Phi}^+}\big)$} 
    \end{split}
    \end{equation}
We now ``bounce'' the GRN channel from the top mode to the bottom mode~\cite{fukui2021alloptical}.
First note that $ \ket{\bar{\Phi}^+} = \projGKP_2 \ket{\text{EPR}}$~\cite{walshe2020continuousvariable}, with the subscript indicating that the GKP projector acts on the second mode, and $\ket{\text{EPR}}$ is a CV EPR state. Using the fact that a displacement operator bounces from one mode to the other according to $\op{D}_1(\alpha) \ket{\text{EPR}} = \op{D}_2(-\alpha^*) \ket{\text{EPR}} $~\cite{walshe2020continuousvariable}, we get $\op{D}_1(\alpha) \ket{\bar{\Phi}^+} = \projGKP_2 \op{D}_2(-\alpha^*) \ket{\text{EPR}} $. Bouncing the displacement operators from the GRN channel on mode 1 onto mode 2 and performing a change of variables $-\alpha^* \rightarrow \alpha$, under which the Gaussian kernel is invariant, allows us to express Eq.~\eqref{GRNbeamsplitterBellpair} as
    \begin{align} \label{eq:GRNGKPbellequiv}
        \mathcal{E}_\text{GRN,2} \circ \projGKP_2 \circ \mathcal{E}_\text{GRN,2} (\outprod{\text{EPR}}{\text{EPR}}).
    \end{align}
In this expression, subscripts indicate that both of the GRN channels and the GKP projector act on the second mode.

Using the GRN GKP Bell pair, Eq.~\eqref{GRNbeamsplitterBellpair}, in the teleportation circuit, we use Eq.~\eqref{eq:GRNGKPbellequiv} and the methods in Ref.~\cite{walshe2020continuousvariable} to find
the conditional CV map for each teleportation step,
    \begin{align} \label{eq:GRNmap}
        \op{P}_{\corrvect{t}} \circ \mathcal{E}_\text{GRN} \circ \projGKP \circ \mathcal{E}_\text{GRN}
        \circ \op D^\dagger(\mu_t),
    \end{align}
with the understanding that these maps are not valid as single trajectories, but must be taken in convex combinations over values of $\mu$ according to the SB POVM.

Qubit maps derived from repeated noisy GKP error correction with GRN proceed just as in Fig.~\ref{fig:general-teleportation-scheme}, with
sequential rounds of teleportation giving composed CV maps of the form in Eq.~\eqref{eq:GRNmap}. 
A chain of GRN teleportation maps can be manipulated with two facts. First, GRN channels commute with displacement operators, $\mathcal{E}_\text{GRN} \circ \op D(\alpha) = \op D(\alpha) \circ \mathcal{E}_\text{GRN} .$
Second, two GRN channels compose to give another GRN channel, with the variance of the composite channel given simply by the sum of the individual variances.
As each GRN channel has variance $\frac{1}{2}\sigma^2$, the composed channel has variance $\sigma^2$. 
Identifying operations between GKP projectors gives the qubit Kraus operators
\begin{align}
    \bar{Q}^\text{GRN}(\mu_t, \alpha_t)
    & \coloneqq
     \sqrt{ \tfrac{1}{\pi} G_{\sigma^2}(\alpha)} \projGKP \op D^{\dagger}(\{\mu_t\}_{\sqrt{\pi}}) \op D(\alpha) \projGKP .\label{eq:GRN-kraus-SB}
\end{align}
We have commuted the SB correction with $\projGKP$ and combined it with the displacement to give Kraus operators that depend only on the fractional parts of the syndrome~\cite{harris2024}. This is equivalent to using the SB POVM in Eq.~\eqref{eq:standbinningPOVM}. 

These qubit Kraus operators are  local and complete (up to an infinite constant)
$        \int d^2\mu_t d^2 \alpha \,  \bar{Q}^\text{GRN}{}^\dagger(\mu_t, \alpha) \bar{Q}^\text{GRN}(\mu_t, \alpha) = \projGKP$,
which is straightforward to show using CV Schur's lemma, see Eq.~\eqref{eq:SchurLemmaCV}.
They give an expression for the \emph{GRN logical channel},
    \begin{align} \label{eq:GRNchannel}
        \bar{\mathcal{E}}^\text{GRN} = \int_\Omega d^2 \mu_t d^2 \alpha \,  
        \bar{Q}^\text{GRN}(\mu_t, \alpha) \cdot \bar{Q}^\text{GRN}{}^\dagger(\mu_t, \alpha)
    \end{align}
Since the qubit Kraus operators depend only on the fractional part of the syndrome, the integral over $\mu_t$ in Eq.~\eqref{eq:GRNchannel}, 
are taken only over a single unit cell, $\Omega = [-\frac{1}{2} \sqrt{ \frac{\pi}{2} }, \frac{1}{2} \sqrt{ \frac{\pi}{2} }) \times [-\frac{1}{2} \sqrt{ \frac{\pi}{2} }, \frac{1}{2} \sqrt{ \frac{\pi}{2} })$, with the extra factor of $\frac{1}{\sqrt{2}}$ in the limits arising from the $\sqrt{2}$ in Eq.~\eqref{eq:knilldisplacement}.

There are several useful representations of qubit maps or channels arise---we focus here on the $4 \times 4$ single-qubit Pauli Transfer Matrix (PTM) $\mat \Gamma$. An overview of the PTM and its relation to the $\chi$-matrix representation can be found in Appendix~\ref{appendix:PTM}. 
Using Eq.~\eqref{eq:PTMelementsdef}, the conditional PTM given syndrome $\mu_t$ has matrix elements given by
\begin{align}
    \Gamma^\text{GRN}_{\vec a,\vec a'}(\mu_t) 
   = \frac{1}{2} \int d^2 \alpha \,
        \text{Tr} \big[ \GKPPauli_{\vec a} 
        \bar{Q}^\text{GRN}{}(\mu_t, \alpha)
        \GKPPauli_{\vec a'} 
        \bar{Q}^\text{GRN}{}^{\dagger}(\mu_t, \alpha)
        \big]
    \end{align}
where $\GKPPauli_{\vec a}$ are the GKP subspace Paulis in Eq.~\eqref{eq:Pauliconnections}.    
An explicit form for the matrix elements is given by a sum over Siegel theta functions, see Appendix~\ref{Appendix:PTM_GRNchannel}. Integrating over $\mu_t$ gives the logical channel.

GRN logical channels are Pauli channels, and as such have diagonal PTMs and $\chi$ matrices. 
In practice, we find the PTMs numerically for different values of $\sigma^2$. Consider a GRN qubit channel with $\sigma^2 = 0.049$, which adds the same variance to the spikes of GKP state as a $\damppar = 0.1$ (10 dB state) damping operator does~[Eq.~\eqref{eq:GRNtoenvelope_comparison}]. The PTM is 
    \begin{align}
    \mat \Gamma^\text{GRN} = 
       \begin{pmatrix}
        1 & 0 & 0 & 0 \\
        0 & 0.9900 & 0 & 0 \\
        0 & 0 & 0.9801 & 0 \\
        0 & 0 & 0 & 0.9900 \\
    \end{pmatrix}
    \end{align}
with Pauli error probabilities $p_I = 0.9900$, $p_X = p_Z = 0.0050$, and $p_Y = 0.0000$ that serve as diagonal elements of the process matrix.

\begin{figure*}[t!]
\label{fig:mixedenvGKP}
    \centering
    \includegraphics[scale=.25]{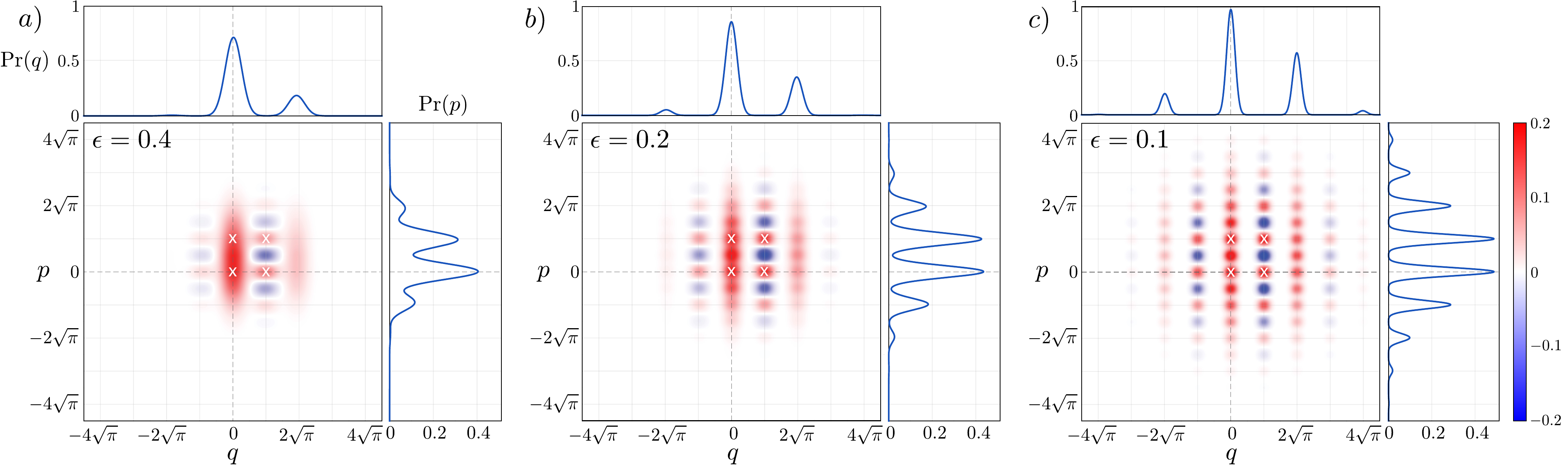}
    \caption{Wigner function and marginal probability distributions for the GKP state $\ket{\bar{0}}$ in a mixture over envelopes, Eq.~\eqref{eq:unbiasedencoding}, for three values of $\damppar = \{0.4, 0.2, 0.1\}$. 
    Asymmetries across the $q=0$ and $p=0$ axes arise from the choice of envelope centers, $S = \{ (0,0), (0,1), (1,0), (1,1)) \}$ in units of $\sqrt{\pi}$, which are marked on the Wigner functions with a white `x'. The mixture probabilities in Eq.~\eqref{eq:unbiasedencoding_physical} for $\bitsvec \in S$ are (a) $\{0.31, 0.31, 0.19, 0.19\}$, (b) $\{0.26, 0.26, 0.24, 0.24\}$, and (c) $\{0.25, 0.25, 0.25, 0.25\}$. 
    }
    \label{fig:mixedenvGKP}
\end{figure*}

\section{Minimal Pauli twirling: Logical channel for finite-energy GKP states} \label{sec:logical_channel}

The preceding section resolves the issues arising from finite-energy qubit maps, Issue~\ref{item1} and Issue~\ref{item2}, by channel twirling over the full stabilizer group motivated by the symmetries of SB decoding.
In this section, we introduce a different approach by channel twirling $\Nop$ using a minimal set of representative Pauli shifts. This approach is more general, as it does not require us to specify a decoding strategy \emph{a priori}, it is at all times physical in that it relies only on finite-energy GKP states and a finite-energy twirling procedure.
The counterbalance is that this minimal Pauli twirling is active: it requires modifying the GKP Bell pairs in the circuit.

\subsection{Minimal Pauli-twirled encoding}

First, we begin with state preparation using a mixed-state GKP encoding based on Pauli twirling the damping operator. 
Consider a pure qubit state $\ket{\psi_\text{qubit}}$ encoded as an ideal GKP state $\ket{ \bar{\psi}}$, and then mixed over shifted envelopes,
\begin{align} \label{eq:unbiasedencoding}
    \op{\rho}_\psi 
    =
    \frac{1}{\twirlnorm} \sum_{\bitsvec \in S}  \op{N}_{\bitsvec}  \outprod{\bar \psi}{\bar \psi} \op{N}_{\bitsvec}
\end{align}
with normalization factor $\mathfrak{n}_\damppar$.
The representative Pauli shifts on the damping operators are chosen from the minimal set 
    \begin{align} \label{eq:shiftvaluesorig}
        S = \{(0,0), (0,1), (1,0), (1,1)\},
    \end{align}
which realizes a Pauli channel twirl of the damping operator $\Nop$. That is, the shifts only adjust the centers of the damping operators but do not affect the logical state.
A result of the twirl is the state independence of the normalization factor,
    \begin{subequations} \label{eq:Paulitwirlnorm}
\begin{align} 
    \twirlnorm &\coloneqq \sum_{\bitsvec \in S}
    \Tr \big[ \op{N}_{\bitsvec}  \outprod{\bar \psi}{\bar \psi} \op{N}_{\bitsvec} \big] \\
    &=\sum_{\bitsvec \in S} \Tr \big[ \op{N} \op{P}^\dagger_{\bitsvec} \outprod{\bar \psi}{\bar \psi} \op{P}_{\bitsvec} \op{N}  \big] 
    \\
    &= 2 \Tr [ \op{N} \projGKP \op{N}  ].
\end{align}
    \end{subequations}
Inside the trace, the channel twirl behaves like a maximum strength depolarizing channel: for some operator $\bar{A}$ with support only in a two-dimensional subspace,
    \begin{equation} \label{eq:Paulitwirl}
    \frac{1}{2} \sum_{\vec{a} \in S} \GKPPauli_{\vec{a}} \bar{A} \GKPPauli_{\vec{a}}
    =
    \Tr[\bar{A}] \projGKP .
\end{equation}
An analytic form for $\twirlnorm$ and the trace of the other damped GKP subspace Paulis can be found in Appendix~\ref{appendix:innerproduct}.

Alternatively, the encoded state in Eq.~\eqref{eq:unbiasedencoding} can be obtained as mixture of normalized, approximate pure states $\ket{\bar \psi_{\damppar,\bitsvec}}=\frac{1}{\sqrt{\mathfrak{n}_{\bitsvec}}} \op{N}_{\bitsvec} \ket{\bar \psi}$,
\begin{align} \label{eq:unbiasedencoding_physical}
    \op{\rho}_{\bar{\psi}} 
    =
    \sum_{\bitsvec \in S}  \frac{\mathfrak{n}_{\bitsvec}}{\twirlnorm} \outprod{\bar \psi_{\damppar,\bitsvec}}{\bar \psi_{\damppar,\bitsvec}},
\end{align} 
with normalization $\mathfrak{n}_{\bitsvec} = \bra{\bar \psi} \op{N}_{\bitsvec}^2 \ket{\bar \psi}$. The coefficients are the mixture probabilities satisfying $\sum_{\vec{b} \in S} \mathfrak{n}_{\bitsvec}/\twirlnorm = 1$.
Example Wigner functions are shown in Fig.~\ref{fig:mixedenvGKP}.

An important property of the Pauli twirled-envelope states in Eq.~\eqref{eq:unbiasedencoding} is that they limit to ideal states in the same way that standard damped states [Eq.~\eqref{eq:finitestates}] do,
\begin{align}
    \lim_{\damppar \rightarrow 0} \op{\rho}_{\bar{\psi}} = \lim_{\damppar \rightarrow 0} \outprod{\bar{\psi}_\damppar}{\bar{\psi}_\damppar}.
\end{align}
Qualitatively, as $\damppar \rightarrow 0$, each displaced envelope converges to a flat envelope, and the weighted sum of the four %
converges to just a single, flat, envelope.

\subsection{Shifted-damping Kraus operators}

Employing the mixed-state GKP encoding above in a teleportation circuit requires modifying the auxiliary states such that the code is preserved at each teleportation step. Modifying the Bell pair by appropriately shifting the input qunaught states before they are entangled on the beams splitter does the trick. This gives a Kraus operator 
    \begin{align} 
        \op{K}_{\bitsvec_{t+1}}(\mu_t) 
         & \propto 
         \Nop_{ \bitsvec_{t+1} + \corrvect{t}} 
         \op{P}_{\corrvect{t}}
         \projGKP \Nop \op{D}^\dagger(\mu_t).
         \label{eq:telKrausOp_shifted_alt}    
    \end{align}
We used the fact that performing a Pauli shift after a shifted damping operator is equal to performing the Pauli shift before a different shifted damping operator,
    \begin{align}
        \op{P}_{\vec \ell'} \op{N}_{\vec \ell} =  \op{N}_{\vec \ell + \vec \ell'} \op{P}_{\vec \ell'}.
    \end{align}
This Kraus operator is different from the original, Eq.~\eqref{eq:telKrausOp_shifted}, only in that the damping operator contains Pauli shifts. Note that the bit vector associated with outcome $\mu_t$ is $\bitsvec_{t+1}$. 
This labeling is a convenience that will make the qubit Kraus operators below cleaner but makes no material difference.
More details, including the normalization factor, can be found in Appendix~\ref{Appendix:MixedBellAncillae}. 

Encoding into a shifted-envelope state, followed by $N$ successive teleportations, and then reading out using the gadget in Eq.~\eqref{fig:povm} gives a global quantum operation. Note that no twirling is required at the readout step.
Just as in the previous section, we adjust our mathematical perspective in order to define qubit maps between GKP projectors, see Eq.~\eqref{eq:correlatedqubitop},
\begin{align} \label{eq:qubitopsshifted}
    \oldQ_{\bitsvec_1}(\mu_{1})
    &=
    \projGKP \Nop \op{D}^{\dagger}(\mu_1) \Nop_{\bitsvec_{1}} \projGKP,
    \\
    \oldQ_{\bitsvec_{t}}(\mu_t, \mu_{t-1})
     &=
    \op{P}_{\corrvect{t}}
    \Nop \op{D}^{\dagger}(\mu_t)
    \Nop_{\bitsvec_{t}+\corrvect{t-1}}
    \projGKP. \label{eq:twotimequbitkraus}
\end{align}
The only difference at this point is that the damping operators include shifts labeled by the bits $\bitsvec_t$.

\subsubsection{Twirl-aware recovery} \label{sec:logPaulitwirl}

After each round of teleportation, we require that the state is in a mixture of envelopes at four locations describing
representative shifts for the GKP Paulis (\emph{i.e.} that the damping operator is Pauli twirled). 
This is achieved by averaging over the bits $\bitsvec_t$ at each time step while using a recovery, described below, that takes $\bitsvec_t$ into account when choosing a representative shift to enact a chosen logical Pauli.

\begin{figure}[b]
    \centering
    \includegraphics[scale=.39]{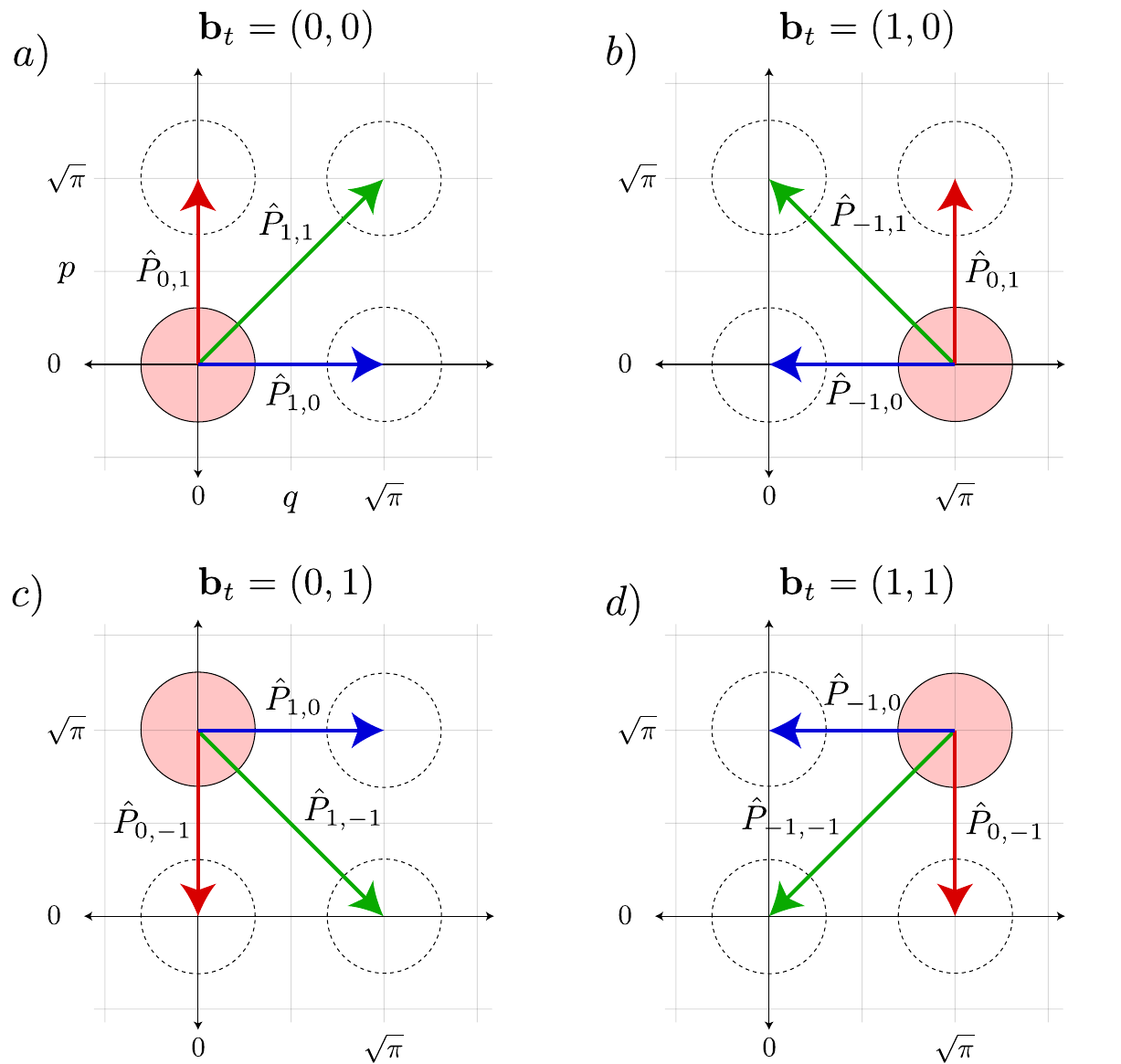}
    \caption{Graphical illustration of the twirl-aware recovery. Each subplot shows locations of the final envelope given particular bit values and corrective shifts chosen from Table~\ref{eq:shiftdecoder}. The pink circle is centered at the location of the final envelope when no correction is applied --- \emph{i.e.} the location of the bit vector $\bitsvec_t$. Arrows depict the shifts that implement $\bar{X}$ (blue), $\bar{Y}$ (green), and $\bar{Z}$ (red) logical corrections. The dotted circles are centered at locations of the final envelope after corrections. The size of the circle has no meaning, it was chosen simply for illustrative purposes.
    }
    \label{fig:bitawarerecovery}
\end{figure}

We require that the recovery (a) applies the intended logical Pauli and (b) ensures that the final shifted damping operator lies in the set $S$, Eq.~\eqref{eq:shiftvaluesorig}. 
Consider the CV Kraus operators in Eq.~\eqref{eq:telKrausOp_shifted_alt} with bit values $\bitsvec_t \in S$, Eq.~\eqref{eq:shiftvaluesorig}.
Given outcomes $\mu_t$, 
we define a correction $\hat P_{\corrvect{t}}$ via the following table,
\begin{equation} \label{eq:shiftdecoder}
    \centering
    \scalebox{1.0}{
    \renewcommand{\arraystretch}{1.3} %
    \begin{tabular}{|c|c|}
        \hline
        \small{desired correction} & \small{shift vector $\corrvect{t}$} \\
        \hline
        \rule{0pt}{2.5ex} %
        $\bar{I}$
        & $(0, 0)$ 
        \\
        $\bar{X}$
        & $\left((-1)^{b_1}, 0\right)$ 
        \\
        $\bar{Y}$
        & $\left((-1)^{b_1}, (-1)^{b_2}\right)$ 
        \\
        $\bar{Z}$
        & $\left(0, (-1)^{b_2}\right)$ \\
        \hline
    \end{tabular}
    }
\end{equation}
which are selections from the minimum-energy SB decoder, Eq.~\eqref{eq:SBdecoder_minshift}.
With these rules, the recovery is a shift that implements the chosen Pauli correction and also places the ``center'' of the final noise operator on one of the four points from the initial set $S$ in Eq.~\eqref{eq:shiftvaluesorig}. 
We illustrate this graphically in Fig.~\ref{fig:bitawarerecovery}.

\begin{figure*}[t]
    \centering
    \includegraphics[scale=.24]{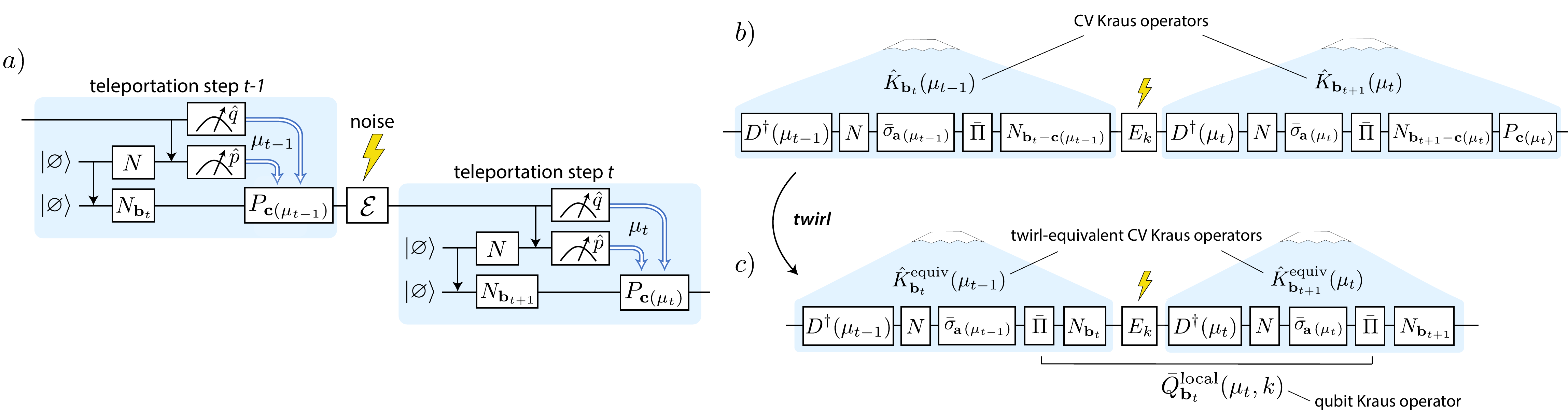}
    \caption{(a) Circuit showing a CPTP channel $\mathcal{E} = \sumint_k \op{E}_k \cdot \op{E}_k^\dagger$ acting between two rounds of teleportation with bit-shifted, damped auxiliary states. (b) The associated chain of operators using the noise channel's Kraus operators $\op{E}_k$. (c) By twirling over the local bits $\bitsvec$ and using the twirl-aware decoder, we can replace the CV Kraus operators with their twirl equivalents, Eq.~\eqref{eq:telKrausOp_shiftednew}. This allows us to identify the qubit Kraus operators in Eq.~\eqref{eq:newQubitops_extranoise}.}
    \label{fig:morenoise}
\end{figure*}

Let us work through an example. Suppose outcome $\mu_t$ is obtained, and the decoder decides that a 
$\bar{Y}$
correction should be applied. It  uses the known bit values $b_1$ and $b_2$ to select a shift using the third row in Eq.~\eqref{eq:shiftdecoder}. In the Kraus operator, Eq.~\eqref{eq:telKrausOp_shifted_alt}, this choice applies the intended correction $\bar{Y}$ followed by a shifted damping operator,
    \begin{align} \label{eq:decodingtable}
        \Nop_{\bitsvec_t + \corrvect{t}} =
        \begin{cases}
     \Nop_{1,1} & \text{if} \quad b_1, b_2 = 0, 0  \\
     \Nop_{1,0} & \text{if} \quad b_1, b_2 = 0, 1  \\
     \Nop_{0,1} & \text{if} \quad b_1, b_2 = 1, 0  \\
     \Nop_{0,0} & \text{if} \quad b_1, b_2 = 1, 1  \\
\end{cases}
    \end{align}
Reading off the subscripts on $\op{N}$, recognize that the shift vectors $\bitsvec_t + \corrvect{t}$ are elements of the set $S$ in Eq.~\eqref{eq:shiftvaluesorig}.
One can verify that the decoder choices for the other Pauli corrections, $\bar{I}$, $\bar{X}$, and $\bar{Z}$, similarly result in shift vectors on the final damping operator that also span $S$. These can also be verified by inspection in Fig.~\ref{fig:bitawarerecovery}.
These facts allow us to take an average over $\bitsvec_t \in S$ (a minimal Pauli twirl). Under this twirl, the intended Pauli correction is applied as is a convex combination of four shifted damping operators selected from $S$. This combination crucially does \emph{not} depend on the outcome
A mixture of this type is the defining feature of the encoding in Eq.~\eqref{eq:unbiasedencoding}, revealing that the final operation at each teleportation step is to re-encode into the unbiased, mixed envelope GKP encoding. 

To complete the story, note that no additional bits or twirling is necessary for the final logical readout. The POVM, Eq.~\eqref{eq:povm}, can be employed as usual.

\subsection{Minimal Pauli-twirled logical channel}

We now have all the tools to construct the logical channel. First, we define a set of \emph{twirl-equivalent CV Kraus operators} $\op{K}^{\te}$ that apply the intended qubit correction but, importantly, have no $\corrvect{t}$-dependence in the final shifted envelope:
 \begin{align} \label{eq:telKrausOp_shiftednew}
        \op{K}^{\te}_{\bitsvec_{t-1}}(\mu_t)
        \propto 
        \Nop_{ \bitsvec_{t-1}} 
        \op{P}_{\corrvect{t}}
        \projGKP \Nop \op{D}^\dagger(\mu_t).
    \end{align}
Because of the twirl-aware recovery, Eq.~\eqref{eq:shiftdecoder}, the twirl-equivalent Kraus operators give rise to the same bit-averaged (\emph{i.e.} twirled) CV map as the original set in Eq.~\eqref{eq:telKrausOp_shifted_alt}:
    \begin{align} \label{eq:telKrausmap_shifted}
        \sum_{\bitsvec_t \in S}  \op{K}^\text{\te}_{\bitsvec_t}(\mu_t) \cdot \op{K}_{\bitsvec_t}^{\te}{}^{ \dagger}(\mu_t) 
        =
      \sum_{\bitsvec_t \in S}
 \op{K}_{\bitsvec_t}(\mu_t) \cdot \op{K}^\dagger_{\bitsvec_t}(\mu_t).
    \end{align}
As long as we promise to perform the average over $\bitsvec_t$, we can use either set of CV Kraus operators to describe the same map. 
As they are local, discussed above, we use the twirl-equivalent qubit maps to identify Kraus operators for a CPTP logical channel as follows.

First, order the CV Kraus operators consecutively from state preparation to readout. Considering just two explicitly,
\begin{align}
    & \cdots \op{K}^{\te}_{\bitsvec_{t+1}}(\mu_{t}) \op{K}^{\te}_{\bitsvec_{t}}(\mu_{t-1}) \cdots
    = \\
   & \cdots \Nop_{ \bitsvec_{t+1}} \projGKP
   \underbrace{\op{P}_{\corrvect{t}} \projGKP \Nop \op{D}^\dagger(\mu_{t}) \Nop_{ \bitsvec_{t}} \projGKP}_{ \newQ_{\bitsvec_{t}}(\mu_{t})}
   \op{P}_{\corrvect{t-1}} \projGKP \Nop \op{D}^\dagger(\mu_{t-1}) \cdots
   \label{eq:bitavgKrausmap},
\end{align}
allows us to identify terms between GKP projectors as qubit operators, just as in Fig.~\ref{fig:general-teleportation-scheme}. Because the final damping operator in the twirl-equivalent CV Kraus operators does not depend on the correction, the qubit Kraus operators are local:
    \begin{align} \label{eq:newQubitops}
        \newQ_{\bitsvec_{t}}(\mu_t) 
        & 
        \coloneqq
        \frac{1}{\twirlnorm \sqrt{\pi}} 
        \GKPPauli_{\corrvect{}}
        \projGKP
        \op{N}
        \op{D}^{\dagger}(\mu_t)
        \op{N}_{\bitsvec_{t}}
        \projGKP.
    \end{align}
To aid in identifying each qubit Kraus operator, we have included an extra projector using 
$\op{P}_{\corrvect{t}} \projGKP = \projGKP \op{P}_{\corrvect{t}} \projGKP$. We label these Kraus operators with $\PT$ to indicate \emph{Pauli-twirled damped} GKP states.

This twirling procedure resolves the issues arising from the untwirled logical maps in Sec.~\ref{sec:logicalmaps} as follows. The operators in Eq.~\eqref{eq:newQubitops} are $\mu_t$-local; they have no dependence on the previous outcome $\mu_{t-1}$. 
Also, these  satisfy completeness when averaged over  the bits and $\mu_t$, as shown in Appendix~\ref{Appendix:completeness}.
This property assures that the local qubit operators form a Kraus representation of a CPTP qubit channel for error correction with twirled, damped GKP auxiliary states,
    \begin{equation} \label{eq:channel}
    \bar{\mathcal{E}}^\PT_t  = 
    \int \text{d}^2 \mu_t 
\sum_{\bitsvec_{t} \in S} \newQ_{\bitsvec_{t}}(\mu_t) \cdot [\newQ_{\bitsvec_{t}}{}(\mu_t)]^\dagger.
\end{equation} 
We emphasize that this channel has decoder freedom (which the GRN channel in the previous section does not) provided the twirl-aware recovery is used to choose the corrective shift.

The qubit Kraus operators describe a bit-averaged map, conditional on outcome $\mu_t$
\begin{align} \label{eq:bitaveragedmap}
   \mathcal{M}_{t}(\mu_t) 
    &=
   \sum_{\bitsvec_{t} \in S} \newQ_{\bitsvec_{t}}(\mu_t) \cdot [\newQ_{\bitsvec_{t}}{}(\mu_t)]^\dagger.
\end{align}
with the average over outcomes giving the channel above, $\bar{\mathcal{E}}_t = \int d^2 \mu_t \mathcal{M}_{t}(\mu_t)$. The conditional maps can be composed to give the global map that depends on the string of syndromes $\vec{\mu}$,
    \begin{align}
        \mathcal{M}_\text{global} &= \mathcal{M}_{R} \circ \mathcal{M}_{t_N} \circ \dots \circ \mathcal{M}_{t_1} \circ \mathcal{M}_{t_0},
    \end{align}
with $\bar{\rho}_\text{out} = \frac{1}{\Pr(\vec{\mu})}\mathcal{M}_\text{global}(\bar{\rho}_\text{in})$ giving the unnormalized qubit state just before projection onto $\ket{\bar{\sigma}_{\vec{a}}}$ in the readout circuit. Note that this global map differs from the one in Eq.~\eqref{eq:globalmap} in that each $\mathcal{M}_t$ here is local in $\mu_t$, and in effect we also obtain $\Pr(\vec{\mu})=\Pr(\mu_{t_n})\cdot \hdots \cdot \Pr(\mu_{t_0})$.  The syndrome averages can also be taken locally to yield
    \begin{align}
        \mathcal{E}_\text{global} &= \mathcal{E}_{R} \circ \mathcal{E}_{t_N} \circ \dots \circ \mathcal{E}_{t_1} \circ \mathcal{E}_{t_0}.
    \end{align}
One could instead also define global decoding routines, in which case this local decomposition does not hold.

The PTM for the Pauli-twirled damped ($\PT$) GKP channel, Eq.~\eqref{eq:channel}, has matrix elements
    \begin{align}\label{eq:TwirledPTMelements}
          \Gamma^\text{\PT}_{\vec a,\vec a'}
        &=
        \int \text{d}^2 \mu_t
        \sum_{\bitsvec_{t} \in S}
        \text{Tr} \frac{1}{2} \left[ \GKPPauli_{\vec a} 
        \newQ_{\bitsvec_{t}}(\mu_t)
        \GKPPauli_{\vec a'} 
        [\newQ_{\bitsvec_{t}}{}(\mu_t)]^\dagger
        \right], 
    \end{align}
We show in Appendix~\ref{Appendix:PTM} that the trace in that formula can be written as a sum of Gaussian-damped Siegel theta functions, Eq.~\eqref{eq:fock-damped-disp2}, weighted by coefficients that depend on the chosen Paulis.
Fully evaluating the expression requires specifying a decoder. Before doing so, we can inspect the qubit channel resulting from syndrome extraction alone---no correction for all $\mu_t$. 
The qubit map averaged over outcomes yields a PTM
\begin{align}
    \mat \Gamma^\text{\PT,syn} = 
       \begin{pmatrix}
        1 & 0 & 0 & 0 \\
        c_{0,1}(\damppar)  & 0 & 0 & 0 \\
        0 & 0 & 0 & 0 \\
        c_{1,0}(\damppar) & 0 & 0 & 0 \\
    \end{pmatrix}
\end{align}
where $c_{\vec{a}} (\damppar) = \frac{2}{ \twirlnorm} \text{Tr} \big[ e^{-\damppar\op{n}} \GKPPauli_{\vec{a}} e^{-\damppar\op{n}} \big]$ 
are proportional to the trace of the damped $\bar X$ and $\bar Z$ GKP Paulis, which can be evaluated as sums over damped theta functions, see Eq.~\eqref{eq:dampedPaulitrace} for more information. 

Now consider SB decoding. In Appendix~\ref{app:standardbinning_PauliChannel}, we show that SB decoding gives rise to a qubit Pauli channel, independent of damping in the GKP states. 
We do not have a compact analytic form for the matrix elements, but they can be evaluated numerically. As an example, using damped GKP states with $\damppar = 0.1$ (corresponding to 10 dB effective squeezing), the PTM is
\begin{align} \label{eq:SBbetapoint1}
    \mat \Gamma^\text{\PT,SB} = 
       \begin{pmatrix}
        1 & 0 & 0 & 0 \\
        0 & 0.9893 & 0 & 0 \\
        0 & 0 & 0.9787 & 0 \\
        0 & 0 & 0 & 0.9893 \\
    \end{pmatrix}.
\end{align}
This gives Pauli error probabilities, $p_I  = 0.9893$, $p_X = p_Z = 0.0053$, and $p_Y  = 0$.

\subsection{Including additional noise and unitaries} \label{sec:additionalnoise}

Consider a CV channel $\mathcal{E}$, which may be noise such as pure loss and dephasing or an intended logical unitary, described by a discrete or continuous Kraus representation,
    \begin{align} \label{eq:newnoise}
        \mathcal{E} = \sumint \op{E}_k \cdot \op{E}_k^\dagger
    \end{align}
satisfying completeness $\sumint \op{E}_k^\dagger  \op{E}_k = \op{I}_\text{CV}$. This channel acts before a state is error corrected, Fig.~\ref{fig:morenoise}, giving the corresponding CV Kraus operators
    \begin{align}
        \op{K}(\mu_t, k)\propto  \Nop_{ \bitsvec_t} \op{P}_{\corrvect{t}}  \projGKP \Nop \op{D}^\dagger(\mu_t) \op{E}_k.
    \end{align}
and qubit Kraus operators
    \begin{align} \label{eq:newQubitops_extranoise}
        \newQ_{\bitsvec_{t}}(\mu_t, k) 
        & 
        \propto
         \op{P}_{\corrvect{t}}
        \projGKP
        \op{N}
        \op{D}^{\dagger}(\mu_t) \op{E}_k
        \op{N}_{\bitsvec_{t}}
        \projGKP\, ,
    \end{align}
where we follow the same bit-averaging procedure to twirl the damping operators. 

The challenge in practice is evaluating the expressions for the matrix elements of the PTM. For the pure-loss channel, one may follow the prescription of Harris \emph{et al.}~\cite{harris2024}, and calculate the process matrix elements directly.
Note that $\mathcal{E}$ can also represent a unitary channel, such as a $\bar T$ gate, or an approximation thereof. We leave these studies for future work.

\section{Average gate fidelity for noisy teleportation} \label{sec:AvgGateFid}

Given that energy constraints necessitate noisy states in GKP teleportation circuit, a primary set of questions are: 
\begin{itemize}
    \item How well does teleportation using damped GKP states preserve encoded logical information?
    \item How close is standard binning decoding to optimal shift decoding?
    \item How well does a teleportation channel with GRN GKP states approximate the damped-state channel?
\end{itemize}
The CPTP qubit channel we constructed by twirling the damping operators, Eq.~\eqref{eq:channel}, allows us answer these questions. 

We use the average fidelity of quantum operation
    \begin{align}
         F_{\text{avg}}(\mathcal{E}) 
        & \coloneqq \int d \psi \bra{\psi} \mathcal{E} \big(\outprod{\psi}{\psi} \big) \ket{\psi} 
    \end{align}
to quantify how well $\mathcal{E}$ quantum information is preserved. $F_{\text{avg}}$ takes values close to one for a map that is close to the identity channel.
For a CPTP channel $\mathcal{E}$ with PTM $\mat \Gamma$, the average fidelity is given by\footnote{An extension to this formula applies even when the operation $\mathcal{E}$ is not trace preserving, see Ref.~\cite{Flammia16}.},
    \begin{align} \label{eq:gatefid}
         F_{\text{avg}}(\mathcal{E}) 
         = \frac{  \text{Tr} [ \mat \Gamma ] + d }{d\left(d+1\right)},  
    \end{align}
which uses the fact that the trace of the PTM is also related to the entanglement fidelity, $\text{Tr} [ \mat \Gamma ] = d^2 F_e(\mathcal{E})$~\cite{Nielsen02}.

\begin{figure}
    \centering
    \includegraphics[scale=.77]{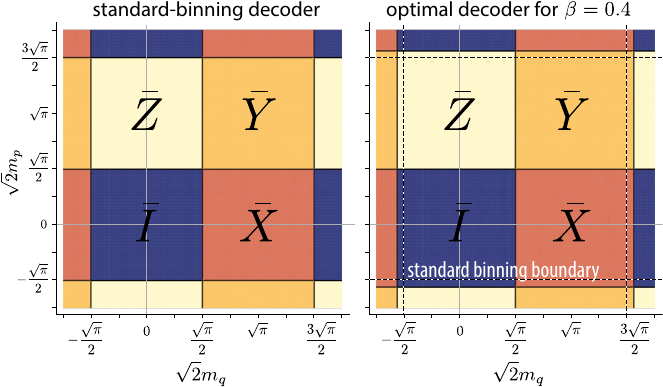}
    \caption{Comparison of the standard binning (SB) and optimal decoders. Representative patch of the SB decoder indicating which logical GKP Pauli to apply for a given syndrome $\{m_q, m_p\}$. A similar patch is shown for the numerically optimized decoder when $\damppar = 0.4$. 
    The key difference is that the decision boundaries have been shifted---the SB boundaries are shown with dashed lines for reference. The decision boundaries for optimal-decoder patches further from the origin are shifted more. 
    }
    \label{fig:decoder_comparison}
\end{figure}

\subsection{Comparing twirling procedures and decoders}

Implementing a correction strategy after syndrome extraction aims to optimize the single-syndrome gate fidelity, which in turn optimizes the average fidelity. For the stabilizer-twirled GRN teleportation in Sec.~\ref{sec:GRNGKPchannel}, the corrections are always SB. 
For the Pauli-twirled teleportation in Sec.~\ref{sec:logical_channel}, the shift to implement a corrective Pauli is selected from the twirl-aware decoder, Eqs.~\eqref{eq:shiftdecoder}, depending on the twirling-bit values $\bitsvec_t$. 
The corrective Pauli can be chosen using SB; however, the formalism above allows for other decoding choices, too. Here, we compare the SB decoder to a numerically optimized decoder that chooses the optimal Pauli for each syndrome to maximize $\Tr [\mat \Gamma]$, which is proportional to the map's overlap with the identity channel. We find (numerically) that as the damping parameter increases, the optimal decoding table, \emph{i.e.} the partition of possible measurement outcomes into correction values, graphically corresponds to an expanded version of the SB decoder --- the outcomes should be binned with respect to a different value $v > \sqrt{\pi}$. A graphical depiction of the SB decoder an optimal decoder is shown in Fig.~\ref{fig:decoder_comparison}. Although the damped-GKP qubit channel with SB decoding gives a Pauli channel, see Appendix~\ref{app:standardbinning_PauliChannel}, with optimal decoding it does not. However, it is very close to one: even for very poor quality GKP states, $\damppar = 0.4$, we find that the largest off-diagonal element in the PTM is $<10^{-4}$. As the damping is reduced, the two PTMs converge. For $\beta = 0.1$, we find that $\mat \Gamma^{\PT, \text{opt}} = \mat \Gamma^{\PT, \text{SB}}$, Eq.~\eqref{eq:SBbetapoint1}, to 4 digits of precision.

\begin{figure}
    \centering
    \includegraphics[scale=.65]{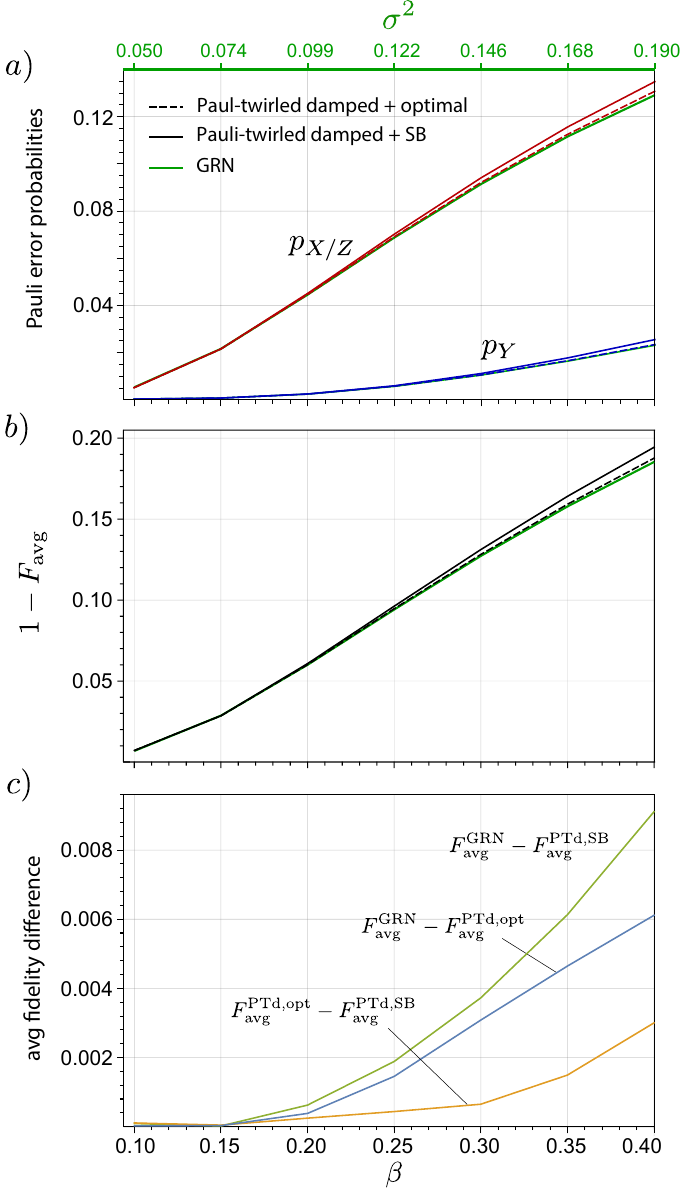}
    \caption{Comparison of qubit channels arising from teleportation using GRN GKP states and with Pauli-twirled damped (PTd) GKP states. For the latter, we compare two decoders: SB and optimal. All figures have the same $x$-axis, which labels $\damppar$ for the damped case and $ \sigma^2 = \frac{1}{2} \tanh \damppar$ for the GRN case (values shown at the top in green). (a) Pauli error probabilities. For each case, the $X$ and $Z$ probabilities are equal, so they are labeled together as $p_{X/Z}$. (b) Average gate infidelity. (c) Difference between average gate fidelities. 
    At $\damppar = 0.1$, limited numerical precision of the PTMs in our simulations gives rise to the slight increase in fidelity difference for the Pauli-twirled cases compared to GRN.
    }
    \label{fig:decoder_performance}
\end{figure}

Below, we compare three settings: the GRN qubit channel (SB decoding built-in by construction), the Pauli-twirled damped channel with SB decoding, and the Pauli-twirled damped channel with optimal decoding.
For the latter two, we take a syndrome-average over the twirled (\emph{i.e.} bit-averaged) conditional maps. 
For the two decoders, we find the conditional PTMs for each syndrome and then numerically average to find the PTMs for the different CPTP channels. Then, we use Eq.~\eqref{eq:gatefid} to compare Pauli error rates and average gate fidelities. 
Before considering the corrections, note that the average fidelity of syndrome extraction with damped GKP states, Eq.~\eqref{appendix:syndromeextraction}, is always $F_{\text{avg}} = \frac{1}{2}$, regardless of the damping parameter in the auxiliary states.  
The Pauli error rates as for increasing damping parameter are shown in Fig.~\ref{fig:decoder_performance}(a) for both decoders along with those for the GRN GKP qubit channel. 

In Fig.~\ref{fig:decoder_performance}(b), we plot the average infidelity $1-F_\text{avg}$ for the damped, Pauli-twirled damped channel using each decoder and also for the GRN GKP channel. The small differences in performance arise from different decoding near the decision boundaries---the optimal decoding patch expands as $\damppar$ increases. Figure~\ref{fig:decoder_comparison}(b) shows an extreme example of $\damppar = 0.4$ to illustrate this expansion. For small $\beta$, these regions contribute little to the average fidelity, because the probability of syndrome values away from multiples of $\sqrt{\pi}$ is Gaussian-suppressed.
Indeed, the relative error is suppressed in $\damppar$, affirming that SB decoding works extremely well, particularly in the regime $\damppar \approx 0.1$ (10 dB), near known fault-tolerance pseudo-thresholds~\cite{menicucci2014fault, bourassa2020blueprint, tzitrin2021fault, walshe2024totl}. In settings where the GKP modes are stitched into fault-tolerant cluster states, a postselection strategy has been suggested that discards qubits for syndromes near decision boundaries, as those states are most prone to error~\cite{fukui2017analog,fukui2019high}. Doing so further removes those areas where the decoders disagree. 
Our results provide further evidence to support the commonplace use of SB decoding for damping noise alone in the fault-tolerance regime. 
Also, the utility of GRN GKP approximations to the damped GKP channel in utility-scale simulations are also apparent, at least at the level of average fidelity. Although we do not analytically prove that they lower bound the infidelity here, the numerics show that over the range we considered, $0.1 \leq \damppar \leq 0.4$ this is indeed the case.

\begin{figure}
    \centering
    \includegraphics[scale=.75]{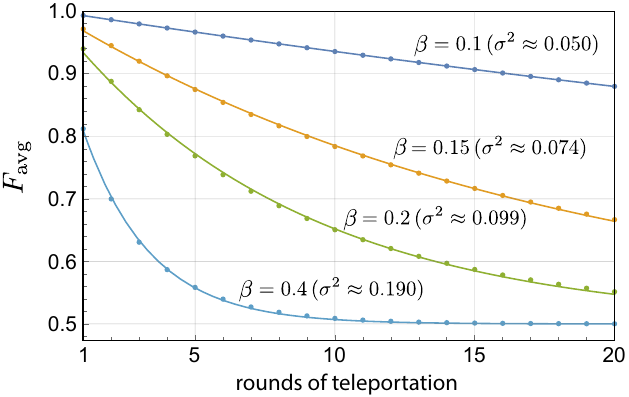}
    \caption{Average gate fidelity after $N$ rounds of noisy teleportation using the Pauli-twirl damped channel with optimal decoding. Shown are fidelities for different damping parameter (dots) and an exponential fit $\frac{1}{2} + a e^{b N}$ (lines) for each. For the values of damping parameter $\damppar = \{0.1, 0.15,0.2,0.4\}$, the exponential decay parameters are $b = \{-0.0138, 
  -0.0552, -0.117, -0.420\}$. For reference, we also give the approximate GRN variances for each $\damppar$.
    }
    \label{fig:Nrounds}
\end{figure}

Finally, in Fig.~\ref{fig:Nrounds} we plot the decay of the average gate fidelity as the number of rounds of teleportation increases for damped GKP states with the optimal decoder. From above, the SB decoder performs slightly worse and the GRN slightly better. Here, we are interested in exponential decay, which is given by the fitted curves for each value of $\damppar$. Regardless of $\damppar$, $F_\text{avg} \rightarrow 0.5$ as $N \rightarrow \infty$. In a quantum computing setting, where the teleportation wires are stitched to one another into a cluster state~\cite{tzitrin2021fault, walshe2024totl}, the role of the foliated code is to deal with the errors introduced by the noisy auxiliary GKP states (in addition to other noise).

\section{Discussion}

In summary, we have provided two recipes to find CPTP logical channels for approximate GKP error correction by channel twirling the damping operators $\Nop$. One recipe involves twirling over the full stabilizer group, and the other over a minimal set of representative Pauli shifts.
These techniques can be applied in a more general setting where the circuit involves teleporting through an auxiliary state $(\mathcal{E}_1 \otimes \mathcal{E}_2 )( \outprod{\bar{\Phi}}{\bar{\Phi}} )$. Here $\mathcal{E}_1$ and $\mathcal{E}_2$ are local but non-identical noises of arbitrary structure (not necessarily damping operators) acting on an ideal Bell pair.
The Bell pair guarantees that a GKP projector appears in the Kraus operator, and the twirling takes care of any biases that the noise might introduce. 
Note that it is possible to create noisy pairs of that form with generic, identical Gaussian CP maps (up to displacements) on each of the input states before the beam splitter.
Extensions to GKP codes on other lattices \cite{shaw2024gates, walshe2024totl} is also straightforward, since the square-lattice GKP code words are related to any other single-mode GKP code via a Gaussian unitary. 

The logical channel identification for minimal Pauli twirling relies on channel twirling the damping operator on the second mode of each GKP Bell pair in the teleportation chain. Additional twirling over the other damping operator (on the first mode) in a fashion that is classically correlated with the final twirl from the previous teleportation step can be used to eliminate logical information from the syndromes.

Although we have focused here solely on coherent noise on the GKP Bell pairs used in teleportation, other sources of decoherence can be studied with the tools in Sec.~\ref{sec:additionalnoise}, including those most relevant in optics: joint loss and dephasing~\cite{Leviant2022quantumcapacity,kyle2023randomcodes} and imperfect state preparation~\cite{arsalan2025stellar}. When other noises are included, it remains to be seen how  GKP quantum information is damaged by these processes and whether SB decoding retains its stellar performance relative to optimal look-up tables.
A next step is to adapt this procedure to multi-mode logical channels for concatenated-GKP codes or in GKP cluster-state settings. A physically sound description of composable logical channels is extremely useful in the  design and benchmarking of large scale quantum computers using the GKP code. 

Finally, we note that the minimal Pauli twirling technique may find applications to codes beyond GKP, both bosonic and qubit-based. However, our derivation relies on two key properties of the GKP code: two qunaughts on a beam splitter create a GKP Bell pair, and two identical damping operators commute through a beam splitter. These properties do not have obvious analogues in other codes, so adaptions would be required. 

\section{Acknowledgments}
The authors thank Pavithran Sridharan for help with the Wigner function plots. We also thank Matthew Stafford, Tom Harris, Takaya Matsuura, Julian Nauth, Nicolas Menicucci, Guo (Jerry) Zheng, and Arsalan Motamedi for discussions. B.Q.B. acknowledges support from the Australian Research Council Centre of Excellence for Quantum Computation and Communication Technology (Project No. CE170100012) and the Japan Science and Technology Agency through the MEXT Quantum Leap Flagship Program (MEXT Q-LEAP).

\notoc
\bibliography{ref}

\clearpage
\newpage

\appendix

{%

\section{Action of the damping operator in phase space} \label{appendix:dampingoperatorphasespace}
}
One Stinespring dilation of the damping operator is the following circuit, which teleports an input state through a two-mode squeezed state,
\begin{align} \label{KnillBScircuitdamping}
        \begin{split}
        \raisebox{-2.2em}{$e^{-\damppar \op{n}} = $}
        \quad \quad \quad \quad
     \Qcircuit @C=0.45cm @R=0.75cm {
    &                      &\qw       &\qw &\bsbal{1} &\rstick{\hspace{-0.25cm}\custommeter[$\op{q}$]{$0$}} \qw  \\
    &\lstick{ \op{S}(r) \ket{\text{vac}} } &\bsbal{1} &\qw &\qw       &\rstick{\hspace{-0.25cm}\custommeter[$\op{p}$]{$0$}} \qw  \\
    &\lstick{ \op{S}^\dagger(r) \ket{\text{vac}} } &\qw       &\qw &\qw       &\qw                   \\
    }
        \end{split}
    \end{align}
\blk
with squeezing parameter $r$ related to the damping parameter via $e^{-\damppar} = \tanh r$. The evolution of input state $\op \rho$ is found by transforming the input Wigner function,
    \begin{align}
        W_\text{in}(\vec{q},\vec{p}) \propto 
        W_{\op \rho}(q_1,p_1) G_{\frac{1}{2}e^{-r}}(q_2,p_2) G_{\frac{1}{2}e^{r}}(q_3,p_3),
    \end{align}
where $G$ is given in Eq.~\eqref{eq:GaussianFunc}. First, transform the Wigner-function arguments via the (inverse) symplectic matrix $\mat{S}^{-1} = \mat{S}_{B_{23}}^{-1} \mat{S}_{B_{12}}^{-1}$. Then, project onto the measurement outcomes by setting $q_1 = 0$ and $p_2 = 0$ and integrate over $p_1$ and $q_2$. This gives
 \begin{align}
        W_\text{out}(q,p) &\propto G_{\frac{1}{2} c_{2r}}(q,p) \int d\tau_1 d\tau_2 W_{\op \rho}(\tau_1,\tau_2) \nonumber \\
        &\quad \times   G_{\frac{1}{2} c_{2r}^{-1} }(\tau_1 - t_{2r}q, \tau_2 -  t_{2r}p)
    \end{align}
with $c_{x} \coloneqq \cosh x$ and $t_x \coloneqq \tanh x$.
Using $e^ {2r} = \coth \frac{\damppar}{2}$, 
which follows from $\coth^{-1} x = \frac{1}{2} \ln \big( \frac{x + 1}{x-1} \big)$, and using a change of variables, one finds Eq.~\eqref{eq:wigenvelope} in the main text.

\section{The Pauli Transfer Matrix (PTM)} \label{appendix:PTM}

A completely-positive (possibly trace decreasing) map for a single qubit can be described 
    \begin{align} \label{eq:qubitmap}
        \mathcal{M} = 
        \frac{1}{2} 
        \sum_{\vec a, \vec a'} \chi_{\vec{a}, \vec{a}'} \GKPPauli_{\vec a} \cdot \GKPPauli_{\vec a'}
    \end{align}
where $\mat \chi$ is the process matrix, and the factor $\tfrac{1}{2}$ comes from the fact that the Pauli matrices are not a normalized operator basis.    
Recall that the labels $\vec{a},\vec{a}'$ are each just a two-bit value specifying one of the four Paulis, Eq.~\eqref{eq:ideal-paulis}.
The diagonal elements give the Pauli error probabilities $p_I$, $p_X$, $p_Y$, and $p_Z$.
Given a Kraus representation, $\mathcal{M} = \sum_n \bar{K}_n \cdot \bar{K}_n^\dagger$, the elements of $\mat{\chi}$ are
    \begin{equation}  \label{eq:chimatelements}
    \chi_{\vec{a}, \vec{a}'} = \frac{1}{2}
     \sum_{n} \tr[ \bar K_n \GKPPauli_{\vec a}] \tr[ \bar K^\dagger_n \GKPPauli_{\vec a'}] ,
    \end{equation}
When $\mathcal{M}$ is trace-preserving, $\Tr[\mat \chi] = 1$  
, and it describes a CPTP quantum channel when the Kraus decomposition resolves the identity, $ \sum_n \bar{K}^\dagger_n \bar{K}_n = \bar{I}$.

An alternate description of $\mathcal{M}$ employs the Pauli-Liouville representation, where each Pauli $\bar{\sigma}_{\vec a}$ is assigned a basis vector $\ket{\vec a}$ in a larger Hilbert space with $\inprod{\vec{a}}{\vec{a}'} = \delta_{\vec{a},\vec{a}'}$. $\mathcal{M}$ is represented as a $4 \times 4$ \emph{Pauli transfer matrix} (PTM), 
    \begin{align} \label{eq:PTM}
        \mat{\Gamma}_\mathcal{M} = \sum_{\vec{a}, \vec{a}'} \Gamma_{\vec{a}, \vec{a}'} \outprod{\vec a}{\vec a'}
    \end{align}
with matrix elements given by
    \begin{equation}\label{eq:PTMelementsdef}
          \Gamma_{\vec{a}, \vec{a}'} = \frac{1}{2} \text{Tr} \big[ \GKPPauli_{\vec a} \mathcal{M}(\GKPPauli_{\vec a'}) \big].
    \end{equation}
The rows/columns are labeled using the correspondence $\vec{a} \in \{ (0,0), (0,1), (1,1), 0,1 \} \rightarrow \{0, 1, 2, 3 \}$.

The Pauli-Liouville representation of a CP map has various useful properties~\cite{Wallman_2014}. The entries of $\mat \Gamma$ are real, and the PTM for the adjoint map $\mathcal{M}^{\dagger}$ is the matrix transpose, $\mat \Gamma_{\mathcal{M}^\dagger} = (\mat \Gamma_{\mathcal{M}})^\tp$.
The composition of two CP maps, $\mathcal{M}_1$ and $\mathcal{M}_2$, is the matrix product of their respective PTMs, $\mat \Gamma_{\mathcal{M}_2\circ \mathcal{M}_1} = \mat \Gamma_{\mathcal{M}_2} \mat \Gamma_{\mathcal{M}_1}$.
In block-matrix form,
\begin{equation}
    \mat \Gamma_{\mathcal{M}}=
    \begin{pmatrix} 
    \text{Pr}_\text{surv}
    & \vec{\beta}\\
    \vec{\alpha} & \mat \gamma
    \end{pmatrix},
\end{equation}
the Pauli-Liouville representation reveals information about $\mathcal{M}$. For a $d$-dimensional system, $\mat \gamma$ is a $(d^2 -1)\times (d^2 -1)$ block,
the $1 \times (d^2 -1)$ vector $\vec{\beta}$ is the \textit{state-dependent leakage} block, and the $(d^2 -1)\times 1$ vector 
$\vec{\alpha}$ is the \emph{non-unital block}.
Note that $\vec{\alpha}=\vec{0}$ if and only if the map is unital, \emph{i.e.}, if the identity is a fixed point of the map.
The \emph{survival probability}, 
        $\text{Pr}_\text{surv} = \frac{1}{d} \text{Tr} \big[ \mathcal{M}( \bar{I} ) \big],$
with $0 \leq \text{Pr}_\text{surv} \leq 1$, quantifies how trace-preserving $\mathcal{M}$ is. For a trace-preserving map, Eq.~(\ref{eq:PTM}) reduces to 
\begin{align} \label{tracepreservingPTM}
    \mat \Gamma_{\mathcal{M}}=
    \begin{pmatrix}
    1 & \vec{0}\\
    \vec{\alpha} & \mat \gamma
    \end{pmatrix}. 
\end{align}
For identity channel $\mathcal I$, the off-diagonal terms are $\vec{0}$, and $\mat \Gamma_\mathcal{I}$ is the $d^2 \times d^2$ identity matrix satisfying $\Tr [\mat \Gamma_\mathcal{I}] = d^2$.

For a CPTP Pauli channel, both the process matrix and the PTM are diagonal. The elements of the PTM are linear combinations of the Pauli error probabilities,
    \begin{subequations}
    \begin{align}
        \Gamma_{II} &= 1,\\
        \Gamma_{XX} &= p_I + p_X - p_Y - p_Z,\\
        \Gamma_{YY} &= p_I - p_X + p_Y - p_Z,\\
        \Gamma_{ZZ} &= p_I - p_X - p_Y + p_Z.
    \end{align}
    \end{subequations}
Similarly, the error probabilities (diagonal elements of the process matrix $\mat{\chi}$) are found from the diagonal elements of the PTM as:
    \begin{subequations}
    \begin{align}
        p_I &= \tfrac{1}{4} \big( \Gamma_{II} + \Gamma_{XX} + \Gamma_{YY} + \Gamma_{ZZ} \big) \\
        p_X &= \tfrac{1}{4} \big( \Gamma_{II} + \Gamma_{XX} - \Gamma_{YY} - \Gamma_{ZZ} \big)\\
        p_Y &= \tfrac{1}{4} \big( \Gamma_{II} - \Gamma_{XX} + \Gamma_{YY} - \Gamma_{ZZ}\big)\\
        p_Z &= \tfrac{1}{4} \big( \Gamma_{II} - \Gamma_{XX} - \Gamma_{YY} + \Gamma_{ZZ}\big).
    \end{align}
    \end{subequations}

\section{Mixed GKP Bell pairs} \label{Appendix:MixedBellAncillae}

To teleport into the mixed-envelope code, we prepare a mixture of displaced enveloped GKP Bell pairs,
    \begin{equation} 
\label{appeq:noisybeamsplitterBellpair_shifted}
 \op{\rho} = \sum_{\bitsvec', \bitsvec \in S} \text{Pr}_{\bitsvec, \bitsvec'} \outprod{\bar{\Phi}_{\damppar,\bitsvec', \bitsvec}^+}{\bar{\Phi}_{\damppar,\bitsvec', \bitsvec}^+}
    \end{equation}
\blk
each of which has the normalized form,
   \begin{equation} \label{appeq:dampedBellpair}
       \ket{\bar{\Phi}_{\damppar,\bitsvec',\bitsvec}^+} = \frac{1}{\sqrt{ \mathfrak{n}_{\text{Bell},\bitsvec', \bitsvec} } }
 \op{N}_{\bitsvec'} \otimes \Nop_{\bitsvec} \ket{\bar{\Phi}^+}. 
    \end{equation}
Here we allow both modes to experience shifted damping. For the channel twirling in main text, this is only required on the second mode; we include the other for completeness.
To find the normalization factor, we note that $\op{N}_{\bitsvec} = \op{P}_{\bitsvec} \op{N} \op{P}^\dagger_{\bitsvec}$ 
and 
$\op{P}_{\bitsvec} = \op{D}\big(\sqrt{\tfrac{\pi}{2}}(b_1 + i b_2)\big)$. Using this in the unnormalized form of Eq.~\eqref{appeq:noisybeamsplitterBellpair_shifted} gives
    \begin{equation} 
    \begin{split}
       \quad \quad 
       \Qcircuit @C=1em @R=1em 
        {
        \lstick{ \ket{\varnothing}} &\bsbal{1} &\gate{N_{\bitsvec'}}  &\qw \\
        \lstick{ \ket{\varnothing}} &\qw       &\gate{N_{\bitsvec}} &\qw 
	} \, 
        &\raisebox{-1em}{\quad = \quad \quad \quad }
        \Qcircuit @C=1em @R=0.7em 
        {
        \lstick{\ket{\varnothing}} &\gate{ {D}(\alpha_+) N {D}^\dagger(\alpha_+)}  &\bsbal{1} &\qw \\
        \lstick{\ket{\varnothing}} &\gate{ {D}(\alpha_-) N {D}^\dagger(\alpha_-)} &\qw       &\qw 
	} 
    \end{split}
    \end{equation}
where $\alpha_\pm \coloneqq \mp \frac{\sqrt{\pi}}{2}(b'_1 \pm b_1 + i(b'_2 \pm b_2))$.\footnote{These displacement amplitudes are not logical Pauli operators with respect to the square-lattice GKP code, although they are with respect to rectangular-lattice codes---for which quanught states are Pauli eigenstates.}
We express the damped state, Eq.~\eqref{appeq:dampedBellpair}, in terms of two shifted, damped qunaughts on a beam splitter with normalizations
    \begin{align}
        \mathfrak{n}_{\varnothing,\pm} \coloneqq \bra{\varnothing} \op{D}(\alpha_\pm) \op{N}^2 \op{D}^\dagger(\alpha_\pm) \ket{\varnothing} \, ,
    \end{align}
and the damped Bell pair is normalized by the square root of their product, $\sqrt{\mathfrak{n}_{\text{Bell},\bitsvec', \bitsvec}} \coloneqq \sqrt{\mathfrak{n}_{\varnothing,+} \mathfrak{n}_{\varnothing,-}}$. 
In order for the state in Eq.~\eqref{appeq:noisybeamsplitterBellpair_shifted} to be written as a mixture over displaced envelopes, we choose $\text{Pr}_{\bitsvec, \bitsvec'} = \mathfrak{n}_{\text{Bell},\bitsvec', \bitsvec} / \sum_{\bitsvec', \bitsvec \in S} \mathfrak{n}_{\text{Bell},\bitsvec', \bitsvec} $.

The Kraus operator applied when teleporting through Eq.~\eqref{appeq:dampedBellpair} and applying a corrective Pauli shift is
    \begin{align} \label{eq:twobitKrausops}
        \op{K}_{\bitsvec, \tilde{\bitsvec}'}(\mu) 
         & = \frac{1}{\sqrt{ \mathfrak{n}_{\text{Bell},\bitsvec', \bitsvec} } } 
         \op{P}_{\corrvect{t}} 
         \Nop_{ \bitsvec} \projGKP \Nop_{\tilde{\bitsvec}'} \op{D}^\dagger(\mu) .
    \end{align}
Take note that the shift vector $\bitsvec'$ has become $ \tilde{\bitsvec}' = (-b'_1, b'_2)$ after bouncing it onto the final mode in the teleportation circuit, since $\op{p}_1 \rightarrow -\op{p}_2$~\cite{walshe2020continuousvariable}. The state used in the main text, Eq.~\eqref{eq:telKrausOp_shifted_alt}, has $\bitsvec' = \tilde{\bitsvec}' = \vec{0}$.

\section{Completeness of the local qubit Kraus operators} \label{Appendix:completeness}

The qubit maps in Eq.~\eqref{eq:qubitopsshifted} satisfy completeness in the logical subspace when averaged over the local bits $\bitsvec_{t}$ arising from the previous teleportation step: 
    \begin{align}
    & \int d^2 \mu_t \sum_{\bitsvec_{t} \in S} \newQ_{\bitsvec_{t}}(\mu_t){}^\dagger \newQ_{\bitsvec_{t}}(\mu_t) \\
    &= 
     \frac{1}{\twirlnorm^2} \int \frac{d^2 \mu_t}{\pi } \sum_{\bitsvec_{t} \in S} 
     \projGKP \op{N}_{\bitsvec_{t}} \op{D}(\mu_t) \op{N} \projGKP \op{N} \op{D}^{\dagger}(\mu_t) \op{N}_{\bitsvec_{t}} \projGKP \\
    &= 
     \frac{1}{2 \twirlnorm}  \sum_{\bitsvec_{t} \in S} 
     \projGKP \op{N}^2_{\bitsvec_{t}}  \projGKP \\
   &= 
     \frac{1}{2 \twirlnorm}  \sum_{\vec{a}} \GKPPauli_{\vec{a}}  
     \projGKP \op{N}^2 \projGKP \GKPPauli_{\vec{a}}  \\  
    &= 
    \projGKP.       
    \end{align}
In the second line, we evaluated the integral over $\mu_t$ using Schur's lemma for continuous variables, 
    \begin{equation} \label{eq:SchurLemmaCV}
        \int \frac{ d^2 \alpha}{\pi} \op{D}^\dagger(\alpha) \op{A} \op{D}(\alpha) = \Tr[\op{A}] \op{I}_\text{CV} ,
    \end{equation}
and used $\Tr[\Nop \projGKP \Nop] = \Tr[\projGKP \Nop^2 \projGKP] = \tfrac{1}{2} \twirlnorm$.\footnote{Let $\phi: G \rightarrow GL(V)$ be an irreducible representation of a finite group G (has finite number of elements). If a matrix $A$ commutes with $\phi(g)$ for all $g \in G$, that is $A \phi(g)=\phi(g) A,~~ \forall g \in G$, then $A=\alpha \hat{\mathds{1}}$ for some $\alpha \in \mathbb{C}$. In other words, if you are given a bunch of matrices, $\phi_1, \dots \phi_n$, that form an irreducible representation of a finite group, only multiples of the identity matrix commute with the $\phi_i$. Because a twirled channel commutes with the elements of the group (or set) over which it was twirled, according to Schur's lemma, it is proportional to the identity map.} Then, we extracted the shifts from the damping operator using Eq.~\eqref{eq:shifteddamping} and pulled them through the projector, turning them into GKP logical Paulis, see Eq.~\eqref{eq:Pauliconnections}. A qubit Pauli twirl, Eq.~\eqref{eq:Paulitwirl}, leads to the final equality.

\widetext
\section{Useful inner products for GKP states} \label{appendix:innerproduct}

The inner product of an operator $\op{O}$ between two ideal GKP computational basis eigenstates can be written as 
    \begin{align} \label{eq:arbGKPinprod}
        \bra{\bar{j}} \op{O} \ket{\bar{k}} 
        & = \frac{1}{\pi}
        \int d^2 \alpha \, \chi_O(\alpha) \, \bra{\bar{j}} \op{D}(\alpha) \ket{\bar{k}} 
    \end{align}
where $\chi_O(\alpha) = \Tr[\op{O} \op{D}^\dagger(\alpha)]$ is the characteristic function for operator $\op{O}$. 
We explicitly evaluate the inner product by expressing the GKP states in the position basis, Eq.~\eqref{eq:defwords}. Then, using $\op{D}(\alpha) \qket{s} = e^{i \alpha_I \alpha_R} e^{i s\sqrt{2} \alpha_I}\qket{s + \sqrt{2}\alpha_R}$ where $\alpha = \alpha_R + i \alpha_I$ and $\inprodsubsub{s}{s'}{q}{q} = \delta(s-s')$, the inner product is
\begin{equation}
    \begin{split}
   \bra{\bar{j}}
   \hat{D}(\alpha)\ket{\bar{k}}
     &=
    \sum_{n,m=-\infty}^{+\infty}
  e^{i \alpha_I \alpha_R}
    e^{i\sqrt{2 \pi}\alpha_I \left(2m+k\right)}
     \delta \big(
     \left(2n-2m+j-k\right)\sqrt{\pi}
     -\sqrt{2}\alpha_R 
     \big)
    \end{split}
\end{equation}
The delta function indicates that $\sqrt{2} \alpha_R \in \sqrt{\pi}\mathbb{Z}$, otherwise the overlap is zero. Further, $\sqrt{2} \alpha_R$ must be an even integer multiple of $\sqrt{\pi}$ if $j=k$ and an odd integer multiple if $j\neq k$. We expect a similar relation for $\alpha_I$.
Because the summation index $n$ only appears inside the delta function, the above expression is invariant under any integer translation $n \rightarrow n + k$, where $k \in \mathbb{Z}$. This allows us to eliminate $m$ from the delta function. 
Then, we use the Fourier series representation of a Dirac comb, Eq.~\eqref{eq:DiracComb}, and $f(x)\delta(x-c)=f(c)\delta(x-c)$ 
to get
\begin{align} \label{eq:usefulinnerproduct}
   \boxed{
   \bra{\bar{j}}
   \hat{D}(\alpha)\ket{\bar{k}}
     =
     \sqrt{\pi} \sum_{n,m=-\infty}^{+\infty}
  e^{i \frac{\pi}{2}m(2n+j+k)}
    \delta \big( \sqrt{\pi} m-\sqrt{2} \alpha_I \big)
     \delta \big(
     \sqrt{\pi}\left(2n+j-k\right)
     -\sqrt{2} \alpha_R 
     \big)} 
\end{align}
The asymmetry in the delta functions is due to the fact that we employed a position-basis description of the GKP code words. Similar calculations can be found in Appendix D of Ref.~\cite{albert2018performance} and Appendix E of Ref.~\cite{Zheng2024pureloss}.

It is instructive to consider two examples. 
First, a displacement with $\alpha_R = \sqrt{\pi/2}$ and $\alpha_I = 0$ implements a GKP Pauli $X$. In Eq.~\eqref{eq:usefulinnerproduct}, this gives $ \delta \left(\sqrt{\pi} m \right) \implies \delta_{m,0} $, rendering the phase trivial, and $\delta\left( \sqrt{\pi} - \sqrt{\pi} \left(2n+k-j\right) \right) \implies \delta_{j-k,2n-1}$. This gives $|j-k| = 1$, as expected. 
Second, a displacement with $\alpha_R = 0$ and $\alpha_I = \sqrt{\pi/2}$ implements a GKP Pauli $Z$. In Eq.~\eqref{eq:usefulinnerproduct}, this gives $ \delta \left(\sqrt{\pi}-\sqrt{\pi} m \right) \implies \delta_{m,1} $, and $\delta\left( - \sqrt{\pi} \left(2n+k-j\right) \right) \implies \delta_{j-k,2n}$. This gives $|j-k|=0$, with the phase $e^{i\pi j}$ being $0$ when $j=k=0$ and $-1$ when $j=k=1$, as expected.

\subsection{Damped displacement operator}

We compute the inner product in Eq.~\eqref{eq:arbGKPinprod} given a ``damped displacement operator'' $\op{O} = e^{-\damppar \hat{n}} \hat{D}(\gamma) e^{-\damppar \hat{n}}$.
Using the characteristic-function expansion of the damping operator~\cite{gottesman2001encoding,noh2019fault}, 
    \begin{align} \label{eq:charfun_damping}
         e^{-\damppar \op{n}} = \frac{1}{\pi} \frac{1}{1-e^{-\damppar}} \int d^2 \, \alpha e^{-\frac{1}{2 \tanh (\damppar/2)}|\alpha|^2} \op{D}(\alpha)
     \end{align}
changing variables, and eliminating integrals with delta functions, we find the characteristic-function description,
\begin{align} \label{eq:env-disp-env}
\boxed {
        e^{-\damppar \hat{n}}
        \hat{D}(\gamma)
        e^{-\damppar \hat{n}}
        =
         \frac{ \tanh(\frac{\damppar}{2})}{\pi (1-e^{-\damppar})^2}
         e^{-\frac{1}{2}\tanh \damppar |\gamma|^2}
        \int \text{d}^2\alpha
        e^{
        -\frac{1}{2} \coth \damppar 
        |\alpha - \gamma \, \text{sech} \damppar  |^2 }
        \hat{D}(\alpha)
        }
\end{align}
which is a two-dimensional Gaussian in $\alpha_R$ and $\alpha_I$ with diagonal covariance matrix:
         $e^{
        -\frac{1}{2} \coth \damppar 
        |\alpha - \gamma\text{sech} \damppar |^2 
        }
        =
         e^{
        -\frac{1}{2\tanh \damppar }
        (\alpha_R - \gamma_R \text{sech} \damppar )^2 
        }
        e^{
        -\frac{1}{2\tanh \damppar }
        (\alpha_I - \gamma_I \text{sech} \damppar )^2
        }.$

Using Eq.~\eqref{eq:env-disp-env} in Eq.~\eqref{eq:arbGKPinprod} and making use of Eq.~\eqref{eq:usefulinnerproduct}, 
the integrals can be solved using Dirac deltas to obtain
\begin{equation}\label{eq:fock-damped-disp-infinite-sum}
    \begin{split}
         \bra{ \bar j}
        e^{-\damppar \hat{n}}
        \hat{D}(\gamma)
        e^{-\damppar \hat{n}}
        \ket{\bar k}
        =&
     \frac{\tanh(\frac{\damppar}{2})e^{-\frac{1}{2} \coth \damppar |\gamma|^2}}{ 2\sqrt{ \pi} (1-e^{-\damppar})^2}
     \sum_{m,n=-\infty}^{+\infty}
         e^{\frac{i\pi}{2} m  (2n+k+j)}
         e^{ -\frac{\coth\damppar}{2} \big[ \frac{\pi}{2} m^2 - \sqrt{2 \pi} m \gamma_I \sech \damppar \big] }
     \\
     &\times 
     e^{-\frac{\coth\damppar}{2} \big[ \frac{\pi}{2} (2n+k-j)^2 - \sqrt{2 \pi}(2n+k-j) \gamma_R \sech \damppar  \big]
     }
     .
    \end{split}
\end{equation}
We can express the sums over $m$ and $n$ in terms of a two-dimensional Riemann-Siegel $\Theta$-function,
\begin{equation} \label{eq:SiegelThetadef}
    \begin{split}
\Theta\left(\mathbf{z},\bm{\tau}\right) 
\coloneqq
\sum_{\mathbf{n}\in \mathbb{Z}^2}
   \text{exp}\left(\pi i \mathbf{n}^\text{T}\bm{\tau}\mathbf{n}
    +2 \pi i \mathbf{n}^\text{T}\mathbf{z}\right),
    \end{split}
\end{equation}
where $\bm{\tau}$ is a complex symmetric matrix in the Siegel upper half plane ($\text{Im } \mat{\tau} > 0$), and $\mathbf{z} \in \mathbb{C}^2$ is a two-dimensional column vector.
Doing so, we get
\begin{equation}\label{eq:fock-damped-disp2}
    \boxed{
    \bra{ \bar j}
        e^{-\damppar \hat{n}}
        \hat{D}(\gamma)
        e^{-\damppar \hat{n}}
        \ket{\bar k}
        =
        \frac{\tanh(\frac{\damppar}{2})
        e^{-\frac{1}{2} \coth \damppar |\gamma|^2}}{ 2 \sqrt{\pi} (1-e^{-\damppar})^2}
       e^{
        -\frac{\pi}{4}(j-k)^2 \coth \damppar
        + (j-k) \sqrt{\frac{\pi}{2}} \gamma_R  \csch \damppar 
        }
        \Theta\big(\vec{z}_{j,k}(\gamma,\damppar),\mat{\tau}(\damppar)\big)
        }
\end{equation}
where 
    \begin{align} \label{eq:taumatrix}
        \mat \tau(\damppar) &=\
            \begin{pmatrix}
                i \coth \damppar & \frac{1}{2} \\
                \frac{1}{2} & i \frac{1}{4} \coth \damppar
            \end{pmatrix},
        \\
        \vec z_{j,k}(\gamma, \damppar) 
         & =
        \big(
            i \tfrac{1}{2}(j - k) \coth \damppar
           - i \tfrac{1}{\sqrt{2\pi}} \gamma_R \csch \damppar  ~,~
           \tfrac{1}{4}(j+k)
            - \tfrac{i}{\sqrt{8 \pi}} \gamma_I \csch \damppar 
        \big)^\tp.
    \end{align}
Another form of this calculation can be found in Albert \emph{et al.}~\cite{albert2018performance}.

For small damping, $\damppar \ll 1$, elements of the $\mat{\tau}$-matrix become quite large. In this case, we can use an alternative form of the Siegel theta function in Eq.~\eqref{eq:SiegelThetadef} arising from Jacobi identities, 
\begin{align} \label{eq:SiegelTheta_alt}
\Theta\left(\mathbf{z},\bm{\tau}\right) 
     & = \sqrt{|i \mat{\tau}^{-1}|} e^{ - i \pi  \vec{z}^\tp \mat{\tau}^{-1} \vec{z}} \Theta \left(\mat{\tau}^{-1} \mathbf{z},- \mat{\tau}^{-1}\right).
\end{align}
Using this form, the $\mat{\tau}$-matrix from Eq.~\eqref{eq:taumatrix} is transformed to
 \begin{align}
        [\mat \tau(\damppar) ]^{-1} &=
            \begin{pmatrix}
                -\frac{i}{2} \tanh 2\damppar & 1 - \sech 2\damppar \\
                1 - \sech 2\damppar & -2 i  \tanh 2\damppar
            \end{pmatrix}
            \xrightarrow{\damppar \ll 1} 
             \begin{pmatrix}
                -i \damppar & 0 \\
                0 & - 4 i \damppar
            \end{pmatrix} + \mathcal{O}(\damppar^2).
    \end{align}
In the small $\damppar$ limit, the off-diagonals of the matrix vanish, and the two-dimensional Siegel theta function can be expressed a product of two one-dimensional Jacobi theta functions.

\subsection{Trace of the damped Paulis} 
For $\gamma = 0$, the formula in Eq.~\eqref{eq:fock-damped-disp2} gives the overlap of damped computational basis states (up to their infinite normalizations)~\cite{matsuura2020equivalence}.
Using a computational-basis decomposition of the GKP subspace Paulis, $\GKPPauli_{\vec{a}} = \sum_{j,k} c_{j,k}^{\vec{a}} \outprod{\bar j}{\bar k}$, where $c_{j,k}^{\vec{a}} = \bra{\bar j} \GKPPauli_{\vec{a}} \ket{\bar k} = (\mat{\sigma}_{\vec{a}})_{j,k}$ are elements of the appropriate Pauli matrix $\mat{\sigma}_{\vec{a}}$, we find
    \begin{align} \label{eq:dampedPaulitrace}
        \Tr[e^{-\damppar \op{n}} \bar \sigma_{\vec{a}} e^{-\damppar \op{n}}] 
          &= \sum_{j,k = 0}^{1} c_{j,k}^{\vec{a}} \bra{\bar{j} } e^{-2\damppar \op{n}} \ket{\bar{k}} 
          =
        \frac{\tanh(\frac{\damppar}{2})
        e^{
        -\frac{\pi}{4}(j-k)^2 \coth \damppar } }{ 2 \sqrt{\pi} (1-e^{-\damppar})^2}
        \sum_{j,k = 0}^{1} c_{j,k}^{\vec{a}}
        \Theta \left( \left(\begin{smallmatrix} 
        i \tfrac{1}{2}(j - k) \coth \damppar \\ \tfrac{1}{4}(j+k)
        \end{smallmatrix} \right) ,\mat{\tau}(\damppar)\right) .
    \end{align}
We can also use this formula to calculate the normalization factor that appears throughout:
    \begin{align}
        \mathfrak{n}_\damppar 
          = \frac{1}{2}\Tr[e^{-\damppar \op{n}} \projGKP e^{-\damppar \op{n}}] 
          = \frac{1}{2} \sum_{j = 0}^{1} \bra{\bar{j} } e^{-2\damppar \op{n}} \ket{\bar{j}} 
         =
            \frac{\tanh(\frac{\damppar}{2})
        }{ 4 \sqrt{\pi} (1-e^{-\damppar})^2}
        \Big[
        \Theta\big( \left(\begin{smallmatrix} 
        0 \\ 0
        \end{smallmatrix} \right) ,\mat{\tau}(\damppar)\big) +  \Theta\big( \left(\begin{smallmatrix} 
        0 \\ 1/2
        \end{smallmatrix} \right),\mat{\tau}(\damppar)\big) \Big]
    \end{align}
There is only one other nontrivial trace due to the properties~\cite{Hastrup2023_LossEC},
    \begin{align}
        \Tr[e^{-\damppar \op{n}} \bar \sigma_{1,1} e^{-\damppar \op{n}}] = 0 \quad \quad \text{and} \quad \quad \Tr[e^{-\damppar \op{n}} \bar \sigma_{1,0} e^{-\damppar \op{n}}] = \Tr[e^{-\damppar \op{n}} \bar \sigma_{0,1} e^{-\damppar \op{n}}] ,
    \end{align}
which are shown by inserting $\op{I} = \op{F}^\dagger \op{F}$, using the cyclic property of the trace, noting that $\op{F}$ commutes with $\op{N}$ and $\op{F} \bar \sigma_{1,1} \op{F} = -\bar  \sigma_{0,1}$ (equivalently, $\bar{H} \bar{Y} \bar{H} = -\bar{Y}$). This is a property of square-lattice GKP and does not apply generally, although a similar argument can be used to show $\Tr[e^{-\damppar \op{n}} \bar \sigma_{1,0} e^{-\damppar \op{n}}] = \Tr[e^{-\damppar \op{n}} \bar \sigma_{1,1} e^{-\damppar \op{n}}] = \Tr[e^{-\damppar \op{n}} \bar \sigma_{0,1} e^{-\damppar \op{n}}]$ for hex-lattice GKP.

\section{Evaluating the PTM for the minimal Pauli twirled channel} \label{Appendix:PTM}

We wish to find the PTM for the conditional qubit map $\mathcal{M}$ given outcome $\mu_t$. To simplify the calculation here, we separate error correction into syndrome extraction and correction $\mathcal{M}(\mu)  = \mathcal{M}_\text{corr}(\mu) \circ \mathcal{M}_\text{syn}(\mu) $. This produces no additional difficulty, since in the derived form of the qubit Kraus operators and their map, the correction has already been isolated from syndrome extraction as a (decoder-dependent) unitary channel for each outcome $\mu$. 
Also, we can compose PTMs for the channels simply; $\mat{\Gamma}(\mu) =\mat{\Gamma}^\text{corr}(\mu) \mat{\Gamma}^\text{syn}(\mu) $, noting that the PTMs for Pauli corrections are diagonal matrices with $\pm 1$ elements: 
\begin{align}
    \mat \Gamma_{\bar{I}} =
       \begin{pmatrix}
        1 & 0 & 0 & 0 \\
        0 & 1 & 0 & 0 \\
        0 & 0 & 1 & 0 \\
        0 & 0 & 0 & 1 \\
    \end{pmatrix},
    \quad      
    \mat \Gamma_{\bar{X}} =
       \begin{pmatrix}
        1 & 0 & 0 & 0 \\
        0 & 1 & 0 & 0 \\
        0 & 0 & -1 & 0 \\
        0 & 0 & 0 & -1 \\
    \end{pmatrix},
    \quad      
          \mat \Gamma_{\bar{Y}} =
       \begin{pmatrix}
        1 & 0 & 0 & 0 \\
        0 & -1 & 0 & 0 \\
        0 & 0 & 1 & 0 \\
        0 & 0 & 0 & -1 \\
    \end{pmatrix},
        \quad  
       \mat \Gamma_{\bar{Z}} =
       \begin{pmatrix}
        1 & 0 & 0 & 0 \\
        0 & -1 & 0 & 0 \\
        0 & 0 & -1 & 0 \\
        0 & 0 & 0 & 1 \\
    \end{pmatrix}
.
\end{align}
 
We obtain the matrix elements for syndrome extraction by substituting the qubit Kraus operators into Eq.~\eqref{eq:PTMelementsdef} and using $\bar{\sigma}_{\vec{a}} \projGKP = \projGKP \bar{\sigma}_{\vec{a}}  = \bar{\sigma}_{\vec{a}} $. The matrix elements for syndrome extraction are,
\begin{align} \label{eq:gamma-a-b}
        \Gamma^\text{syn}_{\vec{a},\vec{a}'}(\mu_t)
        =&
        \frac{1}{\pi \twirlnorm^2}
        \sum_{\bitsvec_t \in S}
        \text{Tr} 
        \big[ 
        \GKPPauli_{\vec{a}} \op{N}
        \op{D}^{\dagger}(\mu_t)
        \op{N}_{\bitsvec_t} \GKPPauli_{\vec{a}'}
        \op{N}_{\bitsvec_t} \op{D}(\mu_t) \op{N}
        \big]\\
        =&
        \frac{1}{\pi\twirlnorm^2}
        \sum_{\bitsvec_t \in S}
        e^{i \phi(\vec{a}',\bitsvec_t)}
        \text{Tr} 
        \big[ 
        \GKPPauli_{\vec{a}} \op{N}
        \op{D}^{\dagger}(\gamma)
        \op{N} \GKPPauli_{\vec{a}'} \op{N} \op{D}(\gamma) \op{N} 
        \big] 
\end{align}
In the above, we first separated $\op{N}_{\bitsvec}$ according to Eq.~\eqref{eq:shifteddamping} then used the cyclic property of the trace to evaluate the conjugated Pauli operator
$\op{P}^\dagger_{\bitsvec} \GKPPauli_{\vec{a}'} \op{P}_{\bitsvec} = e^{i\phi(\vec{a}',\vec{b})}\GKPPauli_{\vec{a}'}$, with the phase taking on $ \pm 1$ depending on $\vec{a}'$ and $\bitsvec$. 
Then, we combine the remaining displacements, $ \op{P}^\dagger_{\bitsvec} \op{D}(\mu_t) = e^{i\theta} \op{D}(\gamma)$, where 
    \begin{equation}
        \gamma \coloneqq \mu_t - \sqrt{\tfrac{\pi}{2}}(b_1 + i b_2),
    \end{equation}
and the phase $\theta$ is inconsequential, as it cancels with its conjugate in the expression above.

Inserting a computational-basis decomposition of the Pauli operators gives
    \begin{equation}\label{eq:PTMmidform}
        \boxed{
        \Gamma^\text{syn}_{\vec{a},\vec{a}'}(\mu_t)
        =
        \frac{1}{\pi \twirlnorm^2}
        \sum_{\bitsvec \in S} e^{i \phi(\vec{a}',\vec{n}_t)}
        \sum_{j,j',k,k' \in \mathbb{Z}_2}
        c^{\vec{a}}_{j,k} c^{\vec{a}'}_{j',k'} \bra{\bar k}
        \op{N}
        \op{D}^{\dagger}(\gamma)
        \op{N} \outprod{\bar j'}{\bar k'} \op{N} \op{D}(\gamma) \op{N} \ket{\bar j} }
    \end{equation}
Each of the inner products can be evaluated in terms of Siegel theta functions using Eq.~\eqref{eq:fock-damped-disp2}. When $\damppar \ll 1$, the alternate form in Eq.~\eqref{eq:SiegelTheta_alt} can be used. This PTM is a compact form for the trace-decreasing qubit map that depends on the damping parameter $\damppar$ and the outcome $\mu_t$.

\subsection{Average PTM for syndrome extraction} \label{appendix:syndromeextraction}

Syndrome extraction has local qubit Kraus operators, Eq.~\eqref{eq:newQubitops}, with correction $\bar{\sigma}(\mu) = \bar{\sigma}_{0,0} = \projGKP$ for all $\mu$. 
We use CV Schur's lemma, Eq.~\eqref{eq:SchurLemmaCV} to find the PTM matrix elements
    \begin{align} \label{eq:avgPTMsydex}
        \Gamma^\text{syn}_{\vec{a},\vec{a}'} 
        &= \int d^2 \mu \,\Gamma^\text{syn}_{\vec{a},\vec{a}'}(\mu_t) 
        =
        \frac{1}{\mathfrak{n}_\damppar^2}
        \text{Tr} 
        \big[ \op{N} \GKPPauli_{\vec{a}} \op{N} \big]
        \text{Tr} 
        \sum_{\bitsvec_t \in S}
        \big[ \op{N}_{\vec  n_t} \GKPPauli_{\vec{a}'} \op{N}_{\bitsvec_t} \big]
        =
        \frac{2}{ \twirlnorm}
        \text{Tr} 
        \big[ \op{N} \GKPPauli_{\vec{a}} \op{N} \big]
        \delta_{\vec{a}',(0,0)}.
    \end{align}
This verifies that $\Gamma^\text{syn}_{0,0} = 1 $ 
and that
$\Gamma^\text{syn}_{0,1} = \Gamma^\text{syn}_{0,2} =\Gamma^\text{syn}_{0,3} = 0 $, which is required for a CPTP channel, Eq.~\eqref{tracepreservingPTM}. However, the syndrome-extraction channel is not in general unital, evident in the PTMs for $\damppar \in \{ 0.4,0.2, 0.1 \}$:
\begin{align}
    \mat \Gamma^\text{syn} = \left\{
       \begin{pmatrix}
        1 & 0 & 0 & 0 \\
        0.2527  & 0 & 0 & 0 \\
        0 & 0 & 0 & 0 \\
        0.2527 & 0 & 0 & 0 \\
    \end{pmatrix},
    \quad      
       \begin{pmatrix}
        1 & 0 & 0 & 0 \\
        0.0374  & 0 & 0 & 0 \\
        0 & 0 & 0 & 0 \\
        0.0374 & 0 & 0 & 0 \\
    \end{pmatrix},
        \quad  
    \begin{pmatrix}
        1 & 0 & 0 & 0 \\
        0.0008  & 0 & 0 & 0 \\
        0 & 0 & 0 & 0 \\
        0.0008 & 0 & 0 & 0 \\
    \end{pmatrix}
    \right\}.
\end{align}
When corrections are included, Eq.~\eqref{eq:avgPTMsydex} does not apply. It was derived using Schur's lemma to average over the syndromes, which cannot be invoked due to additional $\mu_t$-dependent phases arising from the decoder-dependent corrections. Stated another way, the syndrome-averaged matrix elements are those arising from the composed conditional PTM, $ \Gamma_{\vec{a},\vec{a}'} = \int d^2 \mu_t \,  \Gamma_{\vec{a},\vec{a}'}(\mu_t) = \int d^2 \mu_t \, \big[ \mat{\Gamma}^\text{corr}(\mu) \mat{\Gamma}^\text{syn}(\mu_t) \big]_{\vec{a},\vec{a}'}$.

\section{Stabilizer twirling in the teleportation circuit} \label{app:stabilizertwirling}

\subsection{Symmetric observables and stabilizer twirls} \label{app:stabilizertwirling_observables}
Consider decoded logical Pauli operators $\bar{Z}_d \coloneqq \sgn [\cos (\sqrt{\pi}\hat{q}) ]$ and $\bar{X}_d \coloneqq  \sgn [ \cos (\sqrt{\pi}\hat{p} )]$, which are equivalent to standard-binning decoded homodyne measurements (see \emph{e.g.} Ref.~\cite{nathan2024selfcorrectinggkpqubitgates}).
In the language of Eq.~\eqref{eq:standbinningPOVM}, these operators can be identified with $\hat{\Omega}^{q/p}_0-\hat{\Omega}^{q/p}_1$.
The decoded logical Paulis are an example of observables that are invariant under stabilizer shifts --- the entire set of which constitutes the \emph{logical observables} $\{\bar{O}\}$. Their expectation values lie in the $\mathbb{C}$-span of the group generated by the operators above as follows. Let $p(\mathbf{\vec{n}})$ be any probability distribution over $\mathbb{Z}^2$, then the expectation value of stabilizer-invariant observable $\bar{O}$,
\begin{align} \label{appeq:twirlingequivalence1}
    \braket{\bar{O}} &= \Tr [\op \rho \hat{O}]
    = 
    \sum_{\vec{n} \in \mathbb{Z}^2} p(\mathbf{\vec{n}}) \Tr \big[ \op \rho \big(\op{S}_Z\big)^{n_1} \big(\op{S}_X\big)^{n_2} \op O \big(\op{S}^\dagger_X\big)^{n_2} \big(\op{S}_Z\big)^{n_1} \big]
    = 
    \Tr \big[ \tilde \rho  \op{O} \big]
\end{align}
behaves as if they were computed relative to the $p(\vec{n})$-twirled state
\begin{equation} \label{appeq:twirledstate}
    \tilde{\rho} = \sum_{\vec{n} \in \mathbb{Z}^2} p(\vec{n})
    \big(\op{S}^\dagger_X\big)^{n_2} \big(\op{S}_Z^\dagger\big)^{n_1} \op \rho \big(\op{S}_Z\big)^{n_1} \big(\op{S}_X\big)^{n_2}.
\end{equation}
In the limit of a flat distribution, $p(\vec{n})= const.$, we recover the stabilizer (state) twirl.
Additional structure arises when the state can be written as noise acting on an ideal GKP state, \emph{e.g.} $\op \rho = \frac{1}{\mathfrak{n}} \hat{N}\bar{\rho}\hat{N}$. Since $\bar{\rho}$ is itself stabilizer invariant, Eq.~\eqref{appeq:twirledstate} can be written as
    \begin{align} \label{appeq:twirlingequivalence2}
    \tilde \rho & =
    \frac{1}{\mathfrak{n}}
    \sum_{\mathbf{\vec{n}} \in \mathbb{Z}^2} p(\mathbf{\vec{n}}) \Nop_{2 \vec{n}} \bar \rho \Nop_{2 \vec{n}}
\end{align}
where the damping operator is been shifted by the stabilizers
    \begin{equation} \label{appeq:shifteddamping}
        \Nop_{2 \vec{n}} = \op{P}_{2 \vec{n}} \Nop \op{P}^\dagger_{2\vec{n}} = (\op{S}_X )^{n_2} (\op{S}_Z)^{n_1} \Nop (\op{S}^\dagger_Z)^{n_1} (\op{S}^\dagger_X )^{n_2}.
    \end{equation}
In the limit of flat $p(\vec{n})$, Eq.~\eqref{appeq:twirlingequivalence2} can be understood as a channel twirl of $\Nop$ by the stabilizer group,
   \begin{equation} \label{appeq:displacementchannelgeneric}
       \mathcal{E}_\text{damp}  = \Nop \cdot \Nop \quad \longrightarrow \quad \tilde{\mathcal{E}}_\text{damp}  \coloneqq \sum_{\vec{n} \in \mathbb{Z}^2} \Nop_{2\vec{n}} \cdot \Nop_{2\vec{n}} .
    \end{equation}    
These relations are essentially those in Noh \emph{et al.}~\cite{noh2019fault}, although there the authors were not concerned with a physical understanding of when the twirling may be applied. The subtlety we highlight is that the twirling equivalences above apply only for logical observables that have stabilizer symmetries. 
In general, if approximate-GKP state preparation is part of a larger, noisy circuit, the effective operators measured on the state (Heisenberg-picture) may no longer possess these symmetries. 

Note that for the observables we started out with to define a valid qubit-algebra, it is important that the implicit displacements anti-commute. Commutation relations are preserved under symplectic transformations, so it makes sense to \emph{e.g.} squeeze the lattice by a symplectic operations $S$ to tailor to a different noise model, in which case the stabilizer group/twirl would also be changed to $\mathbb{Z}^2 \mapsto S\mathbb{Z}^2$.

\subsection{Stabilizer twirling from the teleportation circuit with SB} \label{appendix:stabtwirling}

How does the above stabilizer twirling arise in a physical setting? Here, we show how it arises naturally in the GKP teleportation circuit when SB is used.
The two-mode EPR measurement in the beam-splitter teleportation circuit can be written in terms of a GKP controlled-X gate, Eq.~\eqref{controlledX}, and squeezing on each mode using an LDU decomposition of the beam splitter~\cite{walshe2020continuousvariable},
\begin{equation} 
\begin{split} 
    \Qcircuit @C=2.0em @R=2.7em  
    {
    &\bsbal{1} &\rstick{\hspace{-0.25cm}\custommeter[$\op{q}$]{$m_q$}} \qw  \\
    &\qw       &\rstick{\hspace{-0.25cm}\custommeter[$\op{p}$]{$m_p$}} \qw 
    } 
    \raisebox{-1.3em}{\qquad  \qquad =  \quad} 
    \Qcircuit @C=1.0em @R=1.0em  
    {
    &\targ &\gate{S(\sqrt{2})} &\rstick{\hspace{-0.25cm}\custommeter[$\op{q}$]{$m_q$}} \qw  \\
    &\ctrlg{-1}{-1} & \gate{S^\dagger(\sqrt{2})}     &\rstick{\hspace{-0.25cm}\custommeter[$\op{p}$]{$m_p$}} \qw 
    } 
     \raisebox{-1.3em}{\qquad  \qquad =  \quad} 
    \Qcircuit @C=1.5em @R=2.4em  
    {
    &\targ           &\rstick{\hspace{-0.25cm}\custommeterwide[$\op{q}$]{$\sqrt{2}m_q$}} \qw  \\
    &\ctrlg{-1}{-1}  &\rstick{\hspace{-0.25cm}\custommeterwide[$\op{p}$]{$\sqrt{2} m_p$}} \qw 
    } 
    \raisebox{-1.5em}{\qquad  \qquad \quad 
    $\xrightarrow{\text{binning}}$} 
     & \Qcircuit @C=1.5em @R=2.5em 
    {
 &&\targ        &\rstick{\hspace{+0cm}\op \Omega_q} \qw  \\
 &&\ctrlg{-1}{-1} &\rstick{\hspace{+0cm}\op \Omega_p} \qw 
		} \, 
\end{split}
\qquad .
\end{equation}
Binning over the outcomes according to Eq.~\eqref{eq:standbinningPOVM} gives the circuit on the right-hand side. Due to the stabilizer invariance of $\op{\Omega}_{q/p}$, we can pull out stabilizers (and their conjugates) from these POVM elements at will and push them through the circuit. For example,
\begin{align}
    \begin{split}
    & \Qcircuit @C=1.2em @R=1.2em 
    {
 &\targ        &\gate{S_X} &\rstick{\hspace{+0cm}\op \Omega_q} \qw  \\
 &\ctrlg{-1}{-1} &\gate{S^\dagger_X} &\rstick{\hspace{+0cm}\op \Omega_p} \qw 
		} \, 
    \raisebox{-1.2em}{\qquad  \quad =  \quad } 
    \Qcircuit @C=1.2em @R=1.8em 
    {
 &\qw        &\targ        &\rstick{\hspace{+0cm}\op \Omega_q} \qw  \\
 &\gate{S_X} &\ctrlg{-1}{-1} &\rstick{\hspace{+0cm}\op \Omega_p} \qw 
		} \, 
\raisebox{-1.3em}{\qquad  \quad and \qquad}
     \Qcircuit @C=1.2em @R=1.8em 
    {
 &\targ        &\qw &\rstick{\hspace{+0cm}\op \Omega_q} \qw  \\
 &\ctrlg{-1}{-1} &\gate{S_Z} &\rstick{\hspace{+0cm}\op \Omega_p} \qw 
		} \, 
    \raisebox{-1.2em}{\qquad  \quad =  \quad } 
    \Qcircuit @C=1.2em @R=1.8em 
    {
 &\qw        &\targ        &\rstick{\hspace{+0cm}\op \Omega_q} \qw  \\
 &\gate{S_Z} &\ctrlg{-1}{-1} &\rstick{\hspace{+0cm}\op \Omega_p} \qw 
		} \,       
    \end{split}
\end{align}
\blk
Similarly, stabilizers can be pulled out of the noisy GKP Bell pairs in the circuit, Eq.~\eqref{noisybeamsplitterBellpair}. For example,
\begin{equation} 
\begin{split} 
   \Qcircuit @C=1.em @R=1.5em  
    {
    \lstick{\ket{\bar{0}}} &\gate{S^\dagger_X} &\targ     &\gate{N} &\qw 
    \\
    \lstick{\ket{\bar{+}}} &\qw &\ctrl{-1} &\gate{N} &\qw  
    } 
    \raisebox{-1.3em}{\quad =  \qquad} 
   \Qcircuit @C=1.em @R=1.5em  
    {
    \lstick{\ket{\bar{0}}} &\targ     &\gate{S^\dagger_X} &\gate{N} &\qw 
    \\
    \lstick{\ket{\bar{+}}} &\ctrl{-1} &\qw        &\gate{N} &\qw  
    } 
    \raisebox{-1.3em}{\quad and \quad \qquad}
    \Qcircuit @C=1.em @R=1.4em  
    {
    \lstick{\ket{\bar{0}}} &\gate{S^\dagger_Z} &\targ     &\gate{N} &\qw 
    \\
    \lstick{\ket{\bar{+}}} &\gate{S^\dagger_Z} &\ctrl{-1} &\gate{N} &\qw  
    } 
    \raisebox{-1.3em}{\quad =  \qquad} 
   \Qcircuit @C=1.em @R=1.5em  
    {
    \lstick{\ket{\bar{0}}} &\targ     &\gate{S^\dagger_Z} &\gate{N} &\qw 
    \\
    \lstick{\ket{\bar{+}}} &\ctrl{-1} &\qw        &\gate{N} &\qw  
    } 
\raisebox{-1.3em}{\quad . }
  \end{split}
\end{equation}
Within the teleportation circuit, the stabilizers we extracted from the above examples gather at and conjugate the first damping operator. This means that when standard binning is used, $\Nop \equiv \op{S}_X \op{S}_Z \Nop \op{S}^\dagger_Z  \op{S}^\dagger_X $ within the circuit. This is the key idea that is used in the detailed construction below for the stabilizer twirled CV channel.

\subsection{Random displacement channel from stabilizer twirling the damping map} \label{Appendix:stabtwirling}

Here we derive the expression for the CV teleportation channel arising from stabilizer twirling the damping operators. This derivation follows closely Appendix A of Noh \emph{et al.}~\cite{noh2019fault}, wherein it is shown that stabilizer twirling a damped GKP state yields a stochastically shifted state under a random displacement channel. We show that this twirling arises naturally when considering the standard binning (SB) POVM and decoder in our circuit, and we use it to find the twirled damping channel that reduces to a GRN channel when damping is small.

Consider the SB POVM, Eq.~\eqref{eq:standbinningPOVM}, with outcomes $j,k$ and SB decoding. Together, these give rise to maps 
    \begin{align}
        \op{E}_{j,k} \label{eq:jkmap}
        & \coloneqq 
          \int_{\Omega_j \times \Omega_k} d^2 \mu \, \op{K}( \mu  ) \cdot \op{K}^\dagger( \mu ) 
        = 
        \op{P}_{\vec{c}(\mu)} \bigg(
     \int_{\Omega_j \times \Omega_k} d^2 \mu \, 
    \,  \Nop \projGKP \Nop \op{D}^\dagger( \mu )
    \cdot 
    \op{D}( \mu  ) \Nop \projGKP \Nop \bigg)
    \op{P}^\dagger_{\vec{c}(\mu)} 
    \end{align}
where 
$\Omega_j = \cup_{n\in \mathbb{Z}} [(2n+j)\sqrt{\frac{\pi}{2}} - \frac{1}{2}\sqrt{\frac{\pi}{2}},(2n+j)\sqrt{\frac{\pi}{2}} + \frac{1}{2}\sqrt{\frac{\pi}{2}})$, $\Omega_k$ 
is defined similarly, and the full CV teleportation channel can be expressed as $\mathcal{E}_\tel = \sum_{jk} \op{E}_{jk} \cdot \op{E}^\dagger_{jk} $.\footnote{The binning intervals differ from the main text by $1/\sqrt{2}$, because we choose to integrate over the raw outcomes here, not the rescaled ones. Either would work; this choice is more convenient here.} 
In Eq.~\eqref{eq:jkmap}, we extracted the SB corrections, because they only depend on $j,k$: $\vec{c}(\mu) = \vec{c}( \lfloor \sqrt{2}\mu \rfloor_{\sqrt{\pi}} \text{ mod } 2) = (j,k)$. 

Because the POVM is invariant with respect to stabilizer shifts as is the GKP projector, we are free to replace $\projGKP \Nop \op{D}^\dagger(\mu)$ with $\projGKP \Nop_{2 \vec{n}} \op{D}^\dagger(\mu)$ in the integral, where $\Nop_{2 \vec{n}}$ is given in Eq.~\eqref{appeq:shifteddamping}. This is equivalent to the circuit description in Sec.~\ref{appendix:stabtwirling} above. For two successive rounds of CV teleportation, we may replace the second damping operator with $\Nop_{2 \vec{n}'}$ by the same argument (one can extract stabilizers as in Sec.~\ref{appendix:stabtwirling} in different ways such that they appear on any wire).
Summing over $\vec{n}$ and $\vec{n}'$ twirls each damping operator, as in Eq.~\eqref{appeq:displacementchannelgeneric}. 
The channel for each round of CV teleportation is described by its (doubly) twirled version,
    \begin{align} \label{eq:twirledchannel}
        \tilde{\mathcal{E}}_\tel
        &\coloneqq \int_\Omega d^2 \mu  \, \op{P}_{\corrvec} \tilde{\mathcal{E}}_\text{damp}\big(  \projGKP \op{D}^\dagger(\mu) \tilde{\mathcal{E}}_\text{damp}(  \cdot  )  \op{D}(\mu) \projGKP \big)  \op{P}^\dagger_{\corrvec} ,
    \end{align}
with the integration over $\mu$ restricted to a single domain $\Omega$ (unit cell) of size $2\sqrt{\pi} \times 2 \sqrt{\pi}$. The twirling removes support in the characteristic function of each damping operator on points not associated with the GKP lattice~\cite{conrad2024chasingshadows}.

Now we explicitly evaluate $\tilde{\mathcal E}_\text{damp}$, following the recipe of Noh and Chamberlain~\cite{noh2019fault} and fixing one mistake therein. Assuming a damping operator of arbitrary structure, $\op{N} = \int d^2 \alpha \, \chi(\alpha) \op{D}(\alpha)$, the channel twirl is
    \begin{align} 
        \tilde{\mathcal E}_\text{damp}(\cdot) 
        & = \sum_{\vec{n} \in \mathbb{Z}} \iint d^2 \alpha \, d^2 \alpha' \, \chi(\alpha) \chi^*(\alpha')  (\op{S}_Z)^{n_1} (\op{S}_X )^{n_2} \op{D}(\alpha) (\op{S}^\dagger_X )^{n_2} (\op{S}^\dagger_Z)^{n_1} \cdot (\op{S}_Z)^{n_1} (\op{S}_X )^{n_2} 
 \op{D}^\dagger(\alpha') (\op{S}^\dagger_X )^{n_2} (\op{S}^\dagger_Z)^{n_1}
        \\
        & = \sum_{\vec{n} \in \mathbb{Z}} \iint d^2 \alpha \, d^2 \alpha' \, \chi(\alpha) \chi^*(\alpha') e^{i 2\sqrt{2\pi}[(\alpha_I-\alpha'_I) n_1 - (\alpha_R - \alpha'_R) n_2]} \op{D}(\alpha)  \cdot  \op{D}^\dagger(\alpha'),
    \end{align}
where we used Eq.~\eqref{eq:displacementbraiding} to find the phase factor arising from conjugating by the stabilizers.\footnote{Reference~\cite{noh2019fault} contains an error in this phase that propagates to their final expression. It does not, however, affect their conclusion in the low damping limit.} Performing the sum over $\vec{n}$ gives delta functions from the Poisson summation formula, Eq.~\eqref{eq:DiracComb},
    \begin{align}
            \sum_{\vec{n} \in \mathbb{Z}^2}
            e^{i 2\sqrt{2\pi}[(\alpha_I-\alpha'_I) n_1 - (\alpha_R - \alpha'_R) n_2]}
            = \sqrt{\frac{\pi}{2}}
            \sum_{\vec{k} \in \mathbb{Z}^2} \delta \Big(\alpha_R - \alpha'_R - k_1 \sqrt{\tfrac{\pi}{2}} \Big) \delta \Big( \alpha_I - \alpha'_I - k_2 \sqrt{\tfrac{\pi}{2}} \Big).
    \end{align}
Using the delta functions to evaluate integrals gives
    \begin{align} 
        \tilde{\mathcal{E}}_\text{damp}( \cdot ) 
        & = \sum_{\vec{k} \in \mathbb{Z}^2}
        \int d^2 \alpha \, \chi(\alpha)\chi^* \big(\alpha - \sqrt{\tfrac{\pi}{2}}(k_1 + i k_2) \big) \op{D}(\alpha) \cdot \op{D}^\dagger \big(\alpha - \sqrt{\tfrac{\pi}{2}}(k_1 + i k_2) \big) 
        \\
        & = \sum_{\vec{k} \in \mathbb{Z}^2} \int d^2 \alpha \, \chi(\alpha)\chi^* \big(\alpha - \sqrt{\tfrac{\pi}{2}}(k_1 + i k_2) \big) e^{i\sqrt{\tfrac{\pi}{2}}(\alpha_I k_1 - \alpha_R k_2)} \op{D}(\alpha) \cdot \op{D}^\dagger(\alpha) \op{P}_{\vec{k}}.
        \label{eq:twirleddampingmap}
    \end{align}
The stabilizer-twirled damping map in Eq.~\eqref{eq:twirleddampingmap} is not in general a random displacement channel, because it is not diagonal in the displacement basis: for $\vec{k} \text{ mod } 2 \neq (0,0)$, logical Pauli shifts are applied. Also, the kernel in the integration is not guaranteed to be real and non-negative for $\vec{k} \neq (0,0)$.

Inserting the characteristic function for $\op{N} = e^{-\damppar \op{n}}$, Eq.~\eqref{eq:charfun_damping}, and simplifying gives
    \begin{align}
        \chi(\alpha)\chi^* \big(\alpha - \sqrt{\tfrac{\pi}{2}}(k_1 + i k_2) \big) 
         & = \frac{1}{\pi^2} \frac{1}{(1-e^{-\damppar})^2}  e^{\frac{-|\alpha|^2}{\tanh (\damppar/2)}}  \sum_{\vec{k} \in \mathbb{Z}^2 } 
        e^{-\frac{\pi |k_1 + i k_2|^2}{ 4 \tanh (\damppar/2)}} e^{\frac{- \sqrt{2\pi} }{ 2 \tanh (\damppar/2)}(\alpha_R k_1 + \alpha_I k_2) }.
    \end{align}
Note that one cannot perform the sum over $\vec{k}$ here to write this expression as a Siegel theta function due to the $\vec{k}$-dependence of the Pauli shifts in Eq.~\eqref{eq:twirleddampingmap}.
However, when $\tanh \frac{\damppar}{2} \ll \frac{\pi}{4}$, the $\vec{k} \neq (0,0)$ terms can be ignored, as they are severely suppressed by the Gaussian in $\vec{k}$. In this case, the twirled damping channel becomes a GRN channel,
   \begin{align} \label{eq:twirleddampchannel}
        \tilde{\mathcal{E}}_\text{damp} \longrightarrow \mathcal{E}_\text{GRN} =
        \int d^2 \alpha \, G_{\sigma^2/2} (\alpha_R, \alpha_I) \op{D}(\alpha) \cdot \op{D}^\dagger(\alpha) ,
    \end{align}
with $\sigma^2 = \tanh \frac{\damppar}{2}$.

\section{PTM for the teleportation with GRN GKP states} \label{Appendix:PTM_GRNchannel}

For the case of GRN, we can find an analytic form for the conditional PTM elements. Since it is a Pauli channel, we need only consider the diagonal elements:
\begin{equation}
    \boxed{
       \Gamma^{\text{GRN}}_{\vec a, \vec a}(\mu) 
        = \frac{e^{-\frac{1}{2 \sigma ^2} |\mu|^2 }}{ 16 \pi \sigma^2}
     \sum_{
j,k,j',k' \in \mathbb Z_2
            }
         c^{\vec{a}}_{j,k} c^{\vec{a}}_{j',k'} 
         e^{-\frac{1}{2 \sigma ^2}
        \left[ \frac{\pi}{2} (j - k')^2 + \sqrt{2 \pi} (j - k') m_1 \right]}
        \Theta\left(\mathbf{z}(\sigma ^2),\bm{\tau}(\sigma ^2)\right) 
         \delta\left(
         j \oplus k \oplus j' \oplus k'
         \right)
         }
\end{equation}
where $\oplus$ denotes addition mod 2, and  the Siegel theta function, Eq.~\eqref{eq:SiegelThetadef}, has parameters
\begin{align}
  \mat{\tau}(\sigma^2)
  &=
  \begin{pmatrix}
   \frac{i}{\sigma^2} & -1 \\
   -1 & \frac{i}{4 \sigma^2} 
  \end{pmatrix}
  \quad \quad \text{and} \quad \quad
    \vec{z}(\sigma^2)
    = 
    \frac{i }{2 \sigma^2} 
    \left(
    (j' - k)
    - \sqrt{ \frac{2}{\pi}} m_1
    ,~
    i \sigma^2 (j - k)
    -
    \frac{1}{\sqrt{2 \pi} }
    m_2
    \right).
\end{align}

\section{Standard binning gives a Pauli channel} \label{app:standardbinning_PauliChannel}

To show that the Pauli-twirled damped-GKP qubit channel $\bar{\mathcal{E}}_t$ in Eq.~\eqref{eq:channel} is a Pauli channel when the SB decoder is used, we show that it commutes with all of the GKP Paulis.
First, recognize that preceding the qubit Kraus operator $\newQ_{\vec{b}_{t}}(\mu_t)$ by a Pauli shift $\op{P}_{\vec \ell}$ %
can be written (normalization ignored)
 \begin{align}
     \newQ_{\vec{b}_{t}}(\mu_t) \op{P}_{\vec \ell}
     & = \op{P}_{\corrvect{t}} \projGKP \op{N} \op{D}^{\dagger}(\mu_t) \op{N}_{\vec{b}_{t}} \projGKP \op{P}_{\vec \ell}
     \\
     & = \op{P}_{\vec{\ell}} \op{P}_{\corrvect{t} - \vec{\ell}} \projGKP \op{N} \op{D}^{\dagger}(\mu'_t) \op{N}_{\vec{b}_{t} - \vec{\ell} } \projGKP  
  \end{align} 
where $\mu'_t = \mu_t - \sqrt{\tfrac{\pi}{2}}(\ell_1 + i \ell_2) $.
Using SB decoding, Eq.~\eqref{eq:SBdecoder}, we rewrite the penultimate shift vector as 
   \begin{align}   
       \corrvect{t} - \vec{\ell} 
       &=    
        \big( \lfloor \sqrt{2} m_q \rfloor_{\sqrt{\pi}} \text{ mod 2} - \ell_1,
        \lfloor \sqrt{2} m_p \rfloor_{\sqrt{\pi}} \text{ mod 2}  - \ell_2 \big) \\
        &=    
        \big( \lfloor \sqrt{2} m'_q + \sqrt{\pi}\ell_1 \rfloor_{\sqrt{\pi}} \text{ mod 2} - \ell_1,
        \lfloor \sqrt{2} m'_p + \sqrt{\pi}\ell_2 \rfloor_{\sqrt{\pi}} \text{ mod 2}  - \ell_2 \big) \\
        &=    
        \big( \lfloor \sqrt{2} m'_q \rfloor_{\sqrt{\pi}} \text{ mod 2}-(\ell_1\text{ mod }2-\ell_1),
        \lfloor \sqrt{2} m'_p \rfloor_{\sqrt{\pi}} \text{ mod 2} -(\ell_2\text{ mod }2-\ell_2)\big) \\
        &= \vec{c}(\mu'_t ) , \label{eq:cvectorcorrespondence}
    \end{align} 
where we used the fact that $ \lfloor x + \sqrt{\pi} n \rfloor_{\sqrt{\pi}} = \lfloor x \rfloor_{\sqrt{\pi}} + \sqrt{\pi} n $ for $n \in \mathbb{Z}$ and that $(\ell_j \text{ mod }2) - \ell_j = 0$.

Now consider the minimal Pauli-twirled CPTP qubit channel in Eq.~\eqref{eq:channel}, which is averaged over bits and syndromes.
For each bit value, we are free to choose an $\vec{\ell}$ (any representative for a specific subspace Pauli $\GKPPauli_{\vec{\ell}}$) such that the bit-averaged map has no $\vec{\ell}$-dependence in the shifted damping operator; this procedure is the same as the twirl-aware recovery above. When integrating over the syndromes, we make a change of variables to $\mu'_t$ and use Eq.~\eqref{eq:cvectorcorrespondence}. Together, these show that the channel commutes with $\GKPPauli_{\vec{\ell}}$: $\bar{\mathcal{E}}_t \circ \GKPPauli_{\vec{\ell}} = \GKPPauli_{\vec{\ell}} \circ \bar{\mathcal{E}}_t$. Since this is true for all GKP Paulis, $\bar{\mathcal{E}}_t$ is itself a Pauli channel.

\section{Logical channels from controlled-gate error correction} \label{Sec:otherflavors}

Many implementations of GKP error correction do not employ beamsplitters as we have above in Eq.~\eqref{Knillcircuitimage}. Rather, they rely on CV controlled-$X$ gates,
    \begin{align} 
        \begin{split} \label{controlledX}
        \raisebox{-1em}{$\CX^{jk}(g) \coloneqq e^{-i g \hat{q}_j \otimes \hat{p}_k} =$}
        \Qcircuit @C=1.5em @R=1.9em 
    {
 &&\ctrlg{g}{1} & \rstick{j} \qw \\
 &&\targ        & \rstick{k} \qw 
		} \, 
        \end{split}
    \end{align}
where mode $j$ is the control, mode $k$ is the target, and $g$ is the gate strength ($g=\pm1$ for square-lattice GKP). The link between the two entangling gates is the LDU decomposition of a beam splitter into two controlled gates and squeezing~\cite{walshe2021streamlined} (or, equivalently, the Bloch-Messiah decomposition of a controlled gate). Here, we review Steane and Knill EC using controlled gates and show (a) that they are equivalent to one another for damped GKP states of the form in Eq.~\eqref{eq:finitestates}, and (b) that they are \emph{inequivalent} to the beamsplitter-based Knill EC used above. Nevertheless, we lay out the path to use the formal ideas in the above sections to find the GKP logical channels for controlled-gate based GKP EC.

\subsection{Steane EC}
Steane EC measures the stabilizers by coupling an auxiliary state to the data mode using CV controlled-$X$ gates Eq.~\eqref{controlledX} and performing a homodyne measurement. Keeping the auxiliary state $\ket{\psi}$ arbitrary for now, the Kraus operators for each of these operations are~\cite{sabapathy20180Nstates, mensen2020phase}
\begin{subequations} \label{eq:controlledgate_Kraus}
\begin{align}
    \begin{split}
       & \Qcircuit @C=1.5em @R=1.9em 
    {
 &                     &\ctrlg{g}{1} & \qw  \\
 & \lstick{\ket{\psi}} &\targ        & \rstick{\hspace{-0.25cm}\custommeter[$\op{q}$]{$m_q$}} \qw 
		} \, 
    \raisebox{-1em}{\qquad  \quad =  $\psi(-g\op{q} + m_q)$} 
    \end{split}
    \quad \quad \quad \text{and} 
    \begin{split}
   & \Qcircuit @C=1.5em @R=1.9em 
    {
 &                     &\targ         & \qw  \\
 & \lstick{\ket{\psi}} &\ctrlg{g}{-1} & \rstick{\hspace{-0.25cm}\custommeter[$\op{p}$]{$m_p$}} \qw 
		} \, 
    \raisebox{-1em}{\qquad  \quad =  $\tilde{\psi}(g\op{p} + m_p)$}
    \end{split}
\end{align}
\end{subequations}
where $\psi(s) = \inprodsubsub{s}{\psi}{q}{}$ and $\tilde{\psi}(t) = \inprodsubsub{t}{\psi}{p}{}$ are the position and momentum wavefunctions of the auxiliary state, related by a Fourier transform, $\tilde{\psi}(t) = \frac{1}{\sqrt{2 \pi}} \int ds \, \psi(s) e^{-i s t}$.
Syndrome extraction uses two back-to-back stabilizer measurements with $g=1$ controlled gates (circuit labels dropped), 
 \begin{align} \label{Steanecircuit}
        \begin{split}
        \Qcircuit @C=0.45cm @R=0.5cm {
    &                      &\ctrl{1} &\qw                   &\qw &\qw &\qw &\qw &\qw                   &\targ     &\qw                   \\
    &\lstick{\ket{\psi_a}} &\targ    &\rstick{\hspace{-0.25cm}\custommeter[$\op{q}$]{$m_q$}} \qw  &    &    &    &    &\lstick{\ket{\psi_b}} &\ctrl{-1} &\rstick{\hspace{-0.25cm}\custommeter[$\op{p}$]{$m_p$}} \qw  \\
    }
        \end{split}
    \end{align}
The Kraus operator for this circuit is 
    \begin{align} \label{eq:SteaneGeneric}
        \op{K}^\text{syn}_\text{Steane}(\mu) & =\tilde{\psi}_b(\op{p} + m_p) \psi_a(-\op{q} + m_q)   
        = \op{D} \big( \tfrac{\mu^*}{\sqrt{2}} \big) \tilde{\psi}_b(\op{p}) \psi_a(-\op{q})  \op{D}^\dagger \big( \tfrac{\mu^*}{\sqrt{2}} \big),
    \end{align}
where the shifts are extracted in the second equality, and $\mu = m_q + i m_p$. The displacement amplitude here differs from the beam splitter case, which contains inherent squeezing and a sign difference on a CX gates in its LDU decomposition~\cite{walshe2020continuousvariable}. Performing the stabilizer measurements in the other order gives a different Kraus operator with the order of $\psi_a(-\op{q})$ and $ \tilde{\psi}_b(\op{p})$ reversed.
 
To see how this works for GKP error correction, consider ideal auxiliary GKP states $\ket{\psi_a} = \ket{\bar{+}}$ and $\ket{\psi_b} = \ket{\bar{0}}$ with respective wavefunctions $\psi_a(s) = \inprodsubsub{s}{\bar{+}}{q}{} \propto \Sh_{\sqrt{\pi}}(s)$ and $\tilde{\psi_b}(t) = \inprodsubsub{t}{\bar{0}}{p}{} \propto \Sh_{\sqrt{\pi}}(t)$.
Each stabilizer measurement projects the state onto an eigenspace of modular position or modular momentum described by the projectors $\Sh_{\sqrt{\pi}}(-\op{q} + m_q)$ and $\Sh_{\sqrt{\pi}}(\op{p} + m_p)$. Together, the resulting operation on the data mode is a shifted GKP projector~\cite{baragiola2019all},
    \begin{align}
        \op{K}^\text{syn}_\text{Steane}(\mu) 
         &= \op{D} \big( \tfrac{\mu^*}{\sqrt{2}} \big) \op{\Pi}_{\GKP} \op{D}^\dagger \big( \tfrac{\mu^*}{\sqrt{2}} \big),
    \end{align} 
where we extracted the shifts, used $\Sh_{\sqrt{\pi}}(\op{q})= \Sh_{\sqrt{\pi}}(-\op{q})$, and then employed Eq.~\eqref{eq:Shacommutation}. 
The order of the stabilizer measurements does not matter for ideal GKP states, since the Kraus operators commute, Eq.~\eqref{eq:Shacommutation}.
After syndrome extraction, the SB decoder implements a non-logical shift by the $\sqrt{\pi}$-modular parts of the syndromes, leaving behind an integer shift corresponding to a logical Pauli: $\op{D}^\dagger \big( \tfrac{\{ m_q \}_{\sqrt{\pi}} - i \{ m_p \}_{\sqrt{\pi}} }{\sqrt{2}} \big) \op{D} \big( \tfrac{\mu^*}{\sqrt{2}} \big) = \op{P}_{\corrvec}$~\cite{harris2024}\blk. A more general decoder can restore the GKP codespace with a full shift $\op{D}^\dagger \big( \tfrac{\mu^*}{\sqrt{2}} \big)$ followed by the logical shift of its choice. 

Now consider damped-GKP states $\ket{\psi_a} = \ket{\bar{+}_\damppar}$ and $\ket{\psi_b} = \ket{\bar{0}_\damppar}$.
For high quality states, the wavefunctions, $ \psi_a(s) = \inprodsubsub{s}{\bar{+}_{\damppar}}{q}{} $ and $\tilde{\psi}_b(t) = \inprodsubsub{t}{\bar{0}_{\damppar}}{p}{}$, are given by a train of Gaussian pulses separated by $\sqrt{\pi}$, each with variance $\Delta^2 = -10\log_{10} \damppar$, multiplied by a broad Gaussian envelope with variance $\Delta^{-2}$. To describe this, we define a function
$\mathfrak{B}_{\Delta}(x) 
    \coloneqq 
    G_{\Delta^{-2}}(x) P_{\Delta^2} (x)$
where the Gaussian $G_{\sigma^2}(x)$ is given in  Eq.~\eqref{eq:GaussianFunc}, and 
\begin{align}
    P_{\sigma^2}(x) 
    & \coloneqq \sum_{n \in \mathbb{Z}} G_{\sigma^2}(x - n \sqrt{\pi} ) 
    =
    \theta_{\sqrt{\pi}} \big( x, 2 \pi i \sigma^2 \big)
\end{align}
is a Jacobi theta function with period $\sqrt{\pi}$ as defined in Ref.~\cite{mensen2020phase}. 
With these GKP states, the Steane syndrome extraction circuit 
sequentially implements approximate projectors onto modular position and momentum via the Kraus operator
\begin{align}  \label{eq:SteaneapproxEC}
    & \op{K}^\text{syn}_\text{Steane}(\mu) \approx \mathfrak{B}_{\Delta} (\hat{p}+m_p)\mathfrak{B}_{\Delta}(-\hat{q}+m_q) 
\end{align}
Each approximate stabilizer measurement describes a simultaneous weak Gaussian measurement and a noisy modular measurement of a quadrature. Consider $\mathfrak{B}_{\Delta}(-\hat{q}+m_q)$. The operator $G_{\Delta^{-2}} ( -\hat{q} + m_q )$ gives a weak ($\Delta \ll 1$) projection of $-\op{q}$ onto $m_q$ with sharpness parametrized by $\Delta^{-2}$. Likewise, $P_{\Delta^2} (-\op{q} + m_q)$ is the noisy $\sqrt{\pi}$-periodic measurement projector along axis $-\op{q}$ with local Gaussian peak-width of $\Delta^2$.
The weak homodyne measurements of each quadrature $(G)$ do not commute with each other, nor do they commute with the conjugate stabilizer measurement ($P$). Interestingly, the noisy modular projectors commute, $[P_{\Delta^2}(\hat{p}), P_{\Delta^2} (\hat{q})] = 0$, indicating that they are compatible observables, see Appendix~\ref{appendix:thetafunctions}. 

When a state traverses the circuit, it experiences an exchange of envelope-sharpening along the $p$-axis accompanied by a peak-broadening along the $q$-axis in the first step and envelope-sharpening along the $q$-axis accompanied by peak-broadening along the $p$-axis in the second step, leading to oscillatory behaviour of effective squeezing parameters as observed in Ref.~\cite{terhal_review}. This effect, and the general inseparability of the ``noise" from the circuit complicates analysis of Steane error correction~\cite{glancy2006}. Nevertheless, we can insert the definition of $\mathfrak{B}_\Delta(x)$ into Eq.~\eqref{eq:SteaneapproxEC} to find 
\begin{align} 
    \op{K}^\text{syn}_\text{Steane}(\mu) 
    & = \op{D} \big( \tfrac{\mu^*}{\sqrt{2}} \big) G_{\Delta^{-2}}(\hat{p}) P_{\Delta^2} (\hat{p}) P_{\Delta^2} (\hat{q}) G_{\Delta^{-2}} (\hat{q})  \label{eq:SteaneapproxEC2}
 \op{D}^\dagger \big( \tfrac{\mu^*}{\sqrt{2}}\big)  
\end{align}
where we have extracted the shifts, swapped the order of the $G$ and $P$ functions in $\op{p}$, and used the parity symmetry of $\mathfrak{B}_{\Delta}(x)$. The noisy modular projectors commute and form a finite-squeezing approximation to the GKP projector, given by a GRN channel acting on $\projGKP$,
    \begin{align}
        P_{\Delta^2} (\hat{p}) P_{\Delta^2} (\hat{q}) = \mathcal{E}_\text{GRN}(\projGKP), 
    \end{align}
 with covariance matrix $\mat \Sigma = \text{diag}( \sigma_q^2/2 , \sigma_p^2/2 )$. See Appendix~\ref{appendix:thetafunctions} for more details.

\subsection{Knill EC}
An alternative to controlled-gate Steane EC is the related Knill scheme \cite{KnillEC}, where the data mode is teleported through a GKP Bell pair. This is just like the beamsplitter-based Knill circuit in Eq.~\eqref{Knillcircuitimage}, except here the circuit uses CV controlled gates.
We consider a generalization of the Bell pair as two arbitrary pure states and a weight-1 gate. The Knill syndrome-extraction circuit for this setting is
\begin{align} \label{Knillcircuit}
        \begin{split}
        \Qcircuit @C=0.45cm @R=0.5cm {
    &                      &\qw      &\qw &\targ     &\rstick{\hspace{-0.25cm}\custommeter[$\op{q}$]{$m_q$}} \qw  \\
    &\lstick{\ket{\psi_a}} &\ctrl{1} &\qw &\ctrl{-1} &\rstick{\hspace{-0.25cm}\custommeter[$\op{p}$]{$m_p$}} \qw  \\
    &\lstick{\ket{\psi_b}} &\targ    &\qw &\qw       &\qw                   \\
    }
        \end{split}
    \end{align}
\blk
Using circuit identities, we show equivalence of Steane and Knill syndrome extraction circuits. First, rearrange the CV SWAP identity~\cite{walshe2023equivalent},
\begin{align} \label{swapcircuitidentity}
    \Qcircuit @C=1.1em @R=1.4em @! 
         {
         & \ctrl{1} &  \qw \\
         & \targ &  \qw
  	 } 
        & \raisebox{-1.1em}{\quad $\, = \, $~~} 
        \Qcircuit @C=0.5cm @R=0.45cm 
        {
        & \qw        & \targ          & \ctrl{1} & \qswap      & \qw \\
        & \gate{F^2} & \ctrlg{-1}{-1} & \targ    & \qswap \qwx & \qw
        }
\end{align}
Measuring one of the modes in either position or momentum gives the equivalences,
\begin{align} \label{swapwithmeasurement}
    \Qcircuit @C=1em @R=1.4em @! 
         {
         &\targ     &\rstick{\hspace{-0.25cm}\custommeter[$\op{q}$]{$m_q$}} \qw \\
         &\ctrl{-1} &\qw
  	 } 
        &\quad \quad \raisebox{-1.1em}{\quad $\, = \, $~~} 
        \Qcircuit @C=0.4cm @R=0.45cm 
        {
        &\gate{F^2} &\ctrlg{-1}{1} &\gate{X(m_q)}           &\qw \\
        &\qw        &\targ         &\rstick{\hspace{-0.25cm}\custommeter[$\op{q}$]{$m_q$}} \qw &
        }
        \quad \quad \raisebox{-1.1em}{\text{and}}  \quad \quad 
       \Qcircuit @C=1em @R=1.4em @! 
         {
         &\ctrl{1} &\rstick{\hspace{-0.25cm}\custommeter[$\op{p}$]{$m_p$}} \qw \\
         &\targ    &\qw
  	 } 
        &\quad \raisebox{-1.1em}{=} \quad 
        \Qcircuit @C=0.4cm @R=0.45cm 
        {
        &\qw        &\targ          &\gate{Z(-m_p)}          &\qw \\
        &\gate{F^2} &\ctrlg{-1}{-1} &\rstick{\hspace{-0.25cm}\custommeter[$\op{p}$]{$m_p$}} \qw &
        }
\end{align}
where $\brasub{s}{q_2} e^{- i g \op{p}_1 \otimes \op{q}_2} = e^{- i g s \op{p}_1} = \op{X}_1(gs)$ and $\brasub{t}{p_2} e^{- i g  \op{q}_1 \otimes \op{p}_2} = e^{- i g t \op{q}_1} = \op{Z}_1(-gt)$.
Re-ordering the commuting controlled gates in the Knill circuit, Eq.~\eqref{Knillcircuit}, and applying these circuit identities one-by-one gives the equivalence between Knill and Steane circuits with a few extra decorations~\cite{conradthesis},
\begin{align} \label{appeq:Knillcircuit_equiv}
    \begin{split}
    & \Qcircuit @C=0.45cm @R=0.5cm {
    &\targ     &\qw &\qw      &\rstick{\hspace{-0.25cm}\custommeter[$\op{q}$]{$m_q$}} \qw \\
    &\ctrl{-1} &\qw &\ctrl{1} &\rstick{\hspace{-0.25cm}\custommeter[$\op{p}$]{$m_p$}} \qw \\
    &\qw       &\qw &\targ    &\qw                   \\
    }
    \quad \quad \quad \quad \raisebox{-1.3em}{ = \quad }
        \Qcircuit @C=0.45cm @R=0.35cm {
    &\qw &\ctrl{1} &\qw                     &\qw &\qw &\qw &\qw &\targ          &\qw &\gate{D^\dagger \big(\frac{\mu^*}{\sqrt{2}} \big)} &\gate{F^2} & \qw \\
    &\qw &\targ    &\rstick{\hspace{-0.25cm}\custommeter[$\op{q}$]{$m_q$}} \qw &    &    &    &   \gate{F^2} &\ctrl{-1} &\rstick{\hspace{-0.25cm}\custommeter[$\op{p}$]{$m_p$}} \qw &                           &           &     \\
    }
    \end{split}
    \end{align}
with $\op{X}(s) \op{Z}(t) = e^{-i st} \op D( \frac{s + it}{\sqrt{2}} )$ (the phase is unimportant), and $ \op{F}^2 = e^{i\pi \op{n}}$ is a parity operator.
Attaching auxiliary states as in Eq.~\eqref{Knillcircuit} and using Eqs.~\eqref{eq:controlledgate_Kraus}, we get the Kraus operator
\begin{align} \label{eq:KnillKrausOp_controlledgates}
    \op{K}^\text{syn}_\text{Knill}(\mu) 
    & = \op{F}^2 \tilde{\psi}_b(-\op{p}) \psi_a(-\op{q}) \op{D}^\dagger \big( \tfrac{\mu^*}{\sqrt{2}} \big) ,
\end{align}
For parity invariant auxiliary state $\ket{\psi_b} = \op{F}^2 \ket{\psi_b}$, \emph{e.g.} a damped GKP state, the Steane and Knill circuits are equivalent up to a final displacement and parity operation. Given auxiliary states with parity symmetry, such as damped GKP states, the Knill EC circuit  $\op{K}^\text{syn}_\text{Knill}(\mu) = \op{F}^2 \op{D}^\dagger \big( \tfrac{\mu^*}{\sqrt{2}} \big) \op{K}^\text{syn}_\text{Steane}(\mu)$. 
The equivalence of the circuits also allows for straightforward adaption of decoders tailored to Steane EC.
Extracting the syndromes using a different Knill circuit with $q$ and $p$ measurements and their respective controlled-$X$ gates swapped gives a different Kraus operator with $\psi_a(\op{q})$ and $ \tilde{\psi}_b(\op{p})$ swapped---just like the equivalent Steane circuit, see above.

The controlled-gate Steane circuit was examined in Ref.~\cite{terhal_review}, where corrective displacements were chosen to exploit the weak measurement effect and actively recenter the state in order to (amongst other things) lower susceptibility to photon-loss errors and Kerr-nonlinearities in addition to returning the state to code space. As shown above, the controlled-gate Knill circuit naturally implements this feature, just like its beamsplitter counterpart in Eq.~\eqref{Knillcircuitimage}. Thus, we expect Knill EC to have lower resource- and time-costs as compared to Steane-EC scheme. 
A consequence of Knill-Steane equivalence is that 
performing Knill EC with the approximate GKP state described above gives effectively the same action on the input state. 
Thus the drawback of this circuit is the same: for damped GKP, finite squeezing in the auxiliary states is interwoven into the Knill EC Kraus operator, see Eq.~\eqref{eq:SteaneapproxEC}. Derivation of a GKP logical channels with approximate GKP auxiliary states using this circuit will be of the same form as for Steane EC.

\subsection{Towards logical channels}
Analysis of a chain of controlled gate Steane EC circuits with damped GKP auxiliary states proceeds similar to Fig.~\ref{fig:general-teleportation-scheme}, with some notable differences. In place of damped projectors, $\Nop \projGKP \Nop$, that arise from beamsplitter teleportation, we find instead the Gaussian-sandwiched, smeared projectors $G_{\Delta^{-2}}(\hat{p}) \mathcal{E}_\text{GRN}(\projGKP) G_{\Delta^{-2}}(\hat{q}) $. Identifying qubit maps arising between GKP projectors is still possible, but further work is necessary to determine the influence of the Gaussian weak measurements and of $\mathcal{E}_\text{GRN}(\projGKP)$, because the shifts from the GRN channel acting on both sides of the projector will lead to classical correlations between adjacent qubit maps. Other techniques to look at the effective logical projection, such as those from Ref.~\cite{marqversen2025}, may also prove useful.

\section{Theta-function commutation relations} \label{appendix:thetafunctions}
Using the relations, $\outprodsubsub{s}{s}{q}{q} = \delta(\op{q} + s) = \op{X}^\dagger(s) \delta(\op{q}) \op{X}(s)$ and $\outprodsubsub{t}{t}{p}{p} = \delta(\op{p} + t) = \op{Z}^\dagger(t) \delta(\op{p}) \op{Z}(t)$, we can express diagonal operators in $\op{q}$ and $\op{p}$ in terms of shift channels with distributions $f(x)$ and $g(x)$,
    \begin{align}
        f(\op q) & 
        = \int ds \, f(s) \outprodsubsub{s}{s}{q}{q}
        = \int ds \, f(s) \op{X}^\dagger(s) \delta( \op q) \op{X}(s) 
        \quad \text{and} \quad
        g(\op p) 
        = \int dt \, g(t) \outprodsubsub{t}{t}{p}{p}
        = \int dt \, g(t) \op{Z}^\dagger(t) \delta( \op p) \op{Z}(t).
    \end{align}
This allows us to express noisy modular projectors (Jacobi theta functions of $\op{q}$ and $\op{p}$) as
\begin{align}
     \theta_{\sqrt{\pi}}(\op{q}, 2 \pi i \sigma_q^2) 
     & = \int ds \, G_{\sigma_q^2}(s) \op{X}^\dagger(s) \Sh_{\sqrt{\pi}}(\op{q}) \op{X}(s) 
     \quad \text{and} \quad
      \theta_{\sqrt{\pi}}(\op{p}, 2 \pi i \sigma_p^2) 
     = \int dt \, G_{\sigma_p^2}(t)  \op{Z}^\dagger(t) \Sh_{\sqrt{\pi}}(\op{p}) \op{Z}(t) \, .
\end{align}
With the help of Eq~\eqref{eq:Shacommutation}, it is straightforward to show that
\begin{align}
    & \theta_{\sqrt{\pi}}(\op{q}, 2 \pi i \sigma_q^2) \theta_{\sqrt{\pi}}(\op{p}, 2 \pi i \sigma_p^2) 
    = \theta_{\sqrt{\pi}}(\op{p}, 2 \pi i \sigma_q^2) \theta_{\sqrt{\pi}}(\op{q}, 2 \pi i \sigma_p^2) 
    = \int d^2 \alpha \, G_{\frac{1}{2}\mat{\Sigma}}(\alpha_R,\alpha_I)  \op{D}^\dagger \big(\alpha \big) \projGKP \op{D} \big(\alpha \big) = \mathcal{E}_\text{GRN} ( \projGKP )
\end{align}
with $\mat \Sigma = \text{diag}(\sigma_q^2 , \sigma_p^2)$.
That is, a product of two noisy modular projectors is equivalent to a GRN channel, Eq.~\eqref{eq:GRN}, acting on an ideal GKP projector $\projGKP$.
This relations hold even when the theta functions have different variances, $\sigma_q^2 \neq \sigma_p^2$, as long as their periods are both $\sqrt{\pi}$. Moreover, the relation does not rely on the distributions being Gaussian. As long as the original diagonal operators in $\op{q}$ and $\op{p}$ can be written individually as displacement channels on $\Sh_{\sqrt{\pi}}(\op{q})$ and $\Sh_{\sqrt{\pi}}(\op{p})$, the resulting operators commute and can be expressed as a displacement channel acting on $\projGKP$.

\newpage

\newcommand{\lr}[1]{\left(#1\right)}
\newcommand{\lrq}[1]{\left[#1\right]}
\newcommand{\Z}{\mathbb{Z}}

\end{document}

%% file: preamble.tex
\newcommandx*\ctrlg[2]{\control \ar @{-}^{#1} [#2,0] \qw}

\let\originalleft\left
\let\originalright\right
\renewcommand{\left}{\mathopen{}\mathclose\bgroup\originalleft}
\renewcommand{\right}{\aftergroup\egroup\originalright}

\theoremstyle{plain}

\theoremstyle{remark}

\DeclareMathOperator{\sgn}{sgn}
\DeclareMathOperator{\tr}{tr}
\DeclareMathOperator{\Tr}{Tr}

\renewcommand{\Im}{\operatorname{Im}}

\DeclareMathOperator{\sech}{sech}
\DeclareMathOperator{\csch}{csch}

\makeatletter

\newcommand{\ringplus}{\mathbin{\text{\@ringplus}}}

\newcommand{\@ringplus}{%
  \ooalign{\hidewidth\raise1.3ex\hbox{\tiny$\circ$}\hidewidth\cr$\m@th+$\cr}%
}

\newcommand{\ringminus}{\mathbin{\text{\@ringminus}}}

\newcommand{\@ringminus}{%
  \ooalign{\hidewidth\raise0.9ex\hbox{\tiny$\circ$}\hidewidth\cr$\m@th-$\cr}%
}
\makeatother

\newcommand{\tp}[0]{\mathrm{T}}

\DeclareFontFamily{U}{wncy}{}
\DeclareFontShape{U}{wncy}{m}{n}{<->wncyr10}{}
\DeclareSymbolFont{mcy}{U}{wncy}{m}{n}
\DeclareMathSymbol{\Sh}{\mathord}{mcy}{"58}

\newcommandx*\bsbal[3][1=black, 3=->]{\ar @[#1]@{#3} [#2,0] \qw}

\newcommandx*\varbs[4][1=black, 3=\theta,4=->]{\ar @[#1]@{#4}^{#3} [#2,0] \qw}

\newcommand{\negspace}{\!}
\newcommand{\lsub}[2]{{\protect\vphantom{#1}}_{#2} \negspace {#1}}
\newcommand{\rsub}[2]{{#1} \negspace {\protect\vphantom{#1}}_{#2}}
\newcommand{\lrsub}[3]{{\protect\vphantom{#1}}_{#2} \negspace {#1} \negspace {\protect\vphantom{#1}}_{#3}}

\newcommand{\ketsub}[2]{\rsub {\ket{#1}} {#2}}
\newcommand{\brasub}[2]{\lsub {\bra{#1}} {#2}}

\newcommand{\pbra}[1]{\brasub{#1} p}
\newcommand{\qbra}[1]{\brasub{#1} q}

\newcommand{\qket}[1]{\ketsub{#1} q}

\newcommand{\inprod}[2]{\left\langle {#1} | {#2} \right\rangle}

\newcommand{\inprodsubsub}[4]{\lrsub {\inprod{#1}{#2}} {#3} {#4}}

\newcommand{\outprod}[2]{\ket {#1}\!\bra {#2}}

\newcommand{\outprodsubsub}[4]{\ketsub {#1}{#3} \brasub{#2}{#4}}

\newcommand{\abs}[1]{\left\lvert{#1}\right\rvert}

\newcommand{\norm}[1]{\left\lVert{#1}\right\rVert}

\newcommand{\op}[1]{\hat{#1}}

\newcommand{\mat}[1]{\bm{\mathrm{#1}}}

\renewcommand{\vec}[1]{\bm{\mathrm{#1}}}

\newcommand{\controlled}[1]{\op{\mathrm{C}}_{#1}}

\newcommand{\CX}[0]{\controlled X}

\newcommand{\blk}{\color{black}}

\usepackage{graphicx}
\graphicspath{{./figures/}}